\documentclass[12pt]{article}
\usepackage{setspace}
\usepackage{caption}

\usepackage{amsthm,amsmath,amssymb,amsfonts,delarray}
\usepackage[dvipdfmx]{graphicx}
\usepackage{fullpage}
\usepackage[round]{natbib}
\usepackage{xcolor}

\usepackage[toc,page,titletoc]{appendix}
\usepackage[all]{xy}
\usepackage{here}
\usepackage{listings}
\usepackage{mathrsfs}
\usepackage{mathtools} 
\mathtoolsset{showonlyrefs}
\usepackage{comment}
\usepackage{framed}
\usepackage{fullpage}
\usepackage[dvipdfmx]{graphicx}
\usepackage{xcolor}
\usepackage{times}
\usepackage{bm}
\usepackage{natbib}
\usepackage{amsmath,amssymb,bbm}
\usepackage{enumerate}
\usepackage{xspace}
\usepackage{array}
\newcolumntype{M}[1]{>{\centering\arraybackslash}m{#1}}

\def\WAIC{\textsc{waic}\xspace}
\def\PCIC{\textsc{pcic}\xspace}
\def\TIC{\textsc{tic}\xspace}
\newcommand{\Tr}{\mathrm{tr \,}}

\newcommand{\Ep}{\mathbf{E}}
\newcommand{\Eppos}{\Ep_{\mathrm{pos}}}
\newcommand{\Covpos}{\mathrm{Cov}_{\mathrm{pos}}}
\newcommand{\Varpos}{\mathrm{Var}_{\mathrm{pos}}}
\newcommand{\Kpos}{{\mathcal{K}_{\mathrm{pos}}}}

\newcommand{\estEppos}{\mathbf{E}_{\mathrm{pos}}^{\Theta^\mathrm{M}}}
\newcommand{\estCovpos}{\mathrm{Cov}_{\mathrm{pos}}^{\Theta^\mathrm{M}}}
\newcommand{\estKpos}{\mathcal{K}_{\mathrm{pos}}^{\Theta^\mathrm{M}}}

\newcommand{\estEpX}{\Ep_{X^n}}
\newcommand{\estCovX}{\mathrm{Cov}_{X^n}}

\newcommand{\EpposP}{{\Ep}^{0}_{\mathrm{pos}}}
\newcommand{\CovposP}{\mathrm{Cov}^{0}_{\mathrm{pos}}}

\newcommand{\EpX}{\Ep_{X}}
\newcommand{\CovX}{\mathrm{Cov}_X}
\newcommand{\VarX}{\mathrm{Var}_X}

\renewcommand{\tilde}{\widetilde}
\newcommand{\logp}[2]{\ell(#1 ; #2)}
\newcommand{\logpstar}[2]{\ell^*(#1 ; #2)}
\newcommand{\sproj}{\mathcal{L}}
\newcommand{\ctext}[1]{\raise0.2ex\hbox{\textcircled{\scriptsize{#1}}}}

\newtheorem{remark}{Remark}

\usepackage{dcolumn}
\usepackage{natbib}
\usepackage{subcaption}
\usepackage{enumerate}
\usepackage{booktabs}

\usepackage{etoolbox}

\begin{document}

\title{\bf W-Kernel and Its Principal Space \\ for Frequentist Evaluation of Bayesian Estimators}

\author{Yukito Iba${}^{1,2}$ \\ 
${}^1$~Zen University (PhiMAS Lab)\\
${}^2$~The Institute of Statistical Mathematics}

\date{}

\maketitle

\noindent E-mails: \verb+pt_yukito_iba@zen.ac.jp, iba@ism.ac.jp+.

\noindent Address for Correspondence:
Zen University, 3-12-11 Shinjuku, Zushi, Kanagawa 249-0007, Japan. 
The Institute of Statistical Mathematics,
10-3 Midori cho, Tachikawa, Tokyo 190-8562, Japan. 

\medskip
\medskip

\noindent {\bf ABSTRACT:}\\
Evaluating the variability of posterior estimates is a key aspect of Bayesian model assessment. In this study, we focus on the posterior covariance matrix $W$, defined through the log-likelihoods of individual observations. Previous studies, notably \cite{MacEachern2002} and \cite{Thomas2018}, examined the role of the principal space of $W$ in Bayesian sensitivity analysis. Here, we show that the principal space of $W$ is also central to frequentist evaluation, using the recently proposed Bayesian infinitesimal jackknife (Bayesian IJ) approximation (\cite{Bayesian_IJK}) as a key tool. We further clarify the relationship between $W$ and the Fisher kernel, showing that a modified version of the Fisher kernel can be viewed as an approximation to $W$. Moreover, the matrix $W$ itself can be interpreted as a reproducing kernel, which we refer to as the $W$-kernel. Based on this connection, we investigate the relation between the $W$-kernel formulation in the data space and the classical asymptotic formulation in the parameter space. We also introduce the matrix $Z$, which is effectively dual to $W$ in the sense of PCA; this formulation provides another perspective on the relationship between $W$ and classical asymptotic theory. In the appendices, we explore approximate bootstrap methods for posterior means and show that projection onto the principal space of $W$ facilitates frequentist evaluation when higher-order terms are included. In addition, we introduce incomplete Cholesky decomposition as an efficient method for computing the principal space of $W$ and discuss the concept of representative subsets of observations.

\medskip

\vfill

\newpage

\section{Introduction}
\label{sec:introduction}

Sensitivity analysis and model checking play crucial roles in Bayesian analysis (\cite{CL3,BDA3}). Herein, let us consider two typical examples for illustration:

\begin{itemize}
\item[] {\bf Assessing sensitivity to changes in observations:}
 The robustness of a Bayesian estimator is quantified based on the response to a hypothetical change in observations. The rate of change caused by the change in weights of the observations is referred to as {\it local case sensitivity} (\cite{ MillarandStewart(2007)}). 

\vspace{0.5ex}

\item[]{\bf Frequentist evaluation via bootstrap replication of observations:}
The performance of a Bayesian estimator is evaluated using artificial sets of data generated by sampling with replacement from the original data. In the frequentist context, each of the artificial sets is considered a surrogate for a novel sample from the original population.
\end{itemize}

A key challenge in these tasks is the repeated computation of estimates for modified datasets. If we compute these estimates using time-consuming methods such as the Markov chain Monte Carlo (MCMC) method, it results in a large computational burden. Thus, approximate methods that replace repeated computations with a single run of the MCMC using the original data are highly desirable. Examples of the studies in which this was the case include \cite{Gustafson_1996, Bradlow_1997, MacEachern_2000, Perez_etal_2006,MillarandStewart(2007), vanderLinde(2007)} for sensitivity analysis, and \cite{Lee_2017} for the bootstrap method. We may also reference \cite{Gelfand_Dey_Chang_1992, P_1997, Watanabe_2010_b,  Vehtari_etal_2017,  Millar, Iba_Yano_arXiv2} for leave-one-out cross-validation, \cite{Efron_2015, Bayesian_IJK} for frequentist covariance and interval estimation, and \cite{McVinish_2013} for the Bayesian p-values.

The aim of this study is to explore a framework that reveals the underlying structure of these problems. In terms of sensitivity analysis, \cite{MacEachern2002} and \cite{Thomas2018} (see also \cite{Bradlow_1997}) proposed that the posterior covariance matrix of log case-deletion weights and its principal space can be a useful tool for summarizing the results of the sensitivity analysis of Bayesian models. The proposed method is termed ``Bayesian explanatory analysis'' in \cite{MacEachern2002}.

In the simplest setting of IID models, the matrix $C$ in \cite{Thomas2018} reduces to the matrix whose components are the posterior covariances between the log-likelihoods of observations, which we denote by $W$ and focus on in the rest of the paper.

In this paper, this idea is extended beyond exploratory analysis. We show that the principal space of the matrix $W$ is not only useful in assessing sensitivity but also in exploring frequentist uncertainty under IID sampling of observations. The key tool for exploring these properties is the Bayesian infinitesimal jackknife approximation (Bayesian IJ), recently proposed in \cite{Bayesian_IJK}.

Another contribution of this paper is to show that the matrix $W$ is approximated by a modified version of the Fisher Kernel (\cite{NIPS1998_db191505,Kernel_Schoelkopf}) and the neural tangent kernel (\cite{NIPS2018_NTK}) in an asymptotic limit; the matrix $W$ itself is also interpreted as a reproducing kernel and termed ''W-kernel'' in this context. The derived approximation explains the eigenvalue structure of the matrix $W$ from the viewpoint of classical asymptotic theory.    

The following sections are organized as follows:
\begin{itemize}
    \item Sec.~\ref{sec:pre} provides an overview of the Bayesian IJ and introduces the common settings and notations.
    \item Sec.~\ref{sec:essential} introduces the $W$ matrix and the $W$-kernel, and shows that projection onto the principal space of $W$ is relevant for the frequentist evaluation procedure, which forms the first main contribution of the paper.
    \item Sec.~\ref{sec:essential_example} illustrates typical eigenvalue structures of the matrix $W$ through examples, including Weibull fitting of life-span data, regression, and smoothing.   
    \item Sec.~\ref{sec:Fkernel} begins with an approximate relation between the $W$-kernel and the Fisher kernel, which leads to the second main result of the paper. It establishes an asymptotic relation between the eigenvalues of $W$ and those of $\hat{\mathcal{J}}^{-1/2}\hat{\mathcal{I}}\hat{\mathcal{J}}^{-1/2}$ in a simple parametric model with weak priors, consistent with the findings of Sec.~\ref{sec:essential_example}.
    \item Sec.~\ref{sec:dual2} introduces the $Z$ matrix, which can be regarded as the dual of $W$ (precisely speaking, of an adjusted form $W^c$ of $W$) in the sense of PCA. Using this, we provide an alternative derivation of the asymptotic relation between the eigenvalues of $W$ and $\hat{\mathcal{J}}^{-1/2}\hat{\mathcal{I}}\hat{\mathcal{J}}^{-1/2}$.
    \item Sec.~\ref{sec:conclusion} concludes with a summary and directions for future research.
    \item Appendix~\ref{sec:app:derivation} derives the formulae used in the main text. Appendix~\ref{sec:kernel_2} provides additional details on the $W$-kernel. Appendix~\ref{sec:cholesky} discusses the Cholesky decomposition as a computational tool for handling the matrix $W$ and introduces the notion of a representative set of observations. Appendix~\ref{sec:app:boot_example} presents examples of the approximate bootstrap and tests the dimensional reduction approach proposed in Sec.~\ref{sec:essential}.
\end{itemize}

\section{Preliminaries}
\label{sec:pre}

\subsection{Settings and Notations}

A set of observations is denoted as $X^{n}=(X_1,\ldots, X_n)$.  
The symbols $X$ and $X^\prime$ without superscripts or subscripts represent generic observations, typically elements sampled from the population or test points in the evaluation of the kernel function. The Bayesian framework is considered with a posterior distribution:
\begin{align}
    p(\theta \mid X^{n} ) = 
    \frac{ \exp\{\sum_{i=1}^{n} \logp{X_{i}}{\theta}\}p(\theta) }
    {\int \exp\{\sum_{i=1}^{n} \logp{X_{i}}{\theta'}\}p(\theta') d\theta'},
    \label{eq:pos}
\end{align}
where
$\logp{x}{\theta}=\log p(x \mid \theta)$ is the log-likelihood of the model, and $p(\theta)$ is a prior density of $\theta$. The posterior mean of a statistic $A(\theta)$ is defined as 
\begin{align}
\Eppos[A(\theta)]=\int A(\theta)p(\theta \mid X^n) d\theta.
\end{align}
The notations $\Covpos$ and $\Varpos$ are used to indicate the posterior covariance and variance, respectively.

To define the local case sensitivity, we also consider a posterior distribution $p_w(\theta ; X^n)$ with weights $(w_1, \ldots, w_n)$ of the observations $X^n=(X_1,\ldots X_n)$ as follows: 
\begin{align}
    p_w(\theta ; X^n) = 
    \frac{ \exp\{\sum_{i=1}^{n} w_i \, \logp{X_{i}}{\theta} \}p(\theta) }
    {\int \exp\{\sum_{i=1}^{n} w_i \, \logp{X_{i}}{\theta'} \}p(\theta') d\theta'}.
    \label{eq:weighted_pos}
\end{align}
When \eqref{eq:weighted_pos} is used, the average of $A(\theta)$ over the distribution $p_w(\theta ; X^n)$ is expressed as 
\begin{align}
\Eppos^w[A(\theta)]=\int A(\theta)p_w(\theta ; X^n) d\theta.
    \label{eq:av_weighted_pos}
\end{align}
Hereafter $w=1$ is used as an abbreviation of $w_j=1, j=1,2,\ldots,n$. 
The local case sensitivity is defined as a derivative of $\Eppos^w[A(\theta)]$ in \eqref{eq:av_weighted_pos} with the weight $w_i$ evaluated at $w=1$. 

We basically assume regularity conditions for the standard asymptotic theory, such that the posterior is well approximated by a multivariate normal distribution when the sample size $n$ is large; hereafter a statistical model that satisfies these conditions is referred to as a {\it regular model}.

As discussed in \cite{Bayesian_IJK}, mathematical aspects of deriving the formula \eqref{eq:cov_g_star} are non-trivial even with such regularity conditions, in that it is necessary to prove that the convergence of the posterior to a normal distribution should comprise some uniformity in data $X^n$. This applies to many of the arguments in this paper. However, the consideration of these details of mathematical rigor in derivations is beyond the scope of this paper and has been left for future studies.       

All of the MCMC computations in this study are performed using the Stan software (\cite{gelman2015stan}) with the ``RStan''package in R.     

\subsection{Local Case Sensitivity Formulae}

Let us introduce a set of formulae that express the sensitivity of the posterior mean $\Eppos[A(\theta)]$ by using posterior cumulants. A basal example of this is the first-order local case sensitivity formula (\citealp{Gustafson_1996, Perez_etal_2006, MillarandStewart(2007), Giordano_etal_2018, Iba_Yano_arXiv2}), which comprises the use of the posterior covariance between the arbitrary statistic $A(\theta)$ and log-likelihood $\logp{X_{i}}{\theta})$ of the observation $X_i$:
\begin{align}
\left. \frac{\partial}{\partial w_i} \Eppos^w[A(\theta)] \right |_{w=1} =\Covpos[A(\theta),  \logp{X_{i}}{\theta}].
 \label{eq:local_1}
\end{align}
The proof is straightforward when an exchange of integration and differentiation is allowed under appropriate regularity conditions.

 A second-order local case sensitivity formula (\cite{Bayesian_IJK, Iba_Yano_arXiv2}) is less known but equally useful. It expresses the second-order derivative using the third-order combined posterior cumulant as follows:
\begin{align}
\left. \frac{\partial^2}{\partial w_i \partial w_j} \Eppos^w[A(\theta)] \right |_{w=1} =
\Kpos[A(\theta), \logp{X_{i}}{\theta}, \logp{X_{j}}{\theta}],
 \label{eq:local_2}
\end{align}
where the third-order combined posterior cumulant $\Kpos[A(\theta),B(\theta),C(\theta)]$ of the arbitrary 
statistics $A(\theta)$, $B(\theta)$, and $C(\theta)$ are defined as
\begin{align}
\Kpos[A(\theta),B(\theta),C(\theta)]=
\Eppos \big [ (A(\theta)-\Eppos[A(\theta)])(B(\theta)-\Eppos[B(\theta)]) (C(\theta)-\Eppos[C(\theta)]) \big ].
 \label{eq:poscum_3}
\end{align}
The derivation of \eqref{eq:local_2} is also straightforward (see \cite{Bayesian_IJK}).  

The expressions on the left-hand sides of these formulae are closely related to the {\it influence functions} in statistics (\cite{Vaart_book, Konishi_Kitagawa_book}); see also the discussion on centering and strong priors in the supplementary material. Applications of influence functions and related concepts in machine learning are discussed in \cite{KohandLiang_2017}, \cite{Giordano_etal_2018}, and \cite{Memory_Perturbation}.  

\subsection{Bayesian Infinitesimal Jackknife}

\cite{Bayesian_IJK}(see also \cite{Giordano_StanCon}) proposed a formula that represented frequentist covariance with the use of the posterior covariance. They referred to the idea behind the formula as ''Bayesian infinitesimal jackknife''(Bayesian IJ). Here, we explain it in a form comprising the second-order terms represented by $\Kpos$. 

If we set $\eta_i=w_i-1$ and use the formulae \eqref{eq:local_1} and \eqref{eq:local_2} to represent the first- and second-order derivatives by $\eta_i$, we have an approximation that
\begin{align}
\Eppos^w [A(\theta)] \, \simeq \, 
\Eppos[A(\theta)]  
  +\sum_{i=1}^n  \Covpos[A(\theta), & \logp{X_{i}}{\theta}] \, \eta_i \label{eq:ijk2} \\
& +\frac{1}{2} \sum_{i,j=1}^n \Kpos[A(\theta), \logp{X_{i}}{\theta}, \logp{X_{j}} {\theta}] \, \eta_i\eta_j,
 \nonumber
\end{align}
where $n$ is the sample size and $A$ represents an arbitrary statistic.

Thus far, we have focused on approximating changes in estimates caused by variations in observation weights. To evaluate the frequentist performance of an estimator, an additional assumption is typically required---namely, that the data consist of independent and identically distributed (IID) observations from a population.

We introduce the following notation:
\begin{align}
\label{eq:1st_star}
\Covpos^*[A(\theta), \logp{X_{i}}{\theta}]  
= \Covpos[A(\theta), \logp{X_{i}}{\theta}]  
- \frac{1}{n} \sum_{j=1}^n \Covpos[A(\theta),\logp{X_{j}}{\theta}],
\end{align}
where $A(\theta)$ is an arbitrary statistic. Assuming that the observations $X_i$ are IID, the following formula approximates the frequentist covariance $\Sigma_{AB}$ between any two statistics $A$ and $B$:
\begin{align}
\label{eq:cov_g_star}
   \hat{\Sigma}^*_{AB}=\sum_{i=1}^n  
   \Covpos^*[A(\theta),  \logp{X_{i}}{\theta}] \, \,
   \Covpos^*[B(\theta),  \logp{X_{i}}{\theta}] 
\end{align}
We refer to equations of this form as \textit{frequentist covariance formulas}.

Alternatively, we express the empirical mean and covariance of a sample $X^n=(X_1,\dots, X_n)$ as follows:
\begin{align}
& \estEpX[f(X)]
= \frac{1}{n} \sum_{i=1}^n f(X_i), \\
& \estCovX[f(X), g(X)]
= \frac{1}{n} \sum_{i=1}^n \left \{
{\left (f(X_i)-\estEpX[f(X)] \right )
\left (g(X_i)-\estEpX[g(X)] \right )} \right \},
\label{eq:estX}
\end{align} 
where $f(X)$ and $g(X)$ are arbitrary statistics of the observation $X$. Using this notation, equation \eqref{eq:cov_g_star} can be rewritten as
\begin{align}
\label{eq:cov_g_star_est}
   \hat{\Sigma}^*_{AB}=\estCovX[\,  
   \Covpos[A(\theta),  \logp{X}{\theta}], \, \,
   \Covpos[B(\theta),  \logp{X}{\theta}]
   \,].
\end{align}
While we generally prefer the form like \eqref{eq:cov_g_star} that explicitly includes each observation $X_i$, the notation \eqref{eq:estX} is useful for discussing the dual roles of frequentist and posterior covariance in later sections.

Rigorous proofs of \eqref{eq:cov_g_star} or \eqref{eq:cov_g_star_est} have been provided by \cite{Bayesian_IJK} using two different methods; an intuitive method of deriving this is to begin with a well-known formula that represents frequentist covariance using the first-order influence functions and replace these influence functions with the derivatives \eqref{eq:local_1}. Another heuristic is to consider a bootstrap estimate of the frequentist covariance and approximate it using the first term of the right-hand side of \eqref{eq:ijk2} (see Section~3.3 of \cite{Bayesian_IJK} and Section~\ref{sec:essential} of the present paper).
 
When the effect of  the prior is weak, the formula \eqref{eq:cov_g_star} is simplified as 
\begin{align}
\label{eq:cov_g}
   \hat{\Sigma}_{AB}=\sum_{i=1}^n  \Covpos[A(\theta),  \logp{X_{i}}{\theta}] \, \Covpos[B(\theta),  \logp{X_{i}}{\theta}].
\end{align}
The condition for the reduction from \eqref{eq:cov_g_star} to \eqref{eq:cov_g} is discussed briefly in Section~\ref{sec:double_center} 
and in more detail in the supplementary material.

It should be noted that the formulae \eqref{eq:cov_g_star} and \eqref{eq:cov_g} reflect a considerable amount of the effect of the prior, while depending on the prior only through the definition of the posterior covariance; this is also analyzed in the supplementary material.

A notable feature of the formulae \eqref{eq:cov_g_star} and \eqref{eq:cov_g} is that they provide a reasonable estimate of the frequentist covariance, even when the model is misspecified; for example, binomial or Poisson likelihoods in the case of over-dispersed count data. This contrasts with a pure Bayesian estimate $\Covpos[A(\theta),  B(\theta)]$ of the covariance, which matches the frequentist covariance only when the likelihood is correct and the prior is weak. 

\begin{remark} \label{rem:VonMises_functional}
Infinitesimal jackknife (IJ) approximation comprises the use of a Taylor series expansion with observation weights for the approximation of the frequentist property of statistics (\cite{Jaeckel_1972,Giordano_etal_2019,Giordano_higher}). The Bayesian IJ is considered as a combination of the IJ and local case sensitivity formulae. The IJ is closely related to a functional expansion introduced by \cite{mises1947asymptotic}; \cite{Bayesian_IJK} applies the expansion in a proof of the frequentist covariance formula.  
\end{remark}

\begin{remark} \label{rem:Watanabe_functional}
\cite{Watanabe_2010_b,Watanabe_book_mtbs} employed functional cumulant expansions similar to  the expansion \eqref{eq:ijk2} in his theory of singular statistical models; they are also used to demonstrate asymptotic equivalence between \WAIC and leave-one-out cross-validation. However, multivariate expansion that corresponds to simultaneous changes of more than one $w_i$ or $\eta_i$ appears to have been used rarely in \cite{Watanabe_2010_b,Watanabe_book_mtbs}, but is essential in the framework proposed in this paper. 
\end{remark}

\section{The Principal Space of $W$ and Its Uses in the Frequentist Evaluation}
\label{sec:essential}

\subsection{The W matrix and W-kernel}
\label{sec:kernel_1}

The matrix $W$, which is a central object in our study, is defined as
\begin{align} \label{eq:W}
W_{ij}= \Covpos[\logp{X_{i}}{\theta}, \logp{X_{j}}{\theta}].
\end{align}
Here $\logp{X_{i}}{\theta}=\log p(X_i \mid \theta)$ is the log-likelihood of the observation $X_i$. As we already mentioned, this is a special case of 
the posterior covariance matrix of log case-deletion weights defined in
\cite{Bradlow_1997}, \cite{MacEachern2002} and \cite{Thomas2018}, but we focus on this specific form because we are interested in its relation to the Bayesian IJ.

We also introduce an empirical version of the W-kernel as 
\begin{align}
\label{eq:K_w}
K_W(X,X^\prime)=
n \, \Covpos[\logp{X}{\theta}, \logp{X'}{\theta}].
\end{align}
$K_W(X_i,X_j)=n W_{ij}$ holds.  The sample size $n$ is introduced on the right-hand side for the comparison to the Fisher kernel in later sections. The expression \eqref{eq:K_w} defines a positive-definite kernel since the right-hand side is a covariance function. A feature vector corresponding to this kernel is 
\begin{align}
    \Phi_\theta(X)=\sqrt{n}\logp{X}{\theta}.
\end{align}
where a continuous-valued parameter $\theta$ plays the role of the index of features. If we define an inner product $ (\,\,, \,\,)_{\mathrm{pos}}$ between the functions of $\theta$ by the covariance $\Covpos$, the W-kernel is expressed as 
\begin{align}
K_W(X,X^\prime)= (\Phi_\theta(X),\Phi_\theta(X^\prime))_{\mathrm{pos}}.
\label{eq:Kphiphi}
\end{align}

A useful variant of the W-kernel is its ``double centered version'' defined as
\begin{align}
\label{eq:K_w_double}
K^c_W(X_i,X_j)=
n \, \Covpos \left [
\logp{X_i}{\theta}
-\frac{1}{n} \sum_{m=1}^n \logp{X_m}{\theta} \, , \, 
\logp{X_j}{\theta}
-\frac{1}{n} \sum_{l=1}^n \logp{X_l}{\theta}
\right ].
\end{align}
Its role in cases with non-negligible effects of priors is discussed in Appendix~\ref{sec:double_center}. Other aspects of the kernel framework, such as the kernel PCA formulation of the eigenvalue problem, Reproducing Kernel Hilbert Space (RKHS) $\mathcal{H}$ associated with $K_W$, and ``population version''  of the kernel are also discussed in Appendix~\ref{sec:kernel_2}. 

\begin{remark}
\cite{jaakkola1999} suggests the use of posterior covariance as an inner product for constructing reproducing kernels, and discusses kernels using posterior information for logistic regression models. The proposed kernel can be viewed as an approximation to a special case of the W-kernel defined in the present work. However, they do not develop a general theory as presented in this paper, nor do they explicitly connect it to their earlier work on the Fisher kernel.
\end{remark}

\subsection{Approximating Posterior Covariances and Cumulants}
\label{sec:essential_posterioraverage}

In this subsection, we show that the posterior covariance and the third-order cumulant can be approximated by its projection onto the principal space of \( W \). 

\subsubsection*{The eigenvalue problem of the $W$ matrix}

We denote the eigenvalues of the symmetric matrix $W$ as $\{\lambda_a, \, a=1,\ldots,n\}$ and the associated unit eigenvectors as $\{\bm{U}_a\}=\{(U^i_a)\}$, and define $\tilde{\eta}_a= \sum_{i=1}^n U^i_a \eta_i$. Then, we have  
\begin{align}
    \sum_{i,j=1}^{n} U^i_a \Covpos[\logp{X_{i}}{\theta},\logp{X_j}{\theta}]U^j_b
    =
    \lambda_a \delta_{ab},
\end{align}
where the Kronecker's delta $\delta_{ab}$ is defined as usual:
\begin{align}
    \delta_{ab}=
    \begin{cases}
        1 \,\, (a=b) \\
        0 \,\, (a \neq b)
    \end{cases}
    .
\end{align}

If we define the projection of the function $\logp{X_{i}}{\theta}$ of $\theta$ to the $a$th eigenvector as 
\begin{align}
\label{eq:proj}
    \sproj_a (\theta)=\sum_{i=1}^n U^i_a \logp{X_{i}}{\theta},
\end{align}
the following relation holds:
\begin{align}
\label{eq:covpos_U}
    \Covpos[\sproj_a(\theta),\sproj_b(\theta)]
    =
    \lambda_a \delta_{ab}.
\end{align}
Since $U=(U^i_a)$ is an orthogonal matrix $(U^{-1})^a_i=U^i_a$, we can reverse the relation \eqref{eq:proj} as 
\begin{align}
    \logp{X_{i}}{\theta}=\sum_{a=1}^n  U^i_a \sproj_a(\theta) 
\end{align}

Based on this preparation, we consider the projection of $\logp{X_{i}}{\theta}$ to a space $\mathcal V$ of $W$ spanned by $U^i_a, \, a=1,\ldots a_M$ using
\begin{align}
\label{eq:star}
    \logpstar{X_{i}}{\theta}=\sum_{a=1}^{a_M}  U^i_a \sproj_a(\theta). 
\end{align}
Using this definition and the relation \eqref{eq:covpos_U}, we can bound the variance of the residuals by the corresponding eigenvalues as
\begin{align}
\label{eq:var_residual}
\Varpos[\logp{X_{i}}{\theta}-\logpstar{X_{i}}{\theta}]
    \leq
    \sum_{a=a_M+1}^n \lambda_a.
\end{align}
The derivation of \eqref{eq:var_residual} is standard and presented in Appendix~\ref{sec:app:PCAvariance}.

Hereafter, we formally refer to the space $\mathcal{V}$ as the principal space of the matrix $W$ (the leading-$a_M$ principal space). When this term is used without specifying $a_M$, it refers to a space where the residuals given by \eqref{eq:var_residual} are sufficiently small.

\subsubsection*{A bound for the error in posterior covariances}

Using \eqref{eq:var_residual}, we can bound the error in $\Covpos[A(\theta),\logpstar{X_{i}}{\theta}]$ caused by the projection to the principal space of $W$ as
\begin{align}
\left |
\Covpos  [A(\theta) , \logp{X_{i}}{\theta}] -\Covpos[A(\theta), \logpstar{X_{i}}{\theta}]
\right |
= \left | 
\Covpos[A(\theta), \logp{X_{i}}{\theta}-\logpstar{X_{i}}{\theta}]
\right |
\\
 \leq
\left ( \, 
\Varpos[A(\theta)] \Varpos[\logp{X_{i}}{\theta}-\logpstar{X_{i}}{\theta}]
\, \right )^{1/2}
=
\left ( \, 
\Varpos[A(\theta)]  
\, \right )^{1/2}
\left ( \, 
\sum_{a=a_M+1}^n \lambda_a  
\, \right )^{1/2},
\label{eq:residual_bound}
\end{align}
where we use a form of the Cauchy-Schwarz inequality
$
    |\mathrm{Cov}[B,C]|\leq \left (\mathrm{Var}[B] \right)^{1/2} \left (\mathrm{Var}[C] \right )^{1/2},  
$
which holds for any $B$ and $C$.

\subsubsection*{A bound for the error in third-order cumulants}
\label{sec:app:bound_cum3}

We can also derive a bound for the error in $\Kpos [A(\theta) , \logp{X_{i}}{\theta},  \logp{X_{j}}{\theta}] $ caused by projecting onto the principal space of the matrix $W$.  In the following, to simplify the notation, we write
$A(\theta) $, $\logp{X_{i}}{\theta}$, and $\logpstar{X_{i}}{\theta}$ as $A$, $\ell_i$, and $\ell^*_i$, respectively. We assume $\sup |A-\Eppos[A]|$, $\Varpos[\ell_i ]$, and $\Varpos[\ell_j ]$ are finite.

We start from the following decomposition:
\begin{align}
\Kpos [A,\ell_i, \ell_j] -\Kpos[A, \ell^*_i, \ell^*_j]
= \Kpos [A,\ell_i, \ell_j-\ell_j^*] +\Kpos [A,\ell_i-\ell_i^*, \ell_j^*] 
\label{eq:decomp_kpos}
\end{align}
Then, the first term in the right-hand side of \eqref{eq:decomp_kpos} is evaluated as follows:
\begin{align}
\big | & \Kpos [A,\ell_i, \ell_j-\ell_j^*] \big |
\\ & = 
\bigg | \Eppos \bigg [\big (A-\Eppos[A]\big )\, \big (\ell_i-\Eppos[\ell_i] \big ) \,
\big (\ell_j-\ell_j^*-\Eppos[\ell_j-\ell_j^*] \big ) \bigg ] \bigg |
\\ & \leq 
\sup |A-\Eppos[A]| \,\,
\Eppos \bigg [ \big |\ell_i-\Eppos[\ell_i] \big | \,
\big |\ell_j-\ell_j^*-\Eppos[\ell_j-\ell_j^*] \big | \bigg ]
\\ & \leq 
\sup |A-\Eppos[A]| \,\,
\Varpos \big [\ell_i \big ]^{1/2} \,\,
\Varpos \big [\ell_j-\ell_j^* \big ]^{1/2} 
\\ & \leq
\sup |A-\Eppos[A]| \,\,
\Varpos \big [\ell_i \big ]^{1/2} \,\,
\left ( \,
\sum_{a=a_M+1}^n \lambda_a  
\, \right )^{1/2},
\shortintertext{
where we use \eqref{eq:var_residual}, along with the Cauchy-Schwarz inequality. In a similar way, the second term in the right-hand side of \eqref{eq:decomp_kpos} is also evaluated as:
}
\big | & \Kpos [A,\ell_i-\ell_i^*, \ell_j^*] \big | 
\\ & \leq
\sup |A-\Eppos[A]| \,\,
\Varpos \big [\ell_j^* \big ]^{1/2} \,\,
\left ( \,
\sum_{a=a_M+1}^n \lambda_a  
\, \right )^{1/2}
\\ & \leq
\sup |A-\Eppos[A]| \,\,
\Varpos \big [\ell_j \big ]^{1/2} \,\,
\left ( \,
\sum_{a=a_M+1}^n \lambda_a  
\, \right )^{1/2},
\end{align}
where we use the relation $\Varpos[\ell_i^*] \leq \Varpos[\ell_i]$ in the second line.

Combining the above bounds, we obtain the following bound for the error in third-order cumulants in terms of the residual variance:
\begin{align}
\big  | 
\Kpos [A , \ell_i ,  \ell_j] -\Kpos[A, \ell_i^*, \ell_j^*]
\big |
\leq  
 \sup |A-\Eppos[A]| \bigg ( \big (\Varpos[\ell_i ] \big )^{1/2}+\big (\Varpos[\ell_j ] \big )^{1/2} \bigg ) 
\left ( \,
\sum_{a=a_M+1}^n \lambda_a  
\, \right )^{1/2}.
\label{eq:bound_3rd}
\end{align}

Although the supremum may be infinite over the whole space, posterior concentration effectively restricts it to a compact set. This is a standard argument and we omit the details.

\subsubsection*{Numerical method}

In applications, an important issue is how to treat an eigenvalue problem of a $n\times n$ matrix $W$ efficiently. If we use the incomplete Cholesky decomposition, which has been used for treating the Gram matrix in Kernel methods, the computational burden is reduced to $O(a_M^2 \times n)$ from $O(n^3)$ in the full matrix diagonalization. This is particularly important because a set of observations obtained using the incomplete Cholesky decomposition is often useful itself; it can be regarded as a representative
set of observations that spans the principal space. These issues are detailed in appendix~\ref{sec:cholesky}.

\subsection{Enter Bayesian IJ: Applications in Frequentist Evaluation}
\label{sec:essential_BIJK}

We next introduce Bayesian IJ as a tool for frequentist evaluation. 
We now discuss the relevance of the principal space of the matrix~$W$ 
for estimating the frequentist uncertainty of a Bayesian estimator. 
Our approach is based on the following observations:

\begin{itemize}
    \item[(1)] The results in the previous section indicate that 
    the posterior covariances and third-order cumulants 
    can be approximated by projecting the log-likelihood 
    onto the principal space of~$W$ using~\eqref{eq:star}.
    \item[(2)] By leveraging the idea of Bayesian IJ, 
    we can approximate the frequentist properties of posterior means 
    in terms of posterior covariances and third-order cumulants. 
\end{itemize}

Combining (1) and (2), we can implement an approximate frequentist evaluation of posterior mean estimators using the projections to the principal space of $W$. Below, we formulate this approach in two different ways. 

\subsubsection*{Frequentist covariance}

A quick way to apply the proposed framework for frequentist uncertainty is to employ the frequentist covariance formula \eqref{eq:cov_g}. Using \eqref{eq:star}, the expression \eqref{eq:cov_g} is approximated as 
\begin{align}
\label{eq:cov_g_ess}
   \hat{\Sigma}^{ess}_{AB}=\sum_{i=1}^n \Covpos[A(\theta), \logpstar{X_{i}}{\theta}] \, \Covpos[B(\theta), \logpstar{X_{i}}{\theta}].
\end{align}
Using \eqref{eq:star} and an orthogonal relation $\sum_i U^a_i U^b_i=\delta_{ab}$, the expression can be written as 
\begin{align}
\label{eq:cov_g_s}
   \hat{\Sigma}^{ess}_{AB}=\sum_{a=1}^{a_M}  \Covpos[A(\theta), \sproj_a(\theta)]  \, \Covpos[B(\theta), \sproj_a(\theta)]. 
\end{align}
These expressions indicate that the frequentist covariance is approximated by the posterior covariance of the projections $\logpstar{X_{i}}{\theta}$ or $\sproj_a(\theta)$ onto the principal space of the matrix $W$. The error of this approximation is controlled by \eqref{eq:residual_bound}.

\begin{remark}
We can proceed in a similar manner when \eqref{eq:cov_g} is replaced by \eqref{eq:cov_g_star}. In this case, however, we can also use the double-centered version $K^c_w$ defined in \eqref{eq:K_w_double} instead of $W$; it may provide a better approximation for strong priors.    
\end{remark}

\subsubsection*{Approximating bootstrap}

Another method of relating the frequentist uncertainty to the principal space of $W$ is to consider an approximate bootstrap, which provides a more intuitive way to obtain an equivalent result. 

Let us assume that a bootstrap replicate 
$
    X^{(B)}=(X^{(B)}_1, \ldots, X^{(B)}_n)
$ 
of the original data $X^{n}=(X_1, \ldots, X_n)$ is generated via resampling with replacement using random numbers $(R_1, \ldots, R_n)$ that follow a multinomial distribution with $n$ trials and equal probability $1/n$ for each outcome. Furthermore, let $A(\theta)$ denote an arbitrary statistic of interest. 

Since the direct application of the bootstrap to evaluate the frequentist properties of $A$ entails a substantial computational burden, approximate bootstrap methods are useful when they require only posterior samples based on the original data. Our task, therefore, is to approximate the posterior mean 
\[
\widehat{A^{(B)}}=\Eppos[A(\theta)\mid X^{(B)}]
\]
defined with a bootstrap replicate $X^{(B)}$, using only the set $\Theta^M=(\Theta^{(1)},\cdots,\Theta^{(M)})$ of parameter draws generated from the posterior based on the original data $X^n$.

We start from the first-order Bayesian IJ approximation to the bootstrap presented in Section~3.3 of \cite{Bayesian_IJK}. This corresponds to retaining only the first-order terms in the $\eta_i$'s on the right-hand side of \eqref{eq:ijk2} and replace $\eta_i=w_i-1$ with $R_i-1$. This leads to the following algorithm for computing an estimate $\widehat{A^{(B)}}$ of $A^{(B)}$ for a given realization of $(R_1, \ldots, R_n)$. 
\begin{align}
\widehat{A^{(B)}}=\Eppos[A(\theta)]+ & \sum_{i=1}^n (R_i-1) 
  \Covpos[A(\theta),\logp{X_{i}}{\theta}]. \label{eq:boot1_IF1}
\end{align}
Here, the notation $\widehat{A^{(B)}}$ is used to distinguish the estimated value from the target quantity $A^{(B)}$. Before applying this formula to various sets of $(R_1, \ldots, R_n)$, we perform a single run of the MCMC for the original data $X^n$ to obtain posterior samples $\theta^M=(\Theta^{(1)},\cdots,\Theta^{(M)})$. These samples are used to estimate $\Eppos[A(\theta)]$ , $\Covpos[A(\theta),\logp{X_i}{\theta}]$,

Then, we can approximate the right-hand side of \eqref{eq:boot1_IF1} using the projection onto the principal space of $W$ as

\begin{align}
\widehat{A^{(B)}}=\Eppos[A(\theta)]+ & \sum_{i=1}^n (R_i-1) 
  \Covpos[A(\theta),\logpstar{X_{i}}{\theta}]. 
  \label{eq:boot1_IF1_proj}
\end{align}
Again, the bound \eqref{eq:residual_bound} provides  the bound of the error in the right-hand side caused by the projection into the principal space. Using \eqref{eq:star}, we obtain
\begin{align}
    \sum_{i=1}^n \Covpos[A(\theta), \, & \logpstar{X_{i}}{\theta}](R_i-1) 
    = \sum_{a=1}^{a_M} \Covpos[A(\theta), \sproj_a(\theta)] \tilde{\eta}_a,
\label{eq:by_tilde_eta}  
\end{align}
where we define 
\begin{align}
   \tilde{\eta}_a=\sum_{i=1}^n U^i_a (R_i-1).
\label{eq:def_tilde_eta}    
\end{align}
On substituting \eqref{eq:by_tilde_eta} and \eqref{eq:def_tilde_eta} into \eqref{eq:boot1_IF1_proj}, we have an alternative form.
\begin{align}
 \widehat{A^{(B)}}=
\Eppos[A(\theta)]  
  +\sum_{a=1}^{a_M} \Covpos[A(\theta), \sproj_a(\theta)]  \tilde{\eta}_a,
\label{eq:boot_local1_essential}
\end{align}
The expression \eqref{eq:boot_local1_essential} indicates that the first-order approximate bootstrap can be implemented only by using projections $(\tilde{\eta}_a)$ of the random number $(R_i)$ to the principal space of the matrix $W$, the total number of which is $a_M$.

\begin{remark}
As discussed in \cite{Bayesian_IJK}, when we identify the bootstrap covariance between $A^{(B)}$ and $B^{(B)}$ with the frequentist covariance, the formula \eqref{eq:boot1_IF1} provides a heuristic derivation of the frequentist covariance formula \eqref{eq:cov_g_star}. If we approximate $(R_i-1)$s in \eqref{eq:boot1_IF1} with mutually independent normal random variates each of which obeys $N(0,1)$, the same argument leads to the simplified form \eqref{eq:cov_g}. 
\end{remark}

\begin{remark}
\cite{Thomas2018} shows that the KL~divergence between the original posterior distribution and the one with weighted observations
\begin{align}
\label{eq:KLdef}
D( p(\theta \mid X^n) \| p_{w}(\theta ; X^n))= 
\Eppos\left [ \log \frac{p(\theta \mid X^n)}{p_{w}(\theta ; X^n)} \right ], 
\end{align}
is approximated as
\begin{align}
\label{eq:KL}
D( p(\theta \mid X^n) \| p_{w}(\theta ; X^n))= 
\frac{1}{2} \sum_{i,j=1}^n \Covpos[\logp{X_{i}}{\theta},\logp{X_j}{\theta}] \eta_i \eta_j +o_p(\|\eta\|^2)
\end{align}
where $\eta_i=w_i-1$ and $\|\eta\|$ denote the $\ell_2$-norm of the vector $\eta=(\eta_i)$. This formula provides the basis for their graphical approach using the principal space of the matrix $W$ (and its generalization) and implicitly suggests that the change in the posterior average can be approximated by the projection onto the principal space. However, the present derivation based on the Bayesian IJ provides an approximation to the right-hand side of the equation~\eqref{eq:ijk2}, which is more explicit and directly connected to the assessment of frequentist uncertainty.

\end{remark}

\subsubsection*{Higher order approximation}

The approximation \eqref{eq:boot_local1_essential} is not particularly useful in reducing computational costs, because the eigenvalue problem for computing the principal space of the matrix $W$ usually requires more computation than a direct iteration of \eqref{eq:boot1_IF1}. On the other hand, the principal space of the matrix $W$ is more useful for the reduction of computational costs in the second-order approximation to the bootstrap. Here, we will discuss this subject as an example of the application of the bound \eqref{eq:bound_3rd} on the error in the third-order posterior cumulant.

Let us define the second-order approximate bootstrap by the formula
\begin{align}
  \widehat{A^{(B)}}=\Eppos[A(\theta)]+ & \sum_{i=1}^n (R_i-1)\, \Covpos[A(\theta),\logp{X_i}{\theta}] \label{eq:boot2_IF2} \\
  & +\frac{1}{2} \sum_{i,j=1}^n(R_i-1)\, (R_j-1)\, \Kpos[A(\theta), \logp{X_i}{\theta}, \logp{X_j}{\theta}].
\end{align}
To derive this, all terms on the right-hand side of \eqref{eq:ijk2} are kept, and $\eta_i$ is replaced by $R_i-1$. As before, prior to applying this formula to various sets of $(R_1,\ldots,R_n)$, we perform a single run of the MCMC for the original data $X^n$ and $\Eppos[A(\theta)]$ , $\Covpos[A(\theta),\logp{X_i}{\theta}]$, and $\Kpos[A(\theta), \logp{X_i}{\theta}, \logp{X_j}{\theta}]$ are estimated from the MCMC samples. In appendix~\ref{sec:app:boot_example_1}, we compare various strategies on approximate bootstrap, including the second-order approximate bootstrap.

A disadvantage of the second-order approximation using \eqref{eq:boot2_IF2} is that constructing the \( n \times n \) matrix \( \Kpos \) requires \( O(n^2 \times M \times p) \) computations for initialization, where \( M \) is the number of posterior samples and \( p \) is the number of target statistics \( A \). Additionally, evaluating \eqref{eq:boot2_IF2} for each bootstrap replicate incurs \( O(n^2 \times N_{b} \times p) \) computations. Hence, computational cost increases as \( n \) becomes large.

The second-order approximation \eqref{eq:boot2_IF2} can be implemented with the computational requirement $O(n \times M \times N_b)$. If we define
$
    \mathsf{L}(\theta; (R_i))=\sum_{i=1}^n (R_i-1)\logp{X_i}{\theta}
$, the second-order term 
\begin{align}
\label{eq:second_boot}
\sum_{i,j=1}^n (R_i-1)(R_j-1)  \, \Kpos[A(\theta),\logp{X_i}{\theta}, \logp{X_j}{\theta}]
\end{align}
in the right-hand side of \eqref{eq:boot2_IF2} is expressed as
\begin{align}
\label{eq:second_alternative}
\Kpos \left [A(\theta), \mathsf{L}(\theta; (R_i)), \mathsf{L}(\theta; (R_j))\right ].
\end{align}
Once we fix the random number $R_i$s, the computation of $\mathsf{L}(\Theta^{(u)}; (R_i))$ for $\Theta^{(u)}, \, m=1, \ldots, M$, requires $O(n \times M)$ computation. Using these, only $O(p)$ computation is required to obtain the empirical estimate of \eqref{eq:second_alternative}.
Hence the total computational time is on the order of $O(n \times M \times N_b)$. This implementation may be a practical choice when $n \times p$ is large, but it can be ineffective when the number $N_b$ of bootstrap replications also becomes large. The main drawback is the need to recompute $\Kpos \left[A(\theta), \mathsf{L}(\theta; (R_i)), \mathsf{L}(\theta; (R_j))\right]$ for each draw of random numbers $(R_i)$.

At this point, we propose approximating the second-order term \eqref{eq:second_boot} by projections onto the principal space of the matrix $W$. This approach is motivated by the bound \eqref{eq:bound_3rd}, where the error in the third-order cumulants introduced by the projection is bounded by the residual variance, under the assumption that \( \sup |A-\Eppos[A]| \) and \( \Varpos[\logp{X_i}{\theta}] \) 
are finite for all \( i \).

The proposed approximation provides us with
\begin{align}
\sum_{\alpha,\beta=1}^{a_M}\tilde{\eta}_\alpha \, \tilde{\eta}_\beta \, \Kpos[A(\theta), \sproj_\alpha (\theta), \sproj_\beta (\theta)], 
\,\,\, 
\sproj_\alpha (\theta)=\sum_{i=1}^n U^i_a \logp{X_{i}}{\theta},
\,\,\, \tilde{\eta}_a=\sum_{i=1}^n U^i_a (R_i-1)
\label{eq:second_proj}
\end{align}
as an approximation to \eqref{eq:second_boot}. The difference from \eqref{eq:second_alternative} is that $\Kpos[A(\theta), \sproj_\alpha (\theta), \sproj_\beta (\theta)]$ is independent of the set $(R_i)$ of random numbers and need not be recomputed for each random draw. If we adopt this approximation in the second-order term and use the incomplete Cholesky decomposition to treat the eigenvalue problem of the matrix $W$, an extra computation of the order $O(a_M^2 \times n)$ is required to initialize the algorithm, where $a_M$ is the dimension of the principal space of the matrix $W$. The computational time to construct the estimate of the matrix $\Kpos$ from MCMC samples reduces to $O(a_M^2 \times M \times p)+O(a_M \times n \times M)$; also, the time to approximate $N_{b}$ bootstrap replications reduces to $O(a_M^2 \times N_{b}\times p )+O(a_M \times n \times N_b)$.

The idea proposed above is implemented and tested in appendix~\ref{sec:app:boot_example_2}. Although the test is only for a single example, it indicates that the projection to the principal space of $W$ is a promising strategy for numerical computation including higher-order terms.

\begin{remark}
The remainder of the numerical problem is the construction of $n \times n$ matrix $W$ using posterior samples; this still requires $O(n^2 \times M)$ computation. In the example in appendix~\ref{sec:app:boot_example_2}, we use a posterior sub-sampling approach to circumvent this difficulty. That is, using the smaller number $M^*$ of posterior samples, the matrix $W$ is constructed, and a rough approximation of the principal space of the matrix $W$ is obtained using the incomplete Cholesky decomposition. Then, using the projections $\sproj_\alpha (\theta), \, \alpha=1,\ldots,a_M$ to this approximate principal space, we construct $\Kpos[A(\theta), \sproj_\alpha (\theta), \sproj_\beta (\theta)]$ using the full set of posterior samples. Using this approach, the amount of computation required for this part of the algorithm can be reduced to \( O(n^2 \times M^*) \).

This posterior sub-sampling approach is imperfect because we still need to store an $n \times n$ matrix in memory, which is difficult for a very large sample size $n$. A promising method of addressing this issue is the use of on-line PCA (\cite{Oja1992, on-line-PCA-review}) to compute the principal space. 
\end{remark}

\section{Examples of the Eigenvalue Structure}
\label{sec:essential_example}

We present examples of the eigenvalue structure of the matrix $W$. \cite{Thomas2018} also discussed the examples of the eigenvalue problem but primarily focused on explanatory data analysis using PCA displays. On the contrary, our main interest here lies in the eigenvalues of $W$, which motivates the study in Section~\ref{sec:Fkernel}. 

Among the three basic examples presented here, the second and third do not involve an IID model. However, the same framework applies when inference is conditioned on $X_i$ or $Z_i$ defined below. Each result is obtained from an MCMC run by performing a full diagonalization of $W$ using the ``eigen'' function in base R.

\subsubsection*{Weibull analysis}

\begin{figure}[htb]
    \centering
    \includegraphics[width=6cm]{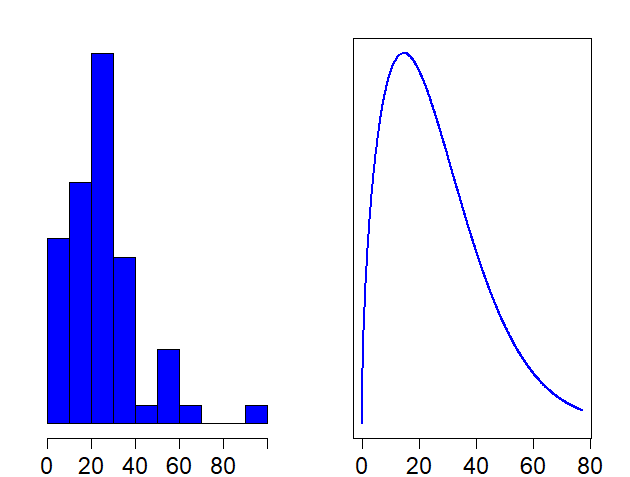}
    \hspace{1cm}
    \includegraphics[width=6cm]{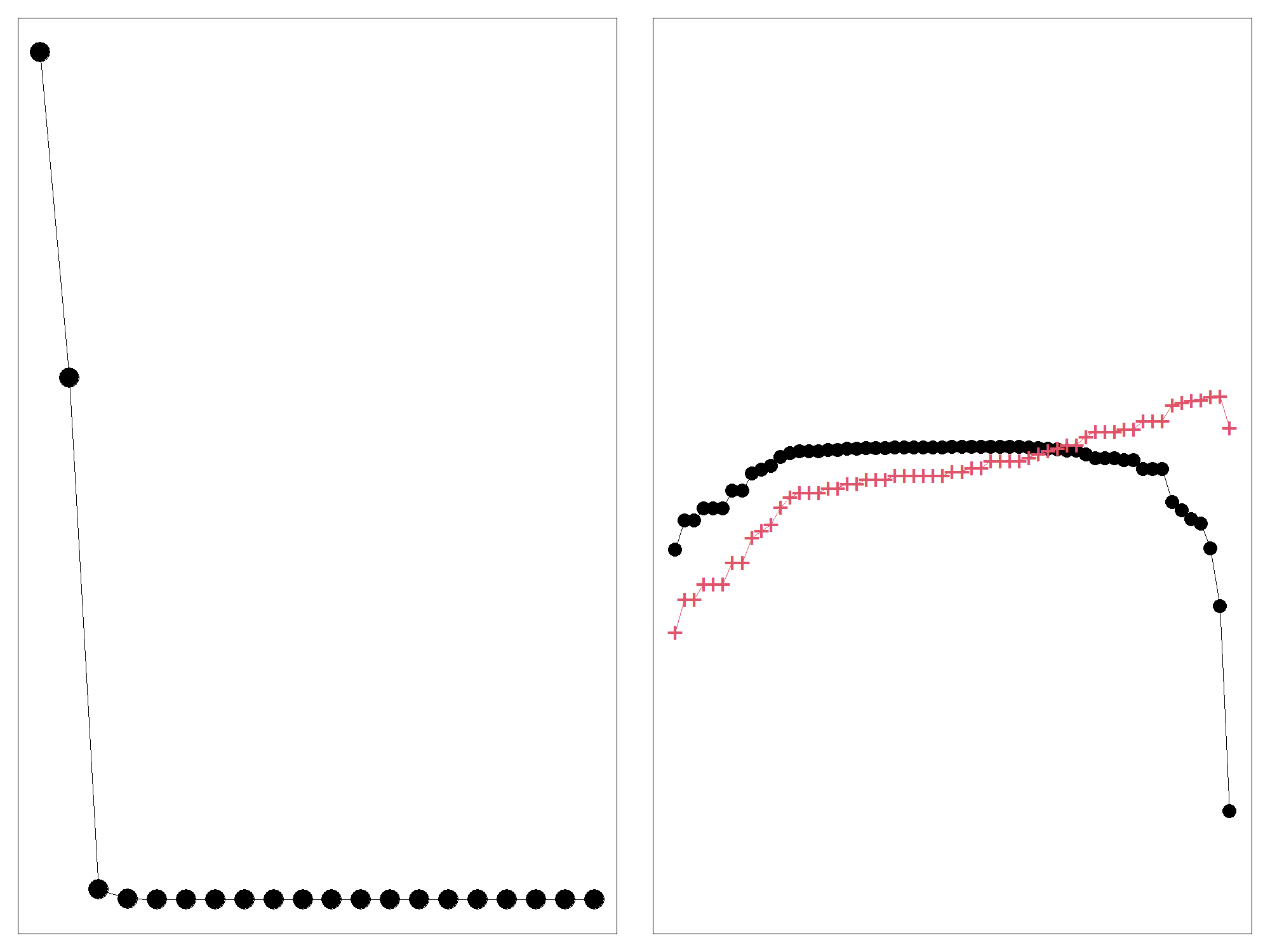} 
    \caption{Weibull analysis. From left to right, the first and second panels present a histogram and a Weibull fitting (MLE) of the data, respectively. The third panel shows the first 20 eigenvalues of $W$ in the decreasing order; the horizontal axis represents the index $1\ldots 59$ of the eigenvalues. The fourth panel shows the eigenvectors corresponds to the first two eigenvalues; the horizontal axis represents the index $1\ldots 59$ of the observations. The black dots correspond to the first eigenvalue, while the red $+$s correspond to the second eigenvalue.}
    \label{fig:age_eigen}
\end{figure}

We first consider the fitting of a non-normal distribution. As an example, let us discuss an analysis of the mortality data $X_i, i=1,\ldots,n$, wherein $X_i$s are the lifespans of 59 females of ancient Egypt estimated from their bones (\cite{Pearson1902}). The histogram of the data is presented in the leftmost panel of Fig.~\ref{fig:age_eigen}. We assume the Weibull distribution to be a likelihood, whose probability density is given by
\begin{align}
    p^{wb}(x \mid \gamma, \lambda)=\frac{\gamma}{\lambda}\left (\frac{x}{\lambda} \right )^{\gamma-1}\exp \left [-\left ( \frac{x}{\lambda} \right)^\gamma \right ], \,\,\, x \geq 0,
    \label{eq:weibull_PDF}
\end{align} 
where an improper prior uniform on $[0,\infty)$ is assumed for each of the parameters $\gamma$ and $\lambda$. It should be noted that this likelihood is not an exponential family in the shape parameter $\gamma$. The density estimated using the maximum likelihood estimator (MLE) (equivalently, the MAP estimate with the above prior) is presented in the second panel in Fig.~\ref{fig:age_eigen}.   

The obtained results are presented in the third and fourth panels in Fig.~\ref{fig:age_eigen}. The first 20 eigenvalues of the matrix $W$ are plotted in the third panel. The first two eigenvalues have a large value, while the others have negligible magnitudes. The dimension of the principal space is two in this case, which is equal to the number of parameters estimated in the model. This is not a coincidence, and is explained in Section~\ref{sec:dual}. 

The normalized eigenvectors corresponding to these two eigenvalues are plotted in the fourth panel against the index of the observations. The amplitude of the eigenvector corresponding to the first eigenvalue has a large value at the high extreme of the age; this means that the estimate is sensitive to the change of the weights in this part of the data.

\subsubsection*{Regression}

Next, we consider an example of the regression. A set of the artificial data is generated as
\begin{align}
X_i=f(z_i)+\tilde{\sigma}\zeta_i, \,\,\,\, \zeta_i \sim t_4, \,\,\,\, i=1,\ldots,n, 
\end{align}
where $t_4$ represents the $t$-distribution with four degrees of freedom. We denote the observations as $X^n = (X_i)$ and the explanatory variable as $(z_i)$. The t-distribution with four degrees of freedom is intentionally introduced to create outliers for a normal model, as defined below. We use a trigonometric curve as a true function $f(z)$ and set $n=30$ and $\tilde{\sigma}=0.3$ in the following. 

We apply a Bayesian regression with a cubic polynomial
\begin{align}
    X_i \sim N(\mu_i,\sigma^2), \,\, \, \mu_i=\beta_0+\beta_1 z_i+\beta_2 z_i^2 +\beta_3 z_i^3, \,\,\,\, i=1,\ldots,n 
\end{align}
to this data set, where an improper prior uniform on $(-\infty,\infty)$ is assumed for each coefficient $\beta_\alpha, \,\, \alpha=0,\dots, 3$. Furthermore, an improper prior uniform on $[0,\infty)$ is assumed for the standard deviation $\sigma$ when we estimate $\sigma$ from the data. When $\sigma$ is provided, we set $\sigma=0.1$. We also test a Student-t distribution with the degree of freedom (df) $\nu=5$ as a distribution of the observational noise, which leads to 
\begin{align}
    X_i \sim \frac{\Gamma((\nu+1)/2)}{\Gamma(\nu/2)}\frac{1}{\sqrt{\nu\pi}\sigma} \left ( 1+ \frac{1}{\nu} \left ( \frac{y-\mu_i}{\sigma}\right )^2 \right )^{-(\nu+1)/2}.
\end{align}
 In this case, priors for the parameters $\beta_\alpha, \,\, \alpha=0,\dots, 3$ and $\sigma$  are the same as those in the normal case, whereas the scale parameter $\sigma$ does not represent the standard deviation in this case. All cases are tested below using the same set of artificial data.

\begin{figure}[htb]
    \centering
    \includegraphics[width=3.5cm]{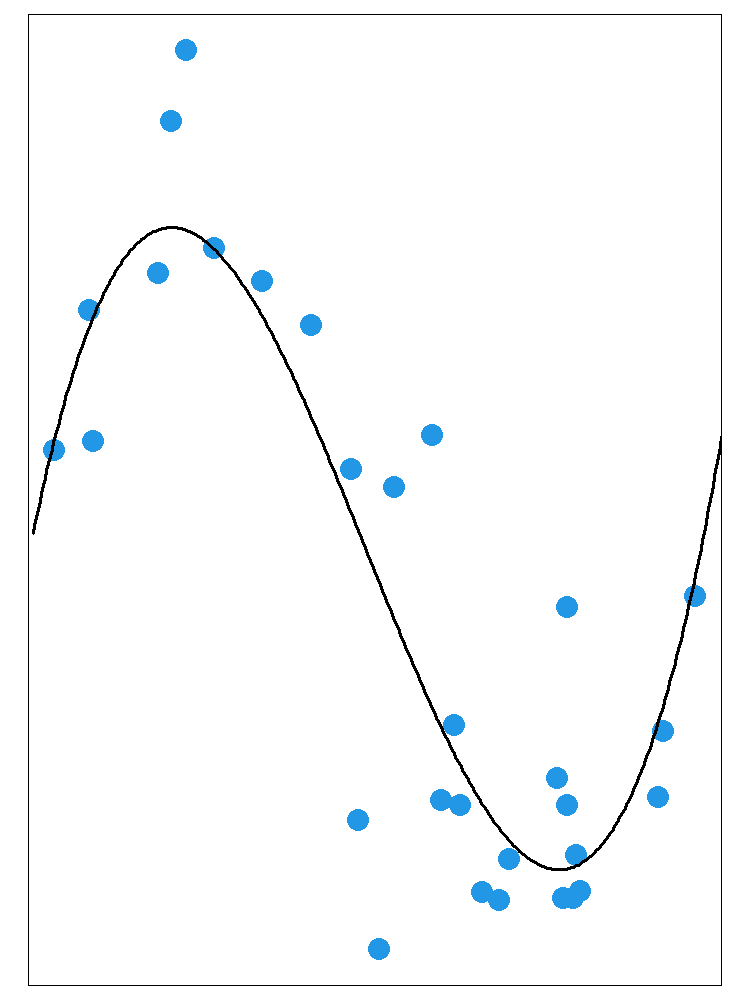}
    \includegraphics[width=3.5cm]{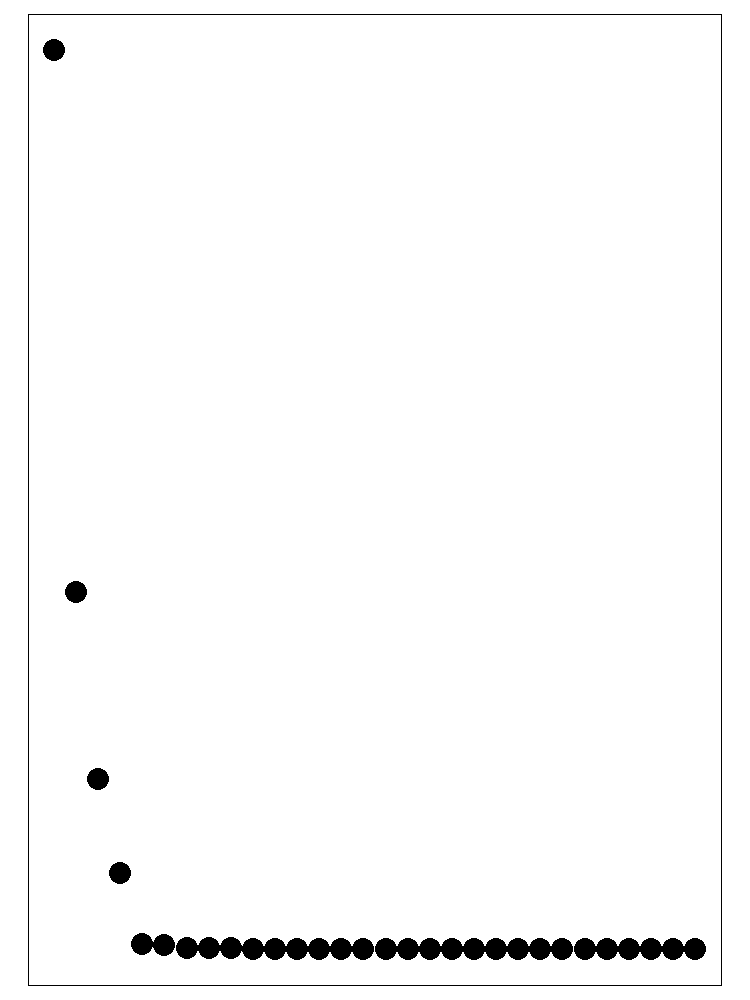}
    \includegraphics[width=3.5cm]{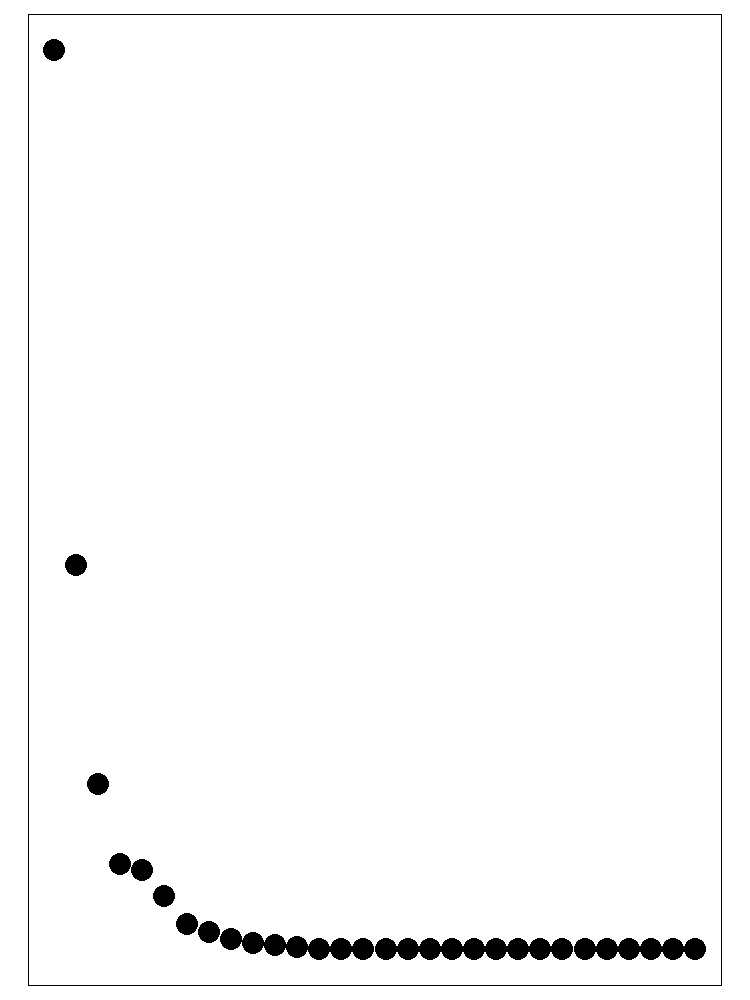}
    \includegraphics[width=3.5cm]{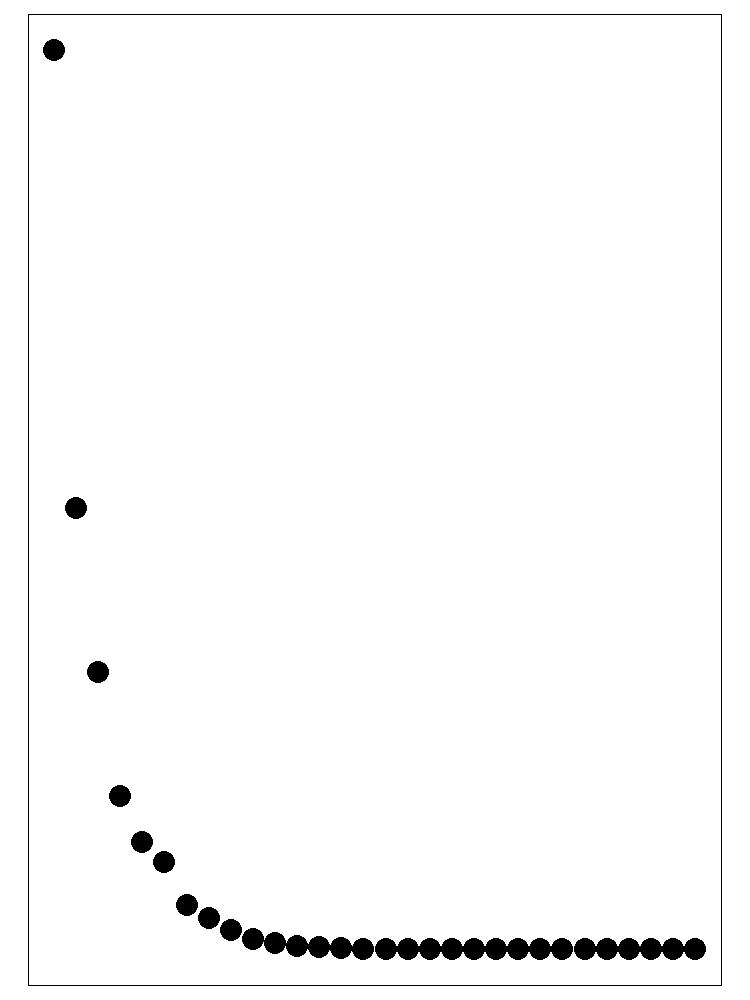}
    \caption{Regression. The first panel presents the data points used in the experiment (blue dot) and the fitted curve; the horizontal axis represents the values of explanatory variable $z$, while the vertical axis represents $X_i$ and the posterior mean of $f(z)$. The second, third, and fourth panel present the eigenvalues of the matrix $W$; from right to left, results with normal ($\sigma$-given), normal($\sigma$-estimated), and Student-t (df=5) likelihood are shown. The horizontal and vertical axes represent an index of eigenvalues and the eigenvalues, respectively. 
    }
    \label{fig:regression_eigen}
\end{figure}

The obtained result is presented in Fig.~\ref{fig:regression_eigen}. The first panel presents data points used in the experiment (blue dots) and the fitted curve assuming the normal likelihood with a given $\sigma$. The second panel presents the eigenvalues in the normal likelihood case with a given $\sigma$; there are four nonzero eigenvalues, and the others are effectively zero; this agrees with the number of estimated parameters. The third panel presents the eigenvalues in the $\sigma$-estimated case; it suggests five or six definitely nonzero eigenvalues, while the number of the estimated parameters is five. However, the cutoff in this case is rather ambiguous; this may be caused by the non-linearity in the estimation of the dispersion parameter $\sigma$. In the fourth panel, the eigenvalues for the Student-t likelihood case are presented; the number of effectively nonzero parameters is also limited in this case. In this case, the details of the result appear to be rather sensitive to the MCMC run, and a relatively long chain was used to obtain stable estimates (five independent runs, each consisting of 50000 iterations, of which the first 1000 were discarded).

\subsubsection*{Smoothing}

The third example employs a more structured model for time-series data. Let us consider the celebrated ``Prussian horse kick data'' $X_{tj},\, t=1,\ldots 20, \, j=1, \ldots 14$; this is a record of deaths by horse kick over 20 years of 14 corps in the Prussian army.  

We treat it as a time-series of length 20 with 14 observations each year and apply a state space model with hidden Markov states $Z_t, t=1,\ldots,20$, wherein a random walk of the state is assumed as a smoothness prior (a local-level model):       
\begin{align}
\label{eq:horse_1}
    Z_t = Z_{t-1} + \zeta_t, \,\,\, t=2, \ldots, 20.
\end{align}
The system noise $\zeta_t$ is an independent variate that obeys a common  normal distribution $N(0,s^2)$; improper priors uniform in $(-\infty,\infty)$ and $(0,\infty)$ are applied for the initial value $Z_1$ and the standard deviation $s$ of the system noise, respectively. Each observation $X_{tj}$ is assumed to be an independent Poisson variate that obeys
\begin{align}
\label{eq:horse_2}
    X_{tj} \sim  Pois (\exp(Z_t)), \,\,\, t=1, \ldots, 20, \,\, j=1, \ldots, 14,
\end{align}
where $Pois(\mu)$ denotes the Poisson distribution of the intensity $\mu$.

For a structured model, wherein part of the model is identified as the likelihood is important; the matrix $W$ depends on this choice. Herein, we define the likelihood using \eqref{eq:horse_2} and identify $Z_t$ as the parameter $\theta$ in the generic framework; the state equation \eqref{eq:horse_1} is considered as a prior for $(Z_t)$ along with an improper prior on $Z_1$. The total number of the observations is $20 \times 14=280$, and $W$ becomes a $280 \times 280$ matrix, which has 280 eigenvalues. 

\begin{figure}[hbt]
    \centering
    \includegraphics[width=5cm]{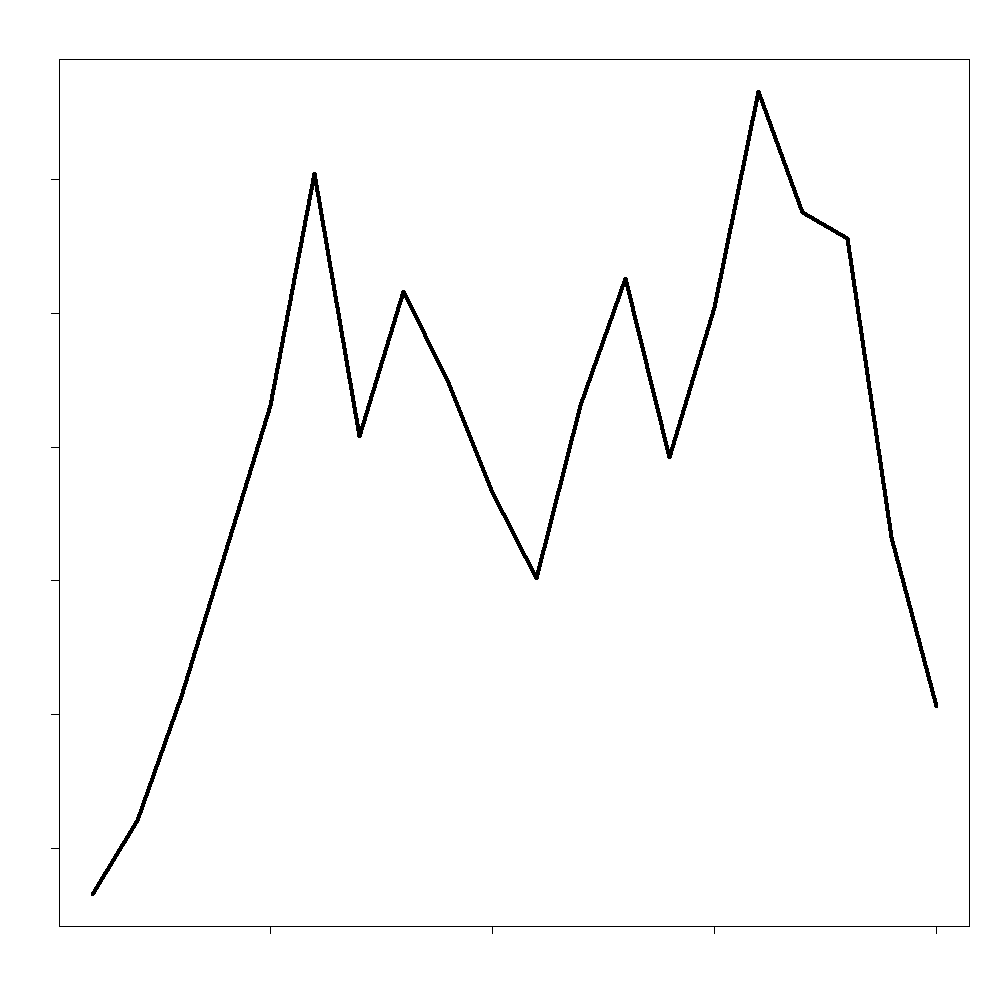}
    \hspace{0.5cm}
    \includegraphics[width=5cm]{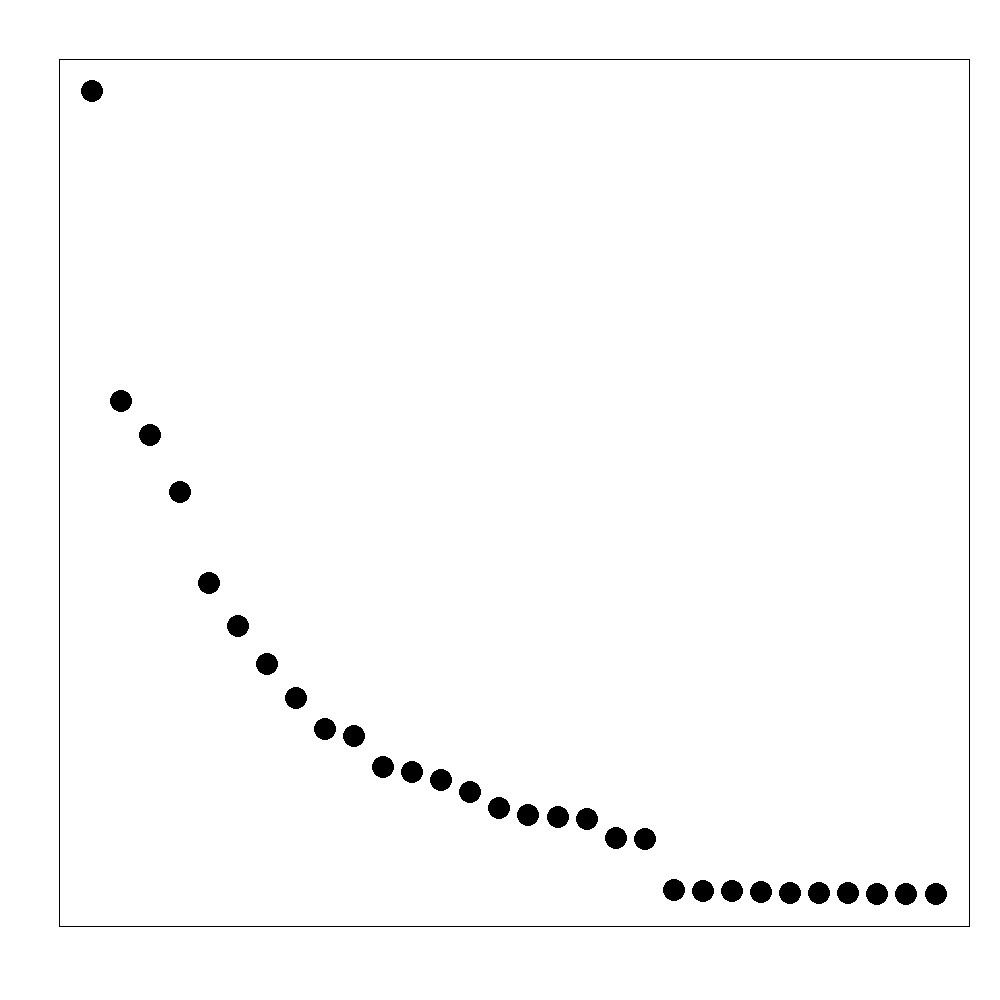}
    \caption{Smoothing. The left panel presents the fitted curve; the horizontal axis corresponds to the year $t=1,\ldots, 20$, while the vertical axis corresponds to the posterior mean of $exp(Z_t)$. The estimated curve is not smooth, because the prior penalizes the first-order difference. The right panel presents the first 30 eigenvalues; the horizontal and vertical axes represent an index of eigenvalues and eigenvalues, respectively. }
    \label{fig:smoothin_eigen}
\end{figure}

The result thus obtained is presented in \eqref{fig:smoothin_eigen}. The left panel shows the fitted curve, while the right panel displays the first 30 eigenvalues. The number of definitely nonzero eigenvalues is 20. Although the nominal number of parameters is 21, only 20 are contained in the likelihood, which again matches the number of definitely nonzero eigenvalues. However, the magnitude of the eigenvalues \( \lambda_i \) decreases smoothly from \( i=1 \) to \( i=20 \). This suggests that the principal space spanned by the first 5 to 10 eigenvalues may be sufficient for obtaining a crude approximation.

\section{\protect{The Relation to the Fisher Kernel and the matrix $\hat{\mathcal{J}}^{-1/2}\hat{\mathcal{I}}\hat{\mathcal{J}}^{-1/2}$}}
\label{sec:Fkernel}

\cite{NIPS1998_db191505} introduced a family of kernels induced from the likelihood functions, which they termed the {\it Fisher kernel} (\cite{Kernel_Schoelkopf,Kernel_Shawe-Taylor,Zhang_2022}). In the following, we show that the Fisher kernel, or more precisely its modified version, is considered an approximation of the W-kernel when the model behind the kernel is regular and the effect of the prior can be ignored. Based on this observation and the relation \eqref{eq:MF_duality_1}--\eqref{eq:MF_duality_2}, we explain why the number of virtually non-zero eigenvalues of the matrix $W$ approximately equals $k$ in a simple parametric model with a noninformative prior, as the sample size $n \to \infty$.

\subsection{Fisher Kernel}

Here, we define information matrices and the Fisher kernel.

\subsubsection*{Information Matrices}

As a preparation, we review a few commonly used definitions and results in asymptotic theory.
We first define the information matrices $\hat{\mathcal{J}}$ and $\mathcal{J}$, whose components are given by
\noeqref{eq:def_J}
\begin{align}
\label{eq:def_J}
 \hat{\mathcal{J}}_{\alpha\beta}=
 -\frac{1}{n}\sum_{i=1}^n
 \frac{\partial^2}{\partial \theta_\alpha \partial \theta_\beta} \logp{X_i}{\theta} \bigg |_{\theta=\hat{\theta}},
\end{align}
where $\hat{\theta}=(\hat{\theta}_\alpha)$ is the MLE based on the data.
A basic relation
\begin{align}
\label{eq:J_pos_basic}
\Eppos[(\theta_\alpha-\hat{\theta}_\alpha)(\theta_\beta-\hat{\theta}_\beta)]
& =\frac{(\hat{\mathcal{J}}^{-1})_{\alpha\beta}}{n}+o_p(1/n)
\end{align}
holds when the effect of the prior can be ignored and the MAP estimate is identified with the MLE $\hat{\theta}$.
We also define another information matrix $\hat{\mathcal{I}}$ as
\noeqref{eq:def_I}
\begin{align}
\label{eq:def_I}
    \hat{\mathcal{I}}_{\alpha\beta}  =
    \frac{1}{n}\sum_{i=1}^n
    \left \{
    \frac{\partial}{\partial \theta_\alpha} \logp{X_i}{\theta} \bigg |_{\theta=\hat{\theta}}
    \frac{\partial}{\partial \theta_\beta} \logp{X_i}{\theta} \bigg |_{\theta=\hat{\theta}}
    \right \}.
\end{align}
Although the symbol $I$ is more common, we use $\mathcal{I}$ to avoid confusion with the identity matrix. A classical result states that, asymptotically, $\hat{\mathcal{I}} \simeq \hat{\mathcal{J}}$ holds when the population $G$ is represented as $p(x \mid \theta_0)$, where $\theta_0$ is the true value of the parameter.

\subsubsection*{Fisher kernel and neural tangent kernel}

The Fisher kernel introduced in \cite{NIPS1998_db191505} is given as
\begin{align}
\label{eq:F_kernel}
    K_{F}(X,X^\prime) & =
    \sum_{\alpha,\beta=1}^k \left \{
    \left. \frac{\partial}{\partial \theta_\alpha} \logp{X}{\theta} \right |_{\theta=\hat{\theta}}
    \left. \frac{\partial}{\partial \theta_\beta} \logp{X^\prime}{\theta} \right |_{\theta=\hat{\theta}}
    (\hat{\mathcal{I}}^{-1})_{\alpha\beta}
    \right \}.
\end{align}

Fisher kernel is closely related to the neural tangent kernel defined in \cite{NIPS2018_NTK}. In its empirical form, it is given as  
\begin{align}
    K_{NTK} (X,X^\prime)=
    \sum_{\alpha,\beta=1}^k \left \{
    \left. \frac{\partial}{\partial \theta_\alpha} \logp{X}{\theta} \right |_{\theta=\hat{\theta}}
    \left. \frac{\partial}{\partial \theta_\beta} \logp{X^\prime}{\theta} \right |_{\theta=\hat{\theta}}
    \right \},
\end{align}
which contains neither $\hat{\mathcal{J}}$ nor $\hat{\mathcal{I}}$. This kernel is also called the {\it plain kernel} in \cite{Kernel_Schoelkopf}.

Kernels based on the likelihood functions are sometimes called as {\it natural kernels}. \cite{Kernel_Schoelkopf} discussed a generic form of the natural kernels parameterized by a matrix $M$; $M=\hat{\mathcal{I}}$ in the Fisher kernel, $M=\hat{\mathcal{J}}$ in the modified Fisher kernel defined in the next subsection, and $M$ equals an identity matrix in the plain kernel.

\subsection{Fisher Kernel as an approximation to W-kernel}

Both the Fisher kernel and the W-kernel are based on likelihood functions, but there is a more direct connection. 

\subsubsection*{Asymptotic relation between the Fisher kernel and the W-kernel}

Let us derive this relation. We assume that the model behind the kernel is regular and the number of parameters is finite and independent of $n$, and the effect of the prior can be ignored. Then, using \eqref{eq:J_pos_basic}, the W-kernel is approximated as
\begin{align} 
& K_W(X,X^\prime)= n \, \Covpos[\logp{X}{\theta}, \logp{X^\prime}{\theta}]
\\ 
& = n \sum_{\alpha,\beta=1}^k \left \{
\left. \frac{\partial}{\partial \theta_\alpha} \logp{X}{\theta} \right |_{\theta=\hat{\theta}}
\left. \frac{\partial}{\partial \theta_\beta} \logp{X^\prime}{\theta} \right |_{\theta=\hat{\theta}}
\Eppos[(\theta_\alpha-\hat{\theta}_\alpha)(\theta_\beta-\hat{\theta}_\beta)]
\right \}+o_p(1)
\\ 
& = \sum_{\alpha,\beta=1}^k \left \{
\left. \frac{\partial}{\partial \theta_\alpha} \logp{X}{\theta} \right |_{\theta=\hat{\theta}}
\left. \frac{\partial}{\partial \theta_\beta} \logp{X^\prime}{\theta} \right |_{\theta=\hat{\theta}}
(\hat{\mathcal{J}}^{-1})_{\alpha\beta}
\right \}+o_p(1),
\end{align}
where $k$ is the number of the parameters. 

Now we define the kernel $K_{MF}$ by
\begin{align}
\label{eq:MF_kernel}
    K_{MF}(X,X^\prime) & =
    \sum_{\alpha,\beta=1}^k \left \{
    \left. \frac{\partial}{\partial \theta_\alpha} \logp{X}{\theta} \right |_{\theta=\hat{\theta}}
    \left. \frac{\partial}{\partial \theta_\beta} \logp{X^\prime}{\theta} \right |_{\theta=\hat{\theta}}
    (\hat{\mathcal{J}}^{-1})_{\alpha\beta}
    \right \},
\end{align}
which approximates $K_W(X,X^\prime)$ as
\begin{align}
\label{eq:MF_kernel_approx}
    K_W(X,X^\prime)=K_{MF}(X,X^\prime)+o_p(1).
\end{align}

The only difference between \eqref{eq:F_kernel} and \eqref{eq:MF_kernel} is that $\hat{\mathcal{I}}$ in \eqref{eq:F_kernel} is replaced by $\hat{\mathcal{J}}$ in \eqref{eq:MF_kernel}. In some literature, the kernel \eqref{eq:MF_kernel}  is also referred to as the Fisher kernel, but we refer to them as the {\it modified Fisher kernel}, because there is considerable difference between \eqref{eq:MF_kernel} and \eqref{eq:F_kernel}, when the model is misspecified. When the population $G$ can be represented by $p(x \mid \theta_0)$, $\hat{\mathcal{I}}$  asymptotically equals $\hat{\mathcal{J}}$. In this case, \eqref{eq:MF_kernel} and  \eqref{eq:F_kernel} are asymptotically identical.

\subsubsection*{Diagonal sums}

An immediate consequence of the asymptotic relation \eqref{eq:MF_kernel_approx} is that the diagonal sums of $K_w$ and $K_{MF}$ are asymptotically identical.

The diagonal sum of the W-kernel divided by the sample size $n$ equals the diagonal sum of the matrix $W$ and is given by
\begin{align}
    \frac{1}{n}\sum_{i=1}^n K_W(X_i,X_i)=
    \Tr[W] =
    \sum_{i=1}^n 
    \Covpos[\logp{X_i}{\theta},\logp{X_i}{\theta}]=
    \sum_{i=1}^n \Varpos[\logp{X_i}{\theta}],
\end{align}
which coincides with the bias correction term of \WAIC (\cite{Watanabe_2010_b,Watanabe_book_mtbs}). 

On the other hand, the diagonal sum of the modified Fisher kernel divided by the sample size $n$ is computed as
\begin{align}
    \frac{1}{n}\sum_{i=1}^n K_{MF}(X_i,X_i)  & =
    \frac{1}{n}\sum_{\alpha,\beta=1}^k  \sum_{i=1}^n \left \{
    \left. \frac{\partial}{\partial \theta_\alpha} \logp{X_i}{\theta} \right |_{\theta=\hat{\theta}}
    \left. \frac{\partial}{\partial \theta_\beta} \logp{X_i}{\theta} \right |_{\theta=\hat{\theta}}
    \right \}
    (\hat{\mathcal{J}}^{-1})_{\alpha\beta}
    \\
    & = \Tr \left [\hat{\mathcal{I}}\hat{\mathcal{J}}^{-1} \right ] = \Tr \left [\hat{\mathcal{J}}^{-1/2}\hat{\mathcal{I}}\hat{\mathcal{J}}^{-1/2} \right ] ,
\end{align}
which coincides with the bias correction term of the \TIC (Takeuchi's information criterion; \cite{Takeuchi_1976, Konishi_Kitagawa_book}). 

This shows that the bias correction terms of \WAIC and \TIC are asymptotically equivalent in regular models when the effect of the prior can be neglected. This equivalence seems known among statisticians, and an informal proof is easily obtained by direct computation. However, it suggests the relation between the matrix $W$ and the matrix $\hat{\mathcal{J}}^{-1/2}\hat{\mathcal{I}}\hat{\mathcal{J}}^{-1/2}$, which is fully explored in the subsequent section.

\subsection{\protect{The Relation to the Matrix {\protect $\hat{\mathcal{J}}^{-1/2}\hat{\mathcal{I}}\hat{\mathcal{J}}^{-1/2}$} via Feature Vector Representation}}
\label{sec:dual}

\subsubsection*{Eigenvalues of the Fisher kernel}

The eigenvalue problems of $K_{MF}(X_i,X_j)$ and the matrix $\hat{\mathcal{J}}^{-1/2}\hat{\mathcal{I}}\hat{\mathcal{J}}^{-1/2}$ are closely related. In fact, the non-zero eigenvalues of both matrices are identical, including their multiplicities, for a regular model with a noninformative prior. To express the relation precisely, we denote the eigenvalues of $\hat{\mathcal{J}}^{-1/2}\hat{\mathcal{I}}\hat{\mathcal{J}}^{-1/2}$ as $\lambda^I_1 \geq \lambda^I_2 \geq \cdots \geq \lambda^I_k$, and the eigenvalues of $\frac{1}{n}K_{MF}$ as $\lambda^{MF}_1 \geq \lambda^{W}_2 \geq \cdots \geq\lambda^{MF}_n$, where $k$ is the number of parameters and $n$ is the number of observations. Assuming $n > k$, the following relation holds:
\begin{align}
\label{eq:dual_MF}
    \lambda^{MF}_j = \lambda^I_j \text{ for $j \leq k$}, \ \ \lambda^{MF}_j = 0 \text{ for $j > k$}.
\end{align}

This relation is already discussed in \cite{Kernel_Schoelkopf}, Section~13.4 and \cite{oliver2000natural} for a generic form of the natural kernels, without concerning the relation to the W-kernel. For convenience, however, we have provided an outline of the derivation here. 

First, we explicitly provide a finite dimensional feature vector that corresponds to the modified Fisher kernels \eqref{eq:MF_kernel}. Using the principal square roots  $\hat{\mathcal{J}}^{-1/2}$ of the matrices $\hat{\mathcal{J}}^{-1}$ as positive-semidefinite matrices that satisfy 
\begin{align}
\label{eq:square_root}
\hat{\mathcal{J}}^{-1}_{\alpha\beta}=
\sum_{\gamma=1}^k
\hat{\mathcal{J}}^{-1/2}_{\alpha\gamma}\hat{\mathcal{J}}^{-1/2}_{\gamma\beta}, 
\end{align}
and we define
\begin{align}
\label{eq:MF_feature}
\Phi^{MF}(\gamma,X)
=\sum_{\alpha=1}^k
  (\hat{\mathcal{J}}^{-1/2})_{\gamma\alpha}
 \frac{\partial}{\partial \theta_\alpha} \logp{X}{\theta} \bigg |_{\theta=\hat{\theta}},
\end{align}

Using the relation
\begin{align}
\,\,\,
\frac{1}{n} K_{MF}(X_i,X_j) & =
\frac{1}{n} \sum_{\gamma=1}^k
\Biggl \{
\sum_{\alpha=1}^k
(\hat{\mathcal{J}}^{-1/2})_{\gamma\alpha}
 \frac{\partial}{\partial \theta_\alpha} \logp{X_i}{\theta} \bigg |_{\theta=\hat{\theta}}
\Biggr \}
\Biggl \{ 
\sum_{\beta=1}^k
(\hat{\mathcal{J}}^{-1/2})_{\gamma\beta}
\frac{\partial}{\partial \theta_\beta} \logp{X_j}{\theta} \bigg |_{\theta=\hat{\theta}}
\Biggr \},
\end{align}
we obtain
\begin{align}
\label{eq:MF_duality_1} 
\frac{1}{n} K_{MF}(X_i,X_j) = \frac{1}{n} \sum_{\gamma=1}^k  \Phi^{MF}(\gamma,X_i)\Phi^{MF}(\gamma,X_j).
\end{align}
This shows that the score function normalized by 
$\hat{\mathcal{J}}$ defined as \eqref{eq:MF_feature} provides a feature vector corresponding to the empirical version of the modified Fisher kernel $K_{MF}$. 

Next, a computation in Appendix~\ref{sec:app:feature_space} derives that the feature vector $\Phi^{MF}(\gamma,X_j)$ satisfies the relation
\begin{align}
\label{eq:MF_duality_2}
\left ( \hat{\mathcal{J}}^{-1/2}\hat{\mathcal{I}}\hat{\mathcal{J}}^{-1/2} \right )_{\gamma\delta} & =
\frac{1}{n} \sum_{i=1}^n  \Phi^{MF}(\gamma,X_i)\Phi^{MF}(\delta,X_i).
\end{align}
Now that we have a dual pair of the relations \eqref{eq:MF_duality_1} and \eqref{eq:MF_duality_2}. Since both of \eqref{eq:MF_duality_1} and \eqref{eq:MF_duality_2} have a form of the Euclidean inner product, an argument standard for the duality in the PCA gives the relation \eqref{eq:dual_MF}.

\subsubsection*{Eigenvalues of the W-kernel}

Thus far, we consider the modified Fisher kernel. Let us turn to the issues of the W-kernel. When the W-kernel $K_w$ is approximated by the modified Fisher kernel $K_{MF}$ as \eqref{eq:MF_kernel_approx}, the relation \eqref{eq:dual_MF} is converted to a relation between 
the eigenvalues of the matrix $W$ and the matrix $\hat{\mathcal{J}}^{-1/2}\hat{\mathcal{I}}\hat{\mathcal{J}}^{-1/2}$. If we denote the eigenvalues of $W$ as $\lambda^{W}_1 \geq \lambda^{W}_2 \geq \cdots \geq\lambda^{W}_n$, and assuming large $n$, this relation is expressed as follows: 
\begin{align}
\label{eq:dual_w}
    \lambda^{W}_j \simeq \lambda^I_j \text{ for $j \leq k$}, \ \ \lambda^{W}_j \simeq 0 \text{ for $j > k$}.
\end{align}
Here we use the matrix $W$ equals $\frac{1}{n} K_W(X_i,X_j)$.

This relation explains why the number of virtually non-zero eigenvalues of the matrix $W$ approximately equals $k$ for a regular model with a noninformative prior, when $n \rightarrow \infty$, as in the first and second examples in Section~\ref{sec:essential_example}.

It also provides an explanation for the dependence and apparent instability of the eigenvectors of the matrix $W$ to the observations $(X_i)$ ; this reflects a common property of the dual PCA in the data space when $n \gg k$. 

We stress that the relation between the W-kernel and the matrix $\hat{\mathcal{J}}^{-1/2}\hat{\mathcal{I}}\hat{\mathcal{J}}^{-1/2}$
discussed here is limited to a parametric model with noninformative priors. We may generalize the expression $\hat{\mathcal{J}}^{-1/2}\hat{\mathcal{I}}\hat{\mathcal{J}}^{-1/2}$ to deal with informative priors and structured models, but the complexity of formulae increases and model-by-model analytical calculation becomes more demanding. In contrast, the W-kernel provides a unified formula for dealing with these cases, which can be automatically computed from posterior samples.  

\begin{remark}
If the model can represent the source population, the matrix $\hat{\mathcal{J}}^{-1/2}\hat{\mathcal{I}}\hat{\mathcal{J}}^{-1/2}$ asymptotically converges to the identity matrix (\cite{Kernel_Schoelkopf}, Section~13.4; \cite{oliver2000natural}). In this case, the nonzero eigenvalues of the matrix $W$ approach unity. However, the computation of the corresponding eigenvectors of $W$ is nontrivial, and their representation in terms of weights on the observations $(X_i)$ depends much on the dataset.
\end{remark}

\subsubsection*{\protect{A function-space perspective from equations \eqref{eq:MF_duality_1}--\eqref{eq:MF_duality_2}}}

While the relations \eqref{eq:MF_duality_1} and \eqref{eq:MF_duality_2} are naturally interpreted within the framework of the classical PCA, it is also instructive to view them from the perspective of function space. This viewpoint will be useful in Section~\ref{sec:comparison_two}.

We first specify the dual pair of function spaces. The primal space is spanned by the modified score functions $\Phi^{MF}(\alpha,X)$ for $\alpha=1,\ldots,k$, 
and its generic element is written as
\begin{align}
    f(X)= \sum_{\alpha=1}^k a_\alpha \, \Phi^{MF}(\alpha,X),
\end{align}
where $a_\alpha$ ($\alpha=1,\ldots,k$) are constants. 
The dual space is spanned by the functions $\Phi^{MF}(X_i,\alpha)$ for $i=1,\ldots,n$, 
which are regarded as functions of $\alpha$, 
and its generic element is written as
\begin{align}
    F(\alpha)= \sum_{i=1}^n c_i \, \Phi^{MF}(\alpha,X_i),
\end{align}
where $c_i$ ($i=1,\ldots,n$) are constants.

The inner product in each space is naturally induced by 
\eqref{eq:MF_duality_1} and \eqref{eq:MF_duality_2}, respectively.
Using these relations, the inner products are explicitly written as
\begin{align}
(f,g)_{p}
&=\sum_{\alpha\gamma=1}^k 
a_\alpha \,
n (\hat{\mathcal{J}}^{-1/2}\hat{\mathcal{I}}\hat{\mathcal{J}}^{-1/2})_{\alpha\gamma} \,
b_\gamma, 
\\
(F,G)_{d}
&=\sum_{ij=1}^n 
c_i\, K_{MF}(X_i,X_j)\, d_j,
\end{align}
where 
$g(X)= \sum_{\alpha=1}^k b_\alpha \, \Phi^{MF}(\alpha,X)$ 
and 
$G(\alpha)= \sum_{i=1}^n d_i \, \Phi^{MF}(\alpha,X_i)$, respectively.

\section{The $Z$ Matrix and Duality in PCA}
\label{sec:dual2}

In this final section, we define the matrix $Z$, which directly provides a form of PCA dual to that defined by the matrix $W$. Using this formulation, we give an alternative derivation of the relation~\eqref{eq:dual_w}, which characterizes the eigenvalues of the  matrix $W$ and of the matrix $\hat{\mathcal{J}}^{-1/2}\hat{\mathcal{I}}\hat{\mathcal{J}}^{-1/2}$.

\subsection{The dual pair $Z$ and $W^c$ of Covariance Matrices}

\subsubsection*{\protect{The $Z$ matrix}}

A convenient way to define the matrix \( Z \) is to use the notation \( \estCovX \) given in \eqref{eq:estX}, which allows for a compact expression. For convenience, we reproduce it here as 
\begin{align}
\estCovX[f(X), g(X)]
= \frac{1}{n} \sum_{i=1}^n \left \{
{\left (f(X_i)-\estEpX[f(X)] \right )
\left (g(X_i)-\estEpX[g(X)] \right )} \right \}.
\end{align} 

Using this, the matrix $Z$ is defined as
\begin{align}
\label{eq:Z_duality}
Z_{rs}= \estCovX \Bigg[
\logp{X}{\Theta^{(r)}}
-\frac{1}{M} \sum_{u=1}^M \logp{X}{\Theta^{(u)}}, \, 
\logp{X}{\Theta^{(s)}}
-\frac{1}{M} \sum_{u=1}^M \logp{X}{\Theta^{(u)}}
\Bigg].
\end{align}

\begin{remark}
In contrast to the W-kernel case, the term 
\[
-\frac{1}{M} \sum_{u=1}^M \logp{X}{\Theta^{(u)}}
\]
in this definition cannot be ignored, even under the assumption of a noninformative prior (see Appendix~\ref{sec:app:Zem}). This is why we define only the double-centered version of the matrix here.      
\end{remark}

\subsubsection*{The {\protect \( W^c \)} matrix}

The covariance matrix $Z$ is essentially the dual counterpart of the matrix $W$ in the sense of PCA. However, by introducing a modified version $W^c$ of $W$, which has parallel characteristics to $Z$, we can describe the duality with $Z$ in a more concise and precise manner. The definition of $W^c$ should satisfy the following properties:
\begin{itemize}
\item[(1)]
Since the definition of $Z$ in \eqref{eq:Z_duality} is given in a double-centered form, the matrix $W^c$ should also be defined as a double-centered version of $W$.
\item[(2)]
Since the form of $Z$ in \eqref{eq:Z_duality} explicitly depends on $(X_i)$, the matrix $W^c$ should be defined in a way that makes explicit use of the posterior samples $\Theta^M=(\Theta^{(1)},\cdots,\Theta^{(M)})$. 
\end{itemize}

The double-centered form required by property~(1) is already given in \eqref{eq:K_w_double}. The requirement in (2) appears newly in this section; up to now, we have implicitly assumed the limit of large posterior samples \( M \to \infty \) in order to define posterior averages and covariances.

To express the finite-posterior-sample version, it is convenient to denote the posterior covariance of arbitrary statistics $A(\theta)$ and $B(\theta)$, estimated from a finite sample, as  
\begin{align}
\estCovpos[A(\theta),B(\theta)]
= \frac{1}{M} \sum_{u=1}^M \left \{
 {\left (A(\Theta^{(u)})-\estEppos[A] \right )
 \left (B(\Theta^{(u)})-\estEppos[B] \right )} \right \}, 
\label{eq:covposfinite}
\end{align}
Using this, the matrix $W^c$ is defined as:
\begin{align}
\label{eq:W_duality}
W^c_{ij}=
\estCovpos \Bigg[
\logp{X_i}{\theta}
-\frac{1}{n} \sum_{l=1}^n \logp{X_l}{\theta}, \, 
\logp{X_j}{\theta}
-\frac{1}{n} \sum_{l=1}^n \logp{X_l}{\theta}
\Bigg],
\end{align}
where $M$ is the number of posterior samples.

The matrix $W^c$ converges to
\begin{align}
\label{eq:Kc/n}
\frac{1}{n} K_W^c(X_i,X_j)=
\Covpos \Bigg[
\logp{X_i}{\theta}
-\frac{1}{n} \sum_{l=1}^n \logp{X_l}{\theta}, \, 
\logp{X_j}{\theta}
-\frac{1}{n} \sum_{l=1}^n \logp{X_l}{\theta}
\Bigg],
\end{align}
in the limit of large posterior samples \( M \to \infty \). As discussed in Appendix~\ref{sec:double_center}, the expression \eqref{eq:Kc/n} is asymptotically equivalent to the matrix $W$ under the assumption that the effect of the prior is negligible. In summary, the matrix $W^c$ is asymptotically equivalent to $W$ when both $n$ and $M$ are large and the effect of the prior is negligible.

\subsubsection*{The duality between the {\protect \( Z \) matrix} and  {\protect \( W^c \) matrix} }

To capture the dual relation between \( Z \) and \( W^c \), we define the matrix \( \mathcal{A} \) as
\begin{align}
\mathcal{A}_{ir}=
\logp{X_i}{\Theta^{(r)}}
-\estEppos[\logp{X_i}{\theta}]
-\estEpX[\logp{X}{\Theta^{(r)}}]
+\estEpX[\estEppos[\logp{X}{\theta}]],
\end{align}
where a direct computation presented in Appendix~\ref{sec:app:WZduality} yields
\begin{align}
     \frac{1}{M}  Z=\frac{1}{nM} \, \mathcal{A}^T \! \mathcal{A}, \qquad
    \frac{1}{n} W^c=\frac{1}{nM} \, \mathcal{A}\mathcal{A}^T,
\label{eq:AA}
\end{align}
where \( T \) denotes matrix transposition.

The relations in \eqref{eq:AA} are the core result of this subsection, showing that the matrix \( Z \) serves as the dual of \( W^c \) in the sense of PCA. By a standard result from ordinary PCA, it follows that \( \frac{1}{M} Z \) and \( \frac{1}{n} W^c \) share the same nonzero eigenvalues, with a correspondence between the associated eigenvectors. Moreover, under the assumption that both $n$ and $M$ are large and the effect of the prior is negligible, the same correspondence asymptotically holds between \( \frac{1}{M} Z \) and \( \frac{1}{n} W \). To state the eigenvalue relation precisely, let the eigenvalues of $Z$ be ordered as $\lambda^Z_1 \geq \lambda^Z_2 \geq \cdots \geq \lambda^Z_M$. Assuming $M > n$, we obtain
\begin{align}
\label{eq:dual_ZW}
     \frac{n}{M} \lambda^{Z}_j =\lambda^W_j \quad \text{for $j \leq n$}, 
    \qquad \lambda^{Z}_j = 0 \quad \text{for $j > n$}.
\end{align}

In practice, the matrix \( Z \) is useful for reducing the computational cost of obtaining the principal space of \( W \) or \( W^c \), particularly when the number of posterior samples \( M \) is smaller than the sample size \( n \). Such situations occur, for example, in multiple imputation for handling missing data, where the number of imputations (i.e., posterior samples) \( M \) is often limited. 

However, our main purpose in defining \( Z \) here is to give an alternative derivation of relation~\eqref{eq:dual_w}, which will be discussed in the following subsection.

\subsubsection*{\protect{A function-space perspective from equation \eqref{eq:AA}}}

In the finite posterior-sample formulation developed here, 
the relations \eqref{eq:AA} are interpreted within the framework of the classical PCA. 
However, we now view them from the perspective of function space, which will be essential in Section~\ref{sec:comparison_two}. 

We first specify the dual pair of function spaces. 
The primal space is spanned by the functions $\logp{X}{\Theta_r}$ for $r=1,\ldots,M$, 
and its generic element is written as
\begin{align}
    f(X)= \sum_{r=1}^M a_r \, \logp{X}{\Theta_r},
\end{align}
where $a_r$ ($r=1,\ldots,M$) are constants. 
The dual space is spanned by the functions $\logp{X_i}{\theta}$ for $i=1,\ldots,n$, 
which are regarded as functions of $\theta$, 
and its generic element is written as
\begin{align}
    F(\theta)= \sum_{i=1}^n c_i \, \logp{X_i}{\theta},
\end{align}
where $c_i$ ($i=1,\ldots,n$) are constants.

From equations \eqref{eq:Z_duality}, \eqref{eq:W_duality}, and \eqref{eq:AA}, 
it is readily observed that both spaces are equipped with 
the frequentist covariance $\estCovX$ defined by \eqref{eq:estX} 
and the posterior covariance $\estCovpos$ defined by \eqref{eq:covposfinite}, 
which constitute a dual pair of inner products. 
Hence, the duality of PCA in this framework takes on the meaning of a Bayesian-Frequentist duality.

\subsection{{\protect Asymptotic relation between the matrices $Z$ and $\hat{\mathcal{J}}^{-1/2}\hat{\mathcal{I}}\hat{\mathcal{J}}^{-1/2}$}}
\label{sec:comparison_two}

Here, we study the correspondence between the principal spaces of the matrices  \( Z \) and $\hat{\mathcal{J}}^{-1/2}\hat{\mathcal{I}}\hat{\mathcal{J}}^{-1/2}$. Using this relation, we give an alternative derivation of the relation~\eqref{eq:dual_w}.

\subsubsection*{Orthogonal embedding}

Now, there are two distinct but related notions of duality: The first is represented by \eqref{eq:MF_duality_1} and \eqref{eq:MF_duality_2}, and the second by the equation~\eqref{eq:AA}. The essential difference between these dual relations lies in the choice of basis (and the corresponding subspace) within the space of functions of the observation $X$. 
For \eqref{eq:MF_duality_1} and \eqref{eq:MF_duality_2}, as explained in Section~\ref{sec:dual}, the basis is formed by the modified score functions $\Phi^{MF}(\alpha,X)$ for $\alpha=1,\ldots,k$, evaluated at $\theta=\hat{\theta}$. For \eqref{eq:AA}, an overcomplete basis is used, consisting of $\logp{X}{\Theta^{(r)}}$ for $r=1,\ldots,M$, where $\Theta^{(r)}$ denotes the $r$th posterior sample. Although each basis spans a different subspace of the function space of $X$, they are asymptotically related in the present setting, as shown below. 

The following derivations mostly rely on Appendix~\ref{sec:app:Zem}. Throughout the following discussion, we assume $k < n \leq M$, 
where $k$, $n$, and $M$ denote the numbers of parameters, observations, and posterior samples, respectively. We also assume that the effect of the prior is negligible.

Let us define the vectors $\Delta L$ with components
\begin{align}
 \Delta L_r=\logp{X}{\Theta^{(r)}}-\frac{1}{M} \sum_{u=1}^M \logp{X}{\Theta^{(u)}},
\end{align}
and $\bm{\Phi}^{MF}$ with components $\bm{\Phi}^{MF}_\alpha=\Phi^{MF}(\alpha,X)$.
Here, we subtract the mean in the definition of  $\Delta L$ for convenience. We also define the matrix \( \tilde{\mathit{\Theta}} \) by
\begin{align}
    \tilde{\mathit{\Theta}} = \sqrt{\frac{n}{M}} \hat{\mathcal{J}}^{1/2}\mathit{\Theta},
\label{eq:itTheta_main}   
\end{align} 
using the matrix \( \mathit{\Theta} \) whose components are $\mathit{\Theta}_{\alpha r}= \Theta^{(r)}_\alpha-\hat{\theta}_\alpha$, where $\hat{\theta}_\alpha$ is the MLE of the parameter $\theta_\alpha$. As shown in Appendix~\ref{sec:app:Zem}, the matrix \( \tilde{\mathit{\Theta}} \) satisfies
\begin{align}
\label{eq:ortho_proj_main}
\lim_{M \rightarrow \infty} 
\tilde{\mathit{\Theta}}\tilde{\mathit{\Theta}}^T =I_k + o_p(1),
\end{align}
where \( I_k \) is the \( k \times k \) identity matrix. The relation \eqref{eq:ortho_proj_main} implies that any singular value of the \( k \times M \) matrix \( \tilde{\mathit{\Theta}} \) asymptotically converges to either \( 0 \) or \( 1 \). Thus, \( \tilde{\mathit{\Theta}} \) can be approximated by a matrix representing an orthogonal embedding of a \( k \)-dimensional space into an \( M \)-dimensional space when both \( n \) and \( M \) are large.

Then, the relation between $\Delta L$ and $\bm{\Phi}^{MF}$ is expressed as
\begin{align}
    \Delta L 
    = \sqrt{\frac{M}{n}}\,\tilde{\mathit{\Theta}}^{\!T}\bm{\Phi}^{MF} 
       + o_p\!\left(\frac{1}{\sqrt{n}}\right),
\label{eq:MFtol}
\end{align}
using the relation \eqref{eq:expand_for_Z}. According to \eqref{eq:MFtol}, the matrix $\tilde{\mathit{\Theta}}$ asymptotically represents the transformation between the two bases, $\bm{\Phi}^{MF}$ and $\Delta L$, in the feature space. 

\subsubsection*{\protect {Relation between the matrices $Z$ and $\hat{\mathcal{J}}^{-1/2}\hat{\mathcal{I}}\hat{\mathcal{J}}^{-1/2}$} and derivation of the relation~\eqref{eq:dual_w}}

The idea of orthogonal embedding suggests that the matrices $Z$ and $\hat{\mathcal{J}}^{-1/2}\hat{\mathcal{I}}\hat{\mathcal{J}}^{-1/2}$ are also naturally connected through a corresponding transformation defined by $ \tilde{\mathit{\Theta}}$. Specifically, the following asymptotic relation holds:
\begin{align}
\frac{n}{M} Z 
= \tilde{\mathit{\Theta}}^T \, 
(\hat{\mathcal{J}}^{-1/2}\hat{\mathcal{I}}
\hat{\mathcal{J}}^{-1/2} ) \,
\tilde{\mathit{\Theta}}
+ o_p(1).
\label{eq:nMZZZ}
\end{align}
The derivation is also in Appendix~\ref{sec:app:Zem}.

Using this relation and the equation \eqref{eq:nMZZZ}, we can show that the nonzero eigenvalues of $\frac{n}{M}Z$ and $\hat{\mathcal{J}}^{-1/2}\hat{\mathcal{I}}\hat{\mathcal{J}}^{-1/2}$ asymptotically coincide when both \( n \) and \( M \) are large and the effect of the prior is negligible. It is stated as:
\begin{align}
\label{eq:ZandMF}
     \frac{n}{M} \lambda^{Z}_j \simeq\lambda^I_j \quad \text{for $j \leq k$}, 
    \qquad \lambda^{Z}_j \simeq 0 \quad \text{for $j > k$}.
\end{align}
If we combine the approximate relation \eqref{eq:ZandMF} with the dual relation \eqref{eq:dual_ZW}, we can re-derive the relation~\eqref{eq:dual_w}.

\subsubsection*{Comparison of two derivations}

In Section~\ref{sec:Fkernel}, we introduced the kernel $K_{MF}$ as a stepping stone and used the dual relations \eqref{eq:MF_duality_1} and \eqref{eq:MF_duality_2} to derive the asymptotic relation between the eigenvalues of the matrices $W$ and $\hat{\mathcal{J}}^{-1/2}\hat{\mathcal{I}}\hat{\mathcal{J}}^{-1/2}$. This route corresponds to the solid arrow $W \rightarrow K_{MF} \rightarrow \hat{\mathcal{J}}^{-1/2}\hat{\mathcal{I}}\hat{\mathcal{J}}^{-1/2}$ in the diagram below.
\begin{align}
\xymatrix{
     W  \ar[r]^{} \ar@{..>}[d]_{} &  K_{MF} \ar[d]_{} 
     \\ 
    Z \ar@{..>}[r]^{} & \hat{\mathcal{J}}^{-1/2}\hat{\mathcal{I}}\hat{\mathcal{J}}^{-1/2}
  } 
\label{eq:diagram}
\end{align}

In contrast, in this section, we take $Z$ as the stepping stone and use the dual relation \eqref{eq:AA} to obtain the same asymptotic relation. This corresponds to the alternative route $W \rightarrow Z \rightarrow \hat{\mathcal{J}}^{-1/2}\hat{\mathcal{I}}\hat{\mathcal{J}}^{-1/2}$, indicated by the dotted arrow in the diagram above.

Note that the vertical arrows in the diagram both represent duality relations. The right arrow corresponds to \eqref{eq:MF_duality_1} and \eqref{eq:MF_duality_2}, 
whereas the left arrow corresponds to \eqref{eq:AA}. 

\section{Summary and Future Work}
\label{sec:conclusion}  

\subsubsection*{Summary}

In this study, we focus on the posterior covariance matrix $W$, defined in terms of the log-likelihood of each observation. The matrix \( W \) can be regarded as a special case of the matrix originally introduced to summarize the results of sensitivity analysis (\cite{Bradlow_1997, MacEachern2002, Thomas2018}). Here, however, we investigate it in the context of the frequentist evaluation of Bayesian estimators.

First, we employ the concept of the Bayesian infinitesimal jackknife (\cite{Bayesian_IJK}) to relate the principal space of the matrix \( W \) to frequentist properties. Using this framework, we show that quantities such as the frequentist covariance and the outcome of the approximate bootstrap can be accurately approximated using projection onto the principal space of \( W \). In particular, we propose an approximate bootstrap scheme in which the second-order terms are replaced by those calculated from the projection onto the principal space; this idea is tested in Appendix~\ref{sec:app:boot_example}.

Next, we examine the relationship between $W$ and the Fisher kernel, showing that a modified form of the Fisher kernel can be viewed as an approximation to $W$. Since the matrix $W$ itself can be interpreted as a reproducing kernel---referred to as the $W$-kernel---this correspondence can also be understood as a relation between kernels. Based on this connection, we establish a link to classical asymptotic theory and show that the matrices $W$ and $\hat{\mathcal{J}}^{-1/2}\hat{\mathcal{I}}\hat{\mathcal{J}}^{-1/2}$ share the same nonzero eigenvalues in an asymptotic regime where the effect of the prior is negligible.

Finally, we introduce the matrix $Z$, which is dual to the matrix $W$ (more precisely, to its double-centered and finite-posterior-sample form $W^c$) in the sense of PCA. Using this formulation, we provide an alternative derivation of the relation to classical asymptotic theory.

\subsubsection*{Future work}

Several important problems remain for future research. First, the applications of the matrix \( W \) and its principal space from a frequentist perspective are still limited, leaving room for further exploration. While we proposed an example of an approximate bootstrap method, many other potential applications remain to be investigated, for instance, in the missing data problem and empirical Bayes methods. Also, a mathematically rigorous proof of the heuristic results derived in this paper remains necessary.

A critical limitation in the proposed and future applications is its reliance on the availability of posterior samples. If an MCMC run is required solely to obtain these samples for applying the proposed method, alternative approaches that do not require MCMC (\cite{Giordano_etal_2018, Giordano_etal_2019, Giordano_higher}, \cite{NIPS1998_db191505,jaakkola1999}) may be preferable. However, if posterior samples have already been obtained from the original data for other purposes, methods based on the W-kernel can be applied with minimal computational cost, making them useful in many applications.

The duality represented by relation~\eqref{eq:AA} is intriguing because it also highlights a correspondence between frequentist and posterior covariances within the proposed framework. Specifically, while $Z=\tfrac{1}{n}\mathcal{A}^T\mathcal{A}$ is expressed in terms of a frequentist covariance, $W^c=\tfrac{1}{M}\mathcal{A}\mathcal{A}^T$ is expressed in terms of a posterior covariance. This suggests a novel form of Bayesian-Frequentist duality that warrants further investigation. As discussed in Section~\ref{sec:comparison_two}, this duality is also related to an apparently different one represented by \eqref{eq:MF_duality_1} and \eqref{eq:MF_duality_2}. These relations may admit geometric interpretations, and it will be important to explore their information-geometric aspects.

In addition to these considerations, we outline additional topics below that merit future research:
\begin{itemize}

\item[]{\bf The  {\protect $Z$} matrix}

The notion of the $Z$ matrix leads to a novel approach to frequentist statistics, implemented through a basis defined using posterior samples. Future work should explore its potential applications in statistical inference, as well as its computational aspects.

An interesting direction for future research is to investigate the connection to the Gaussian process $\xi(u)$, which arises in the singular learning theory developed by Watanabe (\cite{Watanabe_book_mtbs}). The eigenvectors \( (\Xi_\alpha) \) of $Z$ may be closely related to an orthogonal decomposition of $\xi(u)$, which was introduced in his framework to prove a key lemma.

\item[]{\bf Effective dimension}

A common question in this context is whether the trace of the matrix, $\Tr[W] = \sum_{i=1}^n \lambda_i$, or the number of nonzero eigenvalues in $(\lambda_i)$, that is, $\mathrm{rank}(W)$, should be regarded as the ``effective degree of freedom.'' However, there is a difference in meaning between the two. The sum of the eigenvalues, $\Tr[W]$, represents an effective dimension that quantifies the average response to perturbations induced by sampling from the population (in practice, this is approximated using bootstrap sampling of observations). In contrast, $\mathrm{rank}(W)$ indicates the minimum dimension required to represent the response to any perturbation. Another potentially useful quantity derived from the eigenvalues in $(\lambda_i)$ is $\max(\lambda_i)$, which represents the response to perturbation in the most sensitive direction. Further investigation is needed to clarify the theoretical and practical implications of these different notions of effective dimension.

\item[] {\bf Relation to the theory of influence functions} \\
The connection to existing studies on influence functions should be further investigated, for example, the work discussed in \cite{Tangent_van_der_Varrt}.

\item[] {\bf High-dimensional settings} \\
While WAIC has been shown to provide an asymptotically consistent estimator of predictive loss in a certain high-dimensional setting (\cite{okuno_yano_2023}), the general form of the Bayesian IJ may fail to yield asymptotically correct results in such settings (\cite{Bayesian_IJK}). It is therefore important to assess the accuracy of the approximations proposed in this paper under high-dimensional regimes.

\end{itemize}

\section*{Acknowledgements}

I would like to thank Keisuke Yano and Ryan Giordano for the fruitful discussions and useful advice. Also, I sincerely thank the anonymous reviewer for constructive suggestions. I am grateful to Shotaro Akaho, who carefully read earlier versions of the manuscript and provided me significant advice.  I am also grateful to Ryo Karakida, Yusaku Ohkubo, Akifumi Okuno, and Hisateru Tachimori for their kind advice and important suggestions. This work was supported by a JSPS KAKENHI Grant Number JP21K12067 and JP24K15120.

\section*{Note added in proof}

After submission of the final version of this
manuscript, we became aware of several studies discussing the W matrix in the context of Bayesian coresets. The notion of a Bayesian coreset
appears closely related to the representative set of observations discussed in Appendix C. 
The interested reader may wish to consult the following references:\\
Campbell, T.  and Beronov, B. ,
Sparse variational inference: Bayesian coresets from scratch,
Proceedings of the 33rd International Conference on Neural Information
Processing Systems (NeurIPS 2019),
Article No.1028, Pages 11461 - 11472.\\
Naik,C. , Rousseau,J. and Trevor Campbell, T.,
Fast Bayesian coresets via subsampling and quasi-newton refinement,
36th Conference on Neural Information Processing Systems (NeurIPS 2022),
Article No.6, Pages 70 - 83.\\
Chen, N. and Campbell, T.,
Coreset Markov chain Monte Carlo,
arXiv:2310.17063. \\
Campbell, T.,
General bounds on the quality of Bayesian coresets,
arXiv:2405.11780.

\appendix 
\renewcommand{\thesubsection}{A~\arabic{subsection}}

\section{Derivation of Formulae}
\label{sec:app:derivation}

\subsection{The Bound {\protect\eqref{eq:var_residual}} of the Residual Variance}
\label{sec:app:PCAvariance}

Using the relation \eqref{eq:covpos_U}, we obtain
\begin{align}
\label{eq:var_ineq1}
    \Varpos[\logp{X_{i}}{\theta}-\logpstar{X_{i}}{\theta}]
    =\Varpos \left [ \sum_{a=a_M+1}^n  U^a_i \sproj_a(\theta) \right ]
    =
    \sum_{a=a_M+1}^n \lambda_a |U^a_i |^2.
\end{align}
On the other hand, since $U=(U^i_a)$ is an orthogonal matrix 
\begin{align}
    |U^a_i |^2 \leq \sum_{b=1}^n |U^b_i |^2=1,
\end{align}
which results in
\begin{align}
\label{eq:Ueigen_ineq}
    \sum_{a=a_M+1}^n \lambda_a |U^a_i |^2 \leq 
    \sum_{a=a_M+1}^n \lambda_a.
\end{align}
Using \eqref{eq:var_ineq1} and \eqref{eq:Ueigen_ineq}, we obtain a bound of the residuals as
\begin{align}
\Varpos[\logp{X_{i}}{\theta}-\logpstar{X_{i}}{\theta}]
    \leq
    \sum_{a=a_M+1}^n \lambda_a.
\end{align}

\subsection{Derivation of the Relation {\protect \eqref{eq:MF_duality_2}}}
\label{sec:app:feature_space}

From \eqref{eq:MF_feature}, a direct calculation results in 
\begin{align}
\frac{1}{n} \sum_{i=1}^n  \Phi^{MF}(\gamma,X_i)\Phi^{MF}(\delta,X_i)= 
\frac{1}{n} \sum_{i=1}^n
\Biggl \{
\sum_{\alpha=1}^k
(\hat{\mathcal{J}}^{-1/2})_{\gamma\alpha}
 \frac{\partial}{\partial \theta_\alpha} \logp{X_i}{\theta} \bigg |_{\theta=\hat{\theta}}
\Biggr \}
\Biggl \{ 
\sum_{\beta=1}^k
(\hat{\mathcal{J}}^{-1/2})_{\delta\beta}
\frac{\partial}{\partial \theta_\beta} \logp{X_i}{\theta} \bigg |_{\theta=\hat{\theta}}
\Biggr \}
\\
=
\sum_{\alpha,\beta=1}^k
(\hat{\mathcal{J}}^{-1/2})_{\gamma\alpha}\,
\left [
\frac{1}{n}\sum_{i=1}^n
\left \{
\frac{\partial}{\partial \theta_\alpha} \logp{X_i}{\theta} \bigg |_{\theta=\hat{\theta}}
\frac{\partial}{\partial \theta_\beta} \logp{X_i}{\theta} \bigg |_{\theta=\hat{\theta}}
\right \}
\right ] \,
(\hat{\mathcal{J}}^{-1/2})_{\beta\delta}
\\
=
\sum_{\alpha,\beta=1}^k
(\hat{\mathcal{J}}^{-1/2})_{\gamma\alpha}\,
\hat{\mathcal{I}}_{\alpha\beta}
(\hat{\mathcal{J}}^{-1/2})_{\beta\delta}
= \left ( \hat{\mathcal{J}}^{-1/2}\hat{\mathcal{I}}\hat{\mathcal{J}}^{-1/2} \right )_{\gamma\delta}.
\end{align}

\subsection{The Duality between Matrices $W^c$ and $Z$}
\label{sec:app:WZduality}
\allowdisplaybreaks[1]

Our derivation here is based on two different interpretations of the same expression:
\begin{align}
\mathcal{A}_{ir}=
\logp{X_i}{\Theta^{(r)}}
-\estEppos[\logp{X_i}{\theta}]
-\estEpX[\logp{X}{\Theta^{(r)}}]
+\estEpX[\estEppos[\logp{X}{\theta}]].
\end{align}
First, we order the terms as 
\begin{align}
& \mathcal{A}_{ir}=
(\logp{X_i}{\Theta^{(r)}}-\estEppos[\logp{X_i}{\theta}])
-(\estEpX[\logp{X}{\Theta^{(r)}}]-\estEpX[\estEppos[\logp{X}{\theta}]])
\\
& 
= \left ( \logp{X_i}{\Theta^{(r)}}-\frac{1}{M} \sum_{u=1}^M \logp{X_i}{\Theta^{(u)}} \right )
-\estEpX[\text{1st.}],
\\
\shortintertext{where the ``1st.'' indicates a repeat of the  expression in the just preseding parensis $()$. Then, we compute the components of the matrix $\mathcal{A}^T\mathcal{A}$ as}
& (\mathcal{A}^T\!\mathcal{A})_{rs} =\sum_{i=1}^n \mathcal{A}_{ir}\mathcal{A}_{is}
\\ 
& = \sum_{i=1}^n \Biggl [
\left \{ \left ( \logp{X_i}{\Theta^{(r)}}-\frac{1}{M} \sum_{u=1}^M \logp{X_i}{\Theta^{(u)}} \right )
-\estEpX[\text{1st.}]
\right \} \\
& \qquad \qquad \times
\left \{ \left ( \logp{X_i}{\Theta^{(s)}}-\frac{1}{M} \sum_{u=1}^M \logp{X_i}{\Theta^{(u)}} \right )
-\estEpX[\text{1st.}]
\right \} \Biggr ]
\\
& = n \estCovX \left [ 
\logp{X}{\Theta^{(r)}}-\frac{1}{M} \sum_{u=1}^M \logp{X}{\Theta^{(u)}} \,\, , \,\,
\logp{X}{\Theta^{(s)}}-\frac{1}{M} \sum_{u=1}^M \logp{X}{\Theta^{(u)}}
\right ].
\end{align}
Next, we order the terms in a different way as  
\begin{align}
& \mathcal{A}_{ir}=
(\logp{X_i}{\Theta^{(r)}}-\estEpX[\logp{X}{\Theta^{(r)}}])
-(\estEppos[\logp{X_i}{\theta}]-\estEppos[\estEpX[\logp{X}{\theta}]])
\\
& = \left (\logp{X_i}{\Theta^{(r)}}-\frac{1}{n} \sum_{l=1}^n [\logp{X_l}{\Theta^{(r)}}] \right )
-\estEppos[\text{1st.}],
\\
\shortintertext{where we exchange $\estCovpos[\,\,\,]$ and $\estCovX[\,\,\,]$ in the last term. Then, we compute the components of the matrix $\mathcal{A}\mathcal{A}^T$ as}
& (\mathcal{A}\mathcal{A}^T)_{ij}=\sum_{r=1}^M \mathcal{A}_{ir}\mathcal{A}_{jr}
\\ 
& = \sum_{r=1}^M \left [
\left \{ \left ( \logp{X_i}{\Theta^{(r)}}-\frac{1}{n} \sum_{l=1}^n \logp{X_l}{\Theta^{(r)}} \right )
-\estEppos[\text{1st.}]
\right \}
\left \{ \left ( \logp{X_j}{\Theta^{(r)}}-\frac{1}{n} \sum_{l=1}^n \logp{X_l}{\Theta^{(r)}} \right )
-\estEppos[\text{1st.}]
\right \} \right ]
\\
& = M \estCovpos \left [ 
\logp{X_i}{\theta}-\frac{1}{n} \sum_{l=1}^n \logp{X_l}{\theta} \,\, , \,\,
\logp{X_j}{\theta}-\frac{1}{n} \sum_{l=1}^n \logp{X_l}{\theta}
\right ].
\end{align}
 
\subsection{Derivation of the formulae {\protect \eqref{eq:nMZZZ}, \eqref{eq:ortho_proj_main}, and \eqref{eq:itTheta_main}}}

\label{sec:app:Zem}

Here, we derive the formulae \eqref{eq:nMZZZ}, \eqref{eq:ortho_proj_main}, and \eqref{eq:itTheta_main}, which are essential for establishing the relationship between the matrix $Z$ and {\protect $\hat{\mathcal{J}}^{-1/2}\hat{\mathcal{I}}\hat{\mathcal{J}}^{-1/2}$} discussed in the main text.

First we expand $ \logp{X}{\Theta^{(r)}}
-\frac{1}{M} \sum_{u=1}^M \logp{X}{\Theta^{(u)}}$ around the MLE $\hat{\theta}$ as
\begin{align}
\qquad & \logp{X}{\Theta^{(r)}}
-\frac{1}{M} \sum_{u=1}^M \logp{X}{\Theta^{(u)}} \\
& = \logp{X}{\hat{\theta}}
-\frac{1}{M} \sum_{u=1}^M \logp{X}{\hat{\theta}} \\
& \qquad +  \sum_{\alpha=1}^k \left. \frac{\partial}{\partial \theta_\alpha}
\logp{X}{\theta}\right |_{\theta=\hat{\theta}} (\theta_\alpha^{(r)}-\hat{\theta}_\alpha) 
-  \frac{1}{M} \sum_{u=1}^M \sum_{\alpha=1}^k \left. \frac{\partial}{\partial \theta_\alpha}
\logp{X}{\theta}\right |_{\theta=\hat{\theta}} (\Theta^{(u)}_\alpha-\hat{\theta}_\alpha) + o_p\left (\frac{1}{\sqrt{n}} \right )\\
& = \sum_{\alpha=1}^k \left. \frac{\partial}{\partial \theta_\alpha}
\logp{X}{\theta}\right |_{\theta=\hat{\theta}} (\Theta^{(r)}_\alpha-\hat{\theta}_\alpha) + o_p\left (\frac{1}{\sqrt{n}} \right ).
\label{eq:expand_for_Z}
\end{align}
Here, $M$ is the number of posterior samples, and we use the relation
\begin{align}
\lim_{M \rightarrow\infty} \frac{1}{M} \sum_{u=1}^M \Theta^{(u)}_\alpha =\hat{\theta}_\alpha+O_p(1/n).
\end{align}
The term $-\frac{1}{M} \sum_{u=1}^M \logp{X}{\Theta^{(u)}}$ in the definition of $Z$ is essential for the cancellation between the terms $\logp{X}{\hat{\theta}}$ and
$-\frac{1}{M} \sum_{u=1}^M \logp{X}{\hat{\theta}}$ in the expansion. Thus, we cannot remove $-\frac{1}{M} \sum_{u=1}^M \logp{X}{\Theta^{(u)}}$ from the definition of $Z$ even under the setting of noninformative priors. 

Substituting the expression \eqref{eq:expand_for_Z} in \eqref{eq:Z_duality}, we obtain
\begin{align}
& Z_{rs}=
\estCovX \left [
\logp{X}{\Theta^{(r)}}
-\frac{1}{M} \sum_{u=1}^M \logp{X}{\Theta^{(u)}} \,\,  , \,\, 
\logp{X}{\Theta^{(s)}}
-\frac{1}{M} \sum_{u=1}^M \logp{X}{\Theta^{(u)}}
\right ] \\
& = 
\sum_{\alpha,\beta=1}^k 
\estEpX \left [ 
\left. \frac{\partial}{\partial \theta_\alpha}
\logp{X}{\theta}\right |_{\theta=\hat{\theta}} 
\left. \frac{\partial}{\partial \theta_\beta}
\logp{X}{\theta}\right |_{\theta=\hat{\theta}} 
\right ]
(\Theta^{(r)}_\alpha-\hat{\theta}_\alpha)
(\Theta^{(s)}_\beta-\hat{\theta}_\beta) +o_p\left (1/n \right )
\\
& = \sum_{\alpha,\beta=1}^k \hat{\mathcal{I}}_{\alpha\beta} \,\,
(\Theta^{(r)}_\alpha-\hat{\theta}_\alpha)
(\Theta^{(s)}_\beta-\hat{\theta}_\beta) + o_p\left (1/n \right ),
\end{align}
where $k$ denotes the number of the parameters.

Defining the matrix $\mathit{\Theta}$ whose components are
$
    \mathit{\Theta}_{\alpha r}= \Theta^{(r)}_\alpha-\hat{\theta}_\alpha,
$
the above formula is rewritten as
\begin{align}
Z_{rs}=
\sum_{\alpha,\beta=1}^k \hat{\mathcal{I}}_{\alpha\beta} 
\mathit{\Theta}_{\alpha r}\mathit{\Theta}_{\beta s}
+ o_p \left ( 1/n \right ).
\end{align}
Rewriting it in a matrix form, we obtain
\begin{align}
Z = \mathit{\Theta}^T \hat{\mathcal{I}}\mathit{\Theta}
+ o_p(1/n)
= \frac{M}{n} \left (\sqrt{\frac{n}{M}} \hat{\mathcal{J}}^{1/2}\mathit{\Theta} \right )^T 
(\hat{\mathcal{J}}^{-1/2}\hat{\mathcal{I}}
\hat{\mathcal{J}}^{-1/2} )
\left (\sqrt{\frac{n}{M}}\hat{\mathcal{J}}^{1/2}\mathit{\Theta} \right)
+ o_p(1/n),
\label{eq:nMZ}
\end{align}
By multiplying both sides by $\frac{n}{M}$ for convenience
and defining
\begin{align}
    \tilde{\mathit{\Theta}} = \sqrt{\frac{n}{M}} \hat{\mathcal{J}}^{1/2}\mathit{\Theta},
\label{eq:itTheta}   
\end{align} 
we obtain
\begin{align}
\frac{n}{M} Z 
= \tilde{\mathit{\Theta}}^T \, 
(\hat{\mathcal{J}}^{-1/2}\hat{\mathcal{I}}
\hat{\mathcal{J}}^{-1/2} ) \,
\tilde{\mathit{\Theta}}
+ o_p(1).
\label{eq:nMZZ}
\end{align}

From the relation \eqref{eq:J_pos_basic} in the main text, we obtain 
\begin{align}
    \lim_{M \rightarrow \infty} \frac{1}{M}\mathit{\Theta}\mathit{\Theta}^T = \frac{\hat{\mathcal{J}}^{-1}}{n}+ o_p(1/n).
\end{align}
Using the definition \eqref{eq:itTheta} of $\tilde{\mathit{\Theta}}$, this reduces to
\begin{align}
\label{eq:ortho_proj}
\lim_{M \rightarrow \infty} 
\tilde{\mathit{\Theta}}\tilde{\mathit{\Theta}}^T =I_k + o_p(1),
\end{align}
where $I_k$ is the $k \times k$ identity matrix.

\renewcommand{\thesubsection}{B~\arabic{subsection}}   
\setcounter{subsection}{0}

\section{W-kernel}
\label{sec:kernel_2}

\subsection{Reproducing Kernel Hilbert Space Associated with the Kernel $K_W$}
\label{sec:app:RKHS}

This appendix provides a formulation of the Reproducing Kernel Hilbert Space (RKHS) associated with the kernel $K_W$. The inner product structure of $\mathcal{H}$ is directly connected to the posterior covariance structure, allowing us to reinterpret kernel PCA within the framework of $\mathcal{H}$.

The RKHS $\mathcal{H}$ associated with the kernel $K_W$ is defined as the closure of the set of functions on $X$ of the form:
\begin{align}
    f(X)=\sum_{i=1}^k c_i K_W(X,X_i), \qquad c_i \in \mathbb{R}
    \label{eq:fform}
\end{align}
where $X^k=(X_i), \, i=1, \ldots ,k$ is an arbitrary sample from the population with any sample size $k$. Using \eqref{eq:Kphiphi}, $f(X)$ can also be expressed as
\begin{align}
    f(X)=(\sproj(\theta),\Phi_{\theta}(X))_{\mathrm{pos}}, \qquad \sproj(\theta)=\sum_{i=1}^k c_i \Phi_{\theta}(X_i),
    \label{eq:fform2}
\end{align}
where the function $\sproj(\theta)$ can be interpreted as a representation of the coordinate axis determining the projection direction that defines elements in $\mathcal{H}$. The correspondence between $f(X)$ and $\sproj(\theta)$ in \eqref{eq:fform2} is not necessarily one-to-one, but we can identify them modulo a function $h(\theta)$ that satisfies $(h(\theta),\Phi_{\theta}(X))_{\mathrm{pos}}=0$ for any $X$.

Using this correspondence, we can express the contents of Section~\ref{sec:essential} in the language of the RKHS. For example, the projection in \eqref{eq:proj} corresponds to
\begin{align}
    f_a(X)=(\sproj_a (\theta), \Phi_\theta(X))_\mathrm{pos}
    =\sum_{i=1}^n U^i_a K_W(X,X_i),
\end{align}
where $\Phi_\theta(X)=\sqrt{n} \logp{X}{\theta}$. This expression focuses on the function $f_a \in \mathcal{H}$ as a function of $X$, whereas \eqref{eq:proj} represents the same concept in terms of the corresponding function $\sproj_a$ of $\theta$.

The inner product $(\,\,,\,\,)_{\mathcal H}$ in the RKHS is naturally inherited from the posterior inner product $(\,\,, \,\,)_{\mathrm{pos}}$, which is defined for functions of $\theta$ as follows. We define the inner product of the function $f$ and 
\begin{align}
    g(X)=\sum_{i=1}^k d_i K_W(X,X_i), \qquad d_i \in \mathbb{R}
    \label{eq:gform}
\end{align}
in the RKHS as the inner product of the corresponding functions that determine the coordinate axes:
\begin{align}
    (f(X),g(X))_{\mathcal H}=\left (\sum_{i=1}^k c_i\Phi_\theta(X_i),\,\,\sum_{j=1}^k d_j\Phi_\theta(X_j) \right )_{\mathrm{pos}},
\end{align}
which extends naturally to the completion of the space. Using this definition, we obtain the relation
\begin{align}
    (f(X),g(X))_{\mathcal H}=\sum_{i,j=1}^k c_id_i K_W(X_i,X_j),
\end{align}
and hence, the reproducing property holds:
\begin{align}
    (f(\cdot),K_W(\cdot,X))_{\mathcal H}=f(X).
\end{align}

\subsection{A Population Version}

In the definition of $K_W$ given by \eqref{eq:K_w}, the same dataset $X^n = (X_i)$ is used to compute the posterior average. To provide an alternative perspective, it may be useful to define a \textit{population version} $K_{W_0}$, where the posterior is independent of the observed dataset. While this population version cannot be directly applied in data analysis, results obtained with $K_W$ can be regarded as empirical approximations to those derived from $K_{W_0}$.

To define $K_{W_0}$, we need a frequentist assumption that observations are an IID realization from a population $G$. Then, a population version of the posterior with a sample size $n$ is formally introduced as
\begin{align}
    p(\theta ; G,n ) = 
    \frac{\exp\{n \int \logp{x}{\theta} dG(x) \}p(\theta) }
    {\int \exp\{n \int \logp{x}{\theta'} dG(x) \}p(\theta') d\theta'}.
    \label{eq:F_pos}
\end{align}
We recover the original posterior \eqref{eq:pos} when $dG(x)$ is replaced by the empirical version $d\hat{G}_n=(1/n)\sum_{i=1}^n \delta_{X_i}$ with point measures $\delta_{X_i}$ on observations $X^n=(X_1, \ldots, X_n)$.

If we denote a covariance with respect to the population version of the posterior \eqref{eq:F_pos} as $\Covpos^0[\,\,,\,\,]$, we can define a population version of the W-kernel of the sample size $n$ as
\begin{align} 
K_{W_0}(X,X^\prime)=
n \, \CovposP[\logp{X}{\theta}, \logp{X^\prime}{\theta}]
=n \, \CovposP[\log p(X \mid \theta), \log p(X^\prime \mid \theta)].
\end{align}

\subsection{Kernel PCA}

Here, we introduce a kernel PCA (\cite{Kernel_Schoelkopf,Kernel_Shawe-Taylor}). We assume the non-centered kernel $K_W$, characterized by both empirical and infinite-posterior-sample properties, but similar procedures are applied to other versions.

One approach to introducing kernel PCA is to maximize the projection of a unit vector. In our case, this reduces to maximizing
\begin{align}
\label{eq:calF}
    \mathcal{F}(\sproj)=
    \frac{1}{n}\sum_{i=1}^n (\sproj(\theta),\Phi_\theta(X_i))_{\mathrm{pos}}^2=
    \sum_{i=1}^n (\Covpos[\sproj(\theta),\logp{X_i}{\theta}])^2
\end{align}
subject to 
\begin{align}
(\sproj(\theta),\sproj(\theta))_{\mathrm{pos}}
=\Varpos[\sproj(\theta)]=1. 
\end{align}
If we assume the relation \eqref{eq:cov_g}, the function \(\mathcal{F}\) defined in the equation above represents the frequentist variance of \(\Eppos[\sproj(\theta)]\). Thus, our procedure can be interpreted as identifying a projection that maximizes the frequentist variance of the posterior mean while constraining the posterior variance. Introducing a Lagrange multiplier \(\mu\), the problem is further reduced to maximizing
\begin{align}
   \label{eq:proj_axis}
   \tilde{\mathcal{F}}(\sproj)=
   \sum_{i=1}^n \left(\,\Covpos[\sproj(\theta) ,\logp{X_i}{\theta}]\,\right )^2 
   - \mu\Varpos[\sproj(\theta)].  
\end{align}

A common technique in kernel methods is to constrain the form of \(\sproj(\theta)\) as
\begin{align}
    \label{eq:kernel_approx}
    \sproj(\theta)=\sum_{i=1}^n c_i \logp{X_i}{\theta}
\end{align}
where \(c = (c_i), \quad i = 1,\ldots,n\) are unknown coefficients. The key point is that the $X_i$s are not arbitrary samples from the population, but are restricted to the observations in the dataset $X^n$ under analysis. This form can be directly derived by taking the variation of \(\tilde{\mathcal{F}}\) or deduced from the general theory, as discussed in Appendix~\ref{sec:app:RKHS}.

Substituting \eqref{eq:kernel_approx} into \eqref{eq:proj_axis} leads to   
\begin{align}
   \label{eq:proj_axis_c}
   \tilde{\mathcal{F}}(c)=
   \sum_{i,j,\ell=1}^n c_i K_W(X_i,X_j)K_W(X_j,X_\ell) c_\ell 
   - \mu  \sum_{i,j=1}^n c_i K_W(X_i,X_j)c_j, 
\end{align}
where $n\mu$ is redefined as $\mu$. The optimal vector \(c\) is obtained by solving the eigenvalue problem 
\begin{align}
    \sum_{j=1}^n K_W(X_i,X_j) c_j
    = \sum_{j=1}^n\Covpos[\logp{X_i}{\theta},\logp{X_j}{\theta}] c_j  
    = \mu c_i.
\end{align}
This coincides with the eigenvalue problem that defines the principal space of the matrix $W$.

Using the language of RKHS, this is equivalent to finding a function $f(X)$ in $\mathcal{H}$ that maximizes
\begin{align}
\label{eq:calRKHS}
    \mathcal{F}^*(c)=\frac{1}{n}\sum_{i=1}^n (f(X),K(X,X_i))_{\mathcal{H}}^2 
\end{align}
under the constraint $(f(X),f(X))_{\mathcal{H}}=1$, where the function $f(X)$ is defined as
\begin{align}
    f(X)=\sum_{i=1}^n c_i K(X,X_i).
\end{align}
In this formulation, the assumption in \eqref{eq:kernel_approx} naturally follows from the representer theorem (\cite{Kernel_Schoelkopf}).

\subsection{Double Centered Kernels}
\label{sec:double_center}

So far, we have ignored the difference between the original form \eqref{eq:cov_g_star} and the simplified form \eqref{eq:cov_g} of the frequentist covariance formulae. This difference stems from the replacement of the posterior covariance \(\Covpos\) with its ``centered'' form \(\Covpos^*\), which is defined by \eqref{eq:1st_star} and represented here for convenience:  
\begin{align}
\Covpos^*[A(\theta), \logp{X_{i}}{\theta}]  
= \Covpos[A(\theta), \logp{X_{i}}{\theta}]  
- \frac{1}{n} \sum_{j=1}^n \Covpos[A(\theta),\logp{X_{j}}{\theta}],
\end{align}
where \(A(\theta)\) is an arbitrary statistic. The difference between \eqref{eq:cov_g} and \eqref{eq:cov_g_star} appears only in the presence of a strong prior, because the relation  
\begin{align}
\label{eq:cov_zero}
 \Covpos \left [A(\theta), \sum_{i=1}^n\logp{X_{i}}{\theta}  \right ]
 =o_p(1)
\end{align}
holds for a noninformative prior. It is derived in Appendix~\ref{sec:app:cov_zero}. Further details are provided in the supplementary material, along with an approach to defining the notion of a ``strong prior'' in asymptotic theory.

In the present context, this difference is naturally accounted for by using the ``double-centered'' version \(K^c\) of a kernel, which is defined as  
\begin{align}
K^c(X_i,X_j) = K(X_i,X_j)
-\frac{1}{n} \sum_{m=1}^n K(X_i,X_m) 
-\frac{1}{n} \sum_{l=1}^n K(X_l,X_j)
-\frac{1}{n^2} \sum_{l,m=1}^n K(X_l,X_m),
\end{align}
where \(K\) denotes the original kernel. This is a commonly used trick in kernel PCA. In the case of the \(W\)-kernel \(K_W\), the double-centered version is expressed as \eqref{eq:K_w_double} in the main text.  
Using \eqref{eq:cov_zero}, we can show that \eqref{eq:K_w_double} is asymptotically equivalent to $K_W(X_i,X_j)$ for a noninformative prior.

From these expressions, it is clear that using the centered form \(\Covpos^*\) instead of \(\Covpos\) is equivalent to the double centering of the \(W\)-kernel in the proposed kernel framework. Thus, replacing \(K_W\) with \(K^c_W\), we can repeat every step of the derivation of kernel PCA, except that the representation \eqref{eq:cov_g} is replaced by \eqref{eq:cov_g_star}.

\subsection{Derivation of the Relation {\protect\eqref{eq:cov_zero}} }
\label{sec:app:cov_zero}

Here, we outline a rough sketch of the derivation of the relation \eqref{eq:cov_zero} under the assumption that the sample size \( n \) increases while the prior remains fixed.

Let us define
\begin{align}
l_{pos}(\theta)
=\frac{1}{n}\sum_{i=1}^n \logp{X_{i}}{\theta}. 
\end{align}
Using the Cauchy-Schwarz inequality, we obtain  
\begin{align}
\label{eq:A_l_Cauchy_Schwarz}
 \Covpos \left [A(\theta), n l_{pos}(\theta) \right ] 
 \, \leq \,
 n \bigg (\Varpos \left [A(\theta) \right ] \bigg )^{1/2}
 \bigg ( \Varpos \left [l_{pos}(\theta) \right ] \bigg )^{1/2}.
\end{align}
This shows that it is sufficient to estimate the order of the variances on the right-hand side.

We then expand $A(\theta)$ and $l_{pos}(\theta)$ around the MLE $\hat{\theta}$ defined by $l'_{pos}(\theta)=0$ as 
\begin{align}
 A(\theta)
 & = A'(\hat{\theta})(\theta-\hat{\theta})+o(\|\theta-\hat{\theta}\|),
\\
\label{eq:lpos_expand}
 l_{pos}(\theta)
 & = \frac{1}{2} (\theta-\hat{\theta})^T \, l''_{pos}(\hat{\theta}) \, (\theta-\hat{\theta})+o(\|\theta-\hat{\theta}\|^2),
\end{align}
where $A'$ and $l''_{pos}$ indicate the matrices of the first and second derivatives and $\|\,\,\,\,\|$ denotes the $\ell_2$-norm. 
Using \eqref{eq:lpos_expand}, the order of the variances is estimated as
\begin{align*}
 \Varpos \left [A(\theta) \right ]
 & =O_p(\Eppos [(\theta-\hat{\theta})^2])
 =O_p(1/n),
\\
 \Varpos \left [l_{pos}(\theta)\right ]
 & =O_p(\Eppos [(\theta-\hat{\theta})^4])
 =O_p(1/n^2),
\end{align*}
where the behavior of $\Varpos \left [l_{pos}(\theta)\right ]$ is crucial for obtaining the result. Using them and \eqref{eq:A_l_Cauchy_Schwarz}, we obtain
\begin{align*}
 \Covpos & \left [A(\theta), nl_{pos}(\theta) \right ]  \leq O_p(1/\sqrt{n}).
\end{align*}
This leads to the relation \eqref{eq:cov_zero}
\begin{align*}
 \Covpos \left [A(\theta), \sum_{i=1}^n\logp{X_{i}}{\theta}  \right ]
 =
 \Covpos \left [A(\theta), nl_{pos}(\theta) \right ]  =o_p(1).
\end{align*}

\renewcommand{\thesubsection}{C~\arabic{subsection}}   
\setcounter{subsection}{0}

\section{Incomplete Cholesky Decomposition and Representative Set of Observations}
\label{sec:cholesky}

Here, we introduce the incomplete Cholesky decomposition as a tool for obtaining the principal space of the matrix $W$ and related kernels defined in the main text. This naturally leads to the concept of a representative set of observations that effectively spans the principal space of the matrix $W$.

\subsection{Incomplete Cholesky Decomposition}

The incomplete Cholesky decomposition developed in machine learning is an algorithm applied to a positive-semidefinite symmetric matrix; it is often used in kernel methods to treat a large Gram matrix. In our case, it corresponds to the decomposition   
\begin{align}
\label{eq:Incomplete_Cholesky}
    W=  P (LL^T + R) P^T,
\end{align}
where $L$ is a lower triangular matrix that satisfies $L_{ij}=0$ for $j>\tilde{a}_M$, $P$ is a permutation matrix, and $R$ is a residual. 

The incomplete Cholesky decomposition is computed in a sequential manner with a greedy choice of pivot at each step (\cite{Kernel_Shawe-Taylor}; the permutation matrix $P$ is determined by these choices in a post hoc manner. The diagonal sum $\Tr R$ is monitored at each step as an index of residuals (residual variance), and when it becomes negligible, we stop the procedure and consider $P(LL^T)P^T$ as an approximation of $W$ that has sufficient precision to represent the principal space. The number of non-zero columns $\tilde{a}_M$ of $L$ is determined using this procedure. The required computation is in the order of $O(\tilde{a}_M^2 \times n)$. 

Hereafter, we make the following three assumptions to simplify the expressions: (i)~The rows and columns of the matrix $W$ are already sorted simultaneously such that $P$ becomes an identity matrix,  (ii)~The residual $R$ is negligible, (iii)~$\tilde{a}_M=a_M$.
Regarding the point (iii), $\tilde{a}_M$ is usually larger than $a_M$ if we require the same accuracy, because $a_M$ corresponds to the optimal orthogonal basis, but we apply the same notation to them for simplicity.   

\begin{figure}[htb]
    \centering
    \includegraphics[width=10cm]{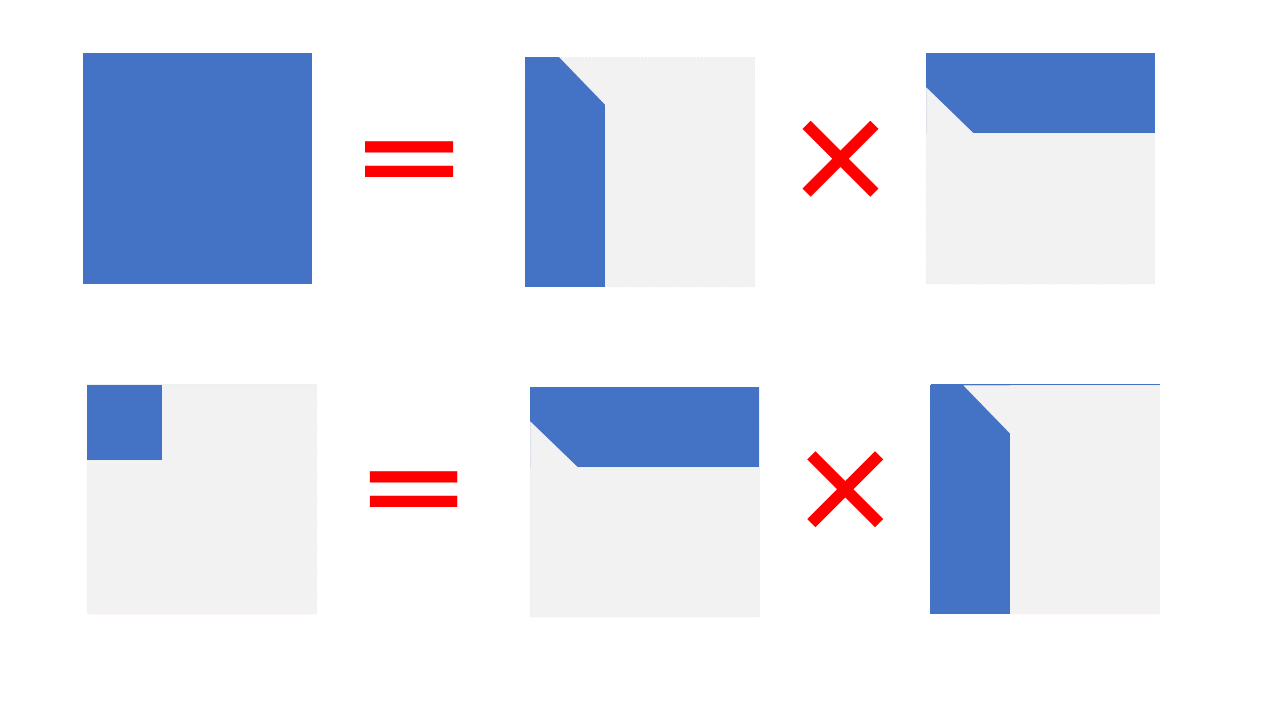}
    \caption{Schismatic view of the incomplete Cholesky decomposition.  The upper and lower panels present $W=LL^T$ and $L^TL$, respectively. Non-zero components of each matrix belong to the dark regions, while the light regions are filled with zeros. The symbol ``$\times$'' indicates a matrix product. We assume that the rows and columns of the matrix $W$ are already sorted and neglect the residual (see the main text for details).}
    \label{fig:ch2}
\end{figure}

On assuming (i) and (ii), the definition \eqref{eq:Incomplete_Cholesky} is simplified to $W= LL^T$. In Fig.~\ref{fig:ch2}, the relation $W=LL^T$ and its dual form $L^TL$ are graphically expressed. 

With these preparations, let us consider an eigenvalue problem of the matrix $L^TL$ as  
\begin{align}
\label{eq:V}
(L^TL) V_a = \lambda^{\mathrm{v}}_a V_a.
\end{align}
Hereafter, we assume that $V_a$ has a unit length and hence $\bm{V}=(V_a)$ is an orthogonal matrix. A generic form of $L^TL$ is a $a_M \times a_M$ dense symmetric matrix, which has eigenvalues $\lambda^{\mathrm{v}}_a, \, a=1,\ldots, a_M$. The $O(a_M^3)$ computation is sufficient to solve this eigenvalue problem.  

On multiplying $L$ on both sides of \eqref{eq:V}, we obtain 
\begin{align}
L(L^TL) V_a = \lambda^{\mathrm{v}}_a L V_a. 
\end{align}
Since $LL^T=W$, the above is rewritten as 
\begin{align}
\label{eq:eigen_V_U}
W (L V_a) = \lambda^{\mathrm{v}}_a (L V_a).
\end{align}
This shows that $\lambda^{\mathrm{v}}_a$, where $a=1,\ldots, a_M$, is equal to the set of non-zero eigenvalues of $W=LL^T$; other eigenvalues of $W$ are (almost) zero. Furthermore, an eigenvector of $W$ corresponding to $\lambda^{\mathrm{v}}_a$ is given by $L V_a$. A unit eigenvector is obtained as 
\begin{align}
\label{eq:eigen_tilde_U}
\tilde{U}_a=L V_a \big / \sqrt{\lambda^{\mathrm{v}}_a},    
\end{align}
because $\tilde{U}^T_a \tilde{U}_a= V^T_a L^T L V_a/\lambda^{\mathrm{v}}_a=1$.

\subsection{Representative Set of the Observations}

Until now, we have considered the incomplete Cholesky decomposition as an efficient computational method for solving a large eigenvalue problem. Another usage of the method is to provide a minimum set of the observations that can effectively represent any perturbation to the data. To observe this, let us define
\begin{align}
\label{eq:dagger_eta}
    \tilde{\eta}^\dagger_b=\sum_{i=1}^n L^T_{bi} \eta_i, \, b=1,\ldots,a_M.
\end{align}
As we assume that the matrix $W$ is already sorted in a manner such that the permutation $P$ becomes an identity matrix, the set of indices $b=1,\cdots,a_M$ corresponds to the observations selected by the incomplete Cholesky decomposition.

The expression \eqref{eq:boot_local1_essential} in the main text shows that the first-order approximate bootstrap can be implemented solely through projections onto the principal space of the matrix $W$. In the same way, any set of perturbations $(\eta_i)$ can be replaced by their projections $(\tilde{\eta}_a)$ onto the principal space, where $\tilde{\eta}_a = \sum_{i=1}^n U^i_a \eta_i$. The perturbed posterior mean of an arbitrary statistic $A(\theta)$ can then be written as
\begin{align}
\label{eq:ijk1s}
\Eppos^w [A(\theta)] \, \simeq \, 
\Eppos[A(\theta)]  
  + \sum_{a=1}^{a_M} \Covpos[A(\theta), \sproj_a(\theta)] \, \tilde{\eta}_a,
  \qquad \tilde{\eta}_a = \sum_{i=1}^n U^i_a \eta_i.
\end{align}

Using \eqref{eq:eigen_tilde_U} and \eqref{eq:dagger_eta}, $\tilde{\eta}_a$ defined in the equation \eqref{eq:ijk1s} is expressed as 
\begin{align}
\label{eq:ess_tilde_dagger}    
    \tilde{\eta}_a
    =\sum_{i=1}^n U^i_a \eta_i
    =\sum_{i=1}^n \tilde{U}^i_a \eta_i
    =\frac{1}{\sqrt \lambda_a} \sum_{b=1}^{a_M} V^b_a \left ( \sum_{i=1}^n L^T_{bi} \eta_i \right )   
    = \frac{1}{\sqrt \lambda_a} \sum_{b=1}^{a_M} V^b_a \tilde{\eta}^\dagger_b,
\end{align}
where we identify $\lambda_a^\mathrm{v}$ and $\tilde{U}^i_a$ with $\lambda_a$ and $U^i_a$.
The expression \eqref{eq:ess_tilde_dagger} shows that the projections $(\tilde{\eta}_a)$ and thus the original perturbations $(\eta_i)$ are represented by $(\tilde{\eta}^\dagger_b)$ defined on a predetermined set $b=1,\cdots,a_M$ of observations of the size $a_M$. 

Based on these results, we refer to a set of observations selected in the incomplete Cholesky decomposition as a {\it representative set of observations}. 
They have only $a_M$ elements, but perturbations to the observations in this set can simulate any perturbation in the effects on the posterior means through \eqref{eq:ess_tilde_dagger} and \eqref{eq:ijk1s}. 

We highlight the following features of the representative set of observations:
\begin{itemize}
    \item This set of observations is independent of the values of the perturbations $\eta_i$s; it is determined in the process of the incomplete Cholesky decomposition of the matrix $W$. 

    \item The choice of the members of a representative set of the observations can be sensitive to the observed data and the numerical accuracy in computing the posterior covariance. A different selection, however, corresponds to a different basis that spans a nearly equal subspace and should be almost equally well in this sense. 
    
    \item  In the special case that the original $\eta_i$ is a random number that obeys $\eta_i \sim N(0,1)$, $\tilde{\eta}^\dagger_b, \,\, b=1,\ldots, a_M$  becomes a set of correlated normal numbers, the covariance of which is given by an $a_M \times a_M$ matrix $L^TL$. Hence, instead of generating $n$ independent normal numbers, we can substitute them for $a_M$ correlated normal numbers defined on the representative set of observations.
\end{itemize}

\subsection{Example}

\begin{figure}[H]
    \centering
    \includegraphics[width=4.5cm]{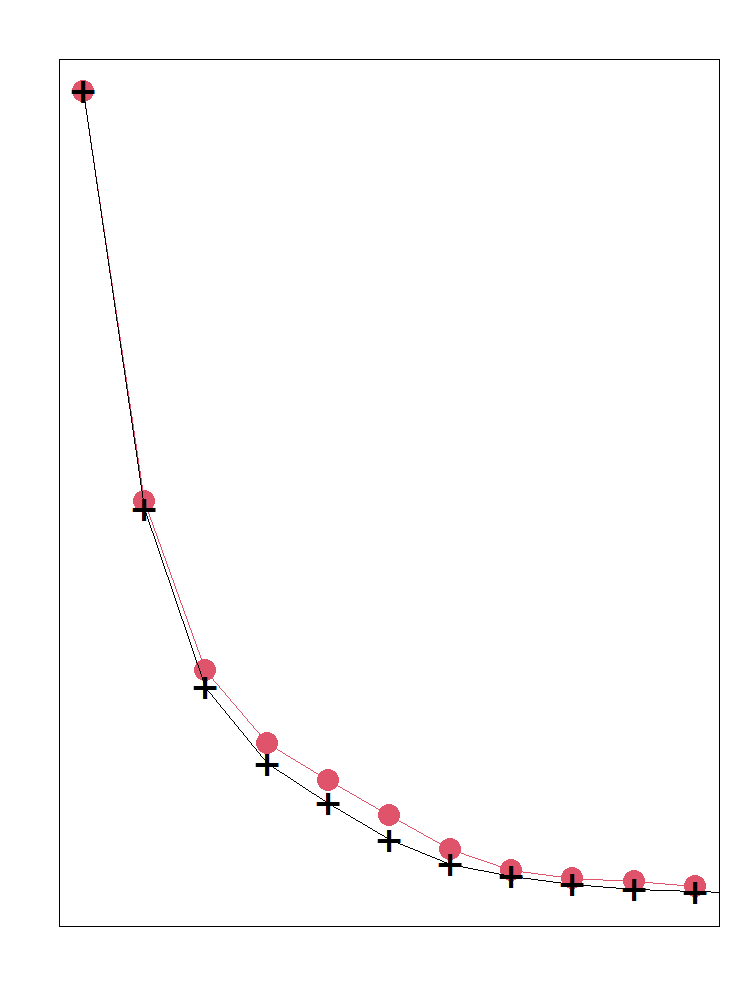} 
    \includegraphics[width=9cm]{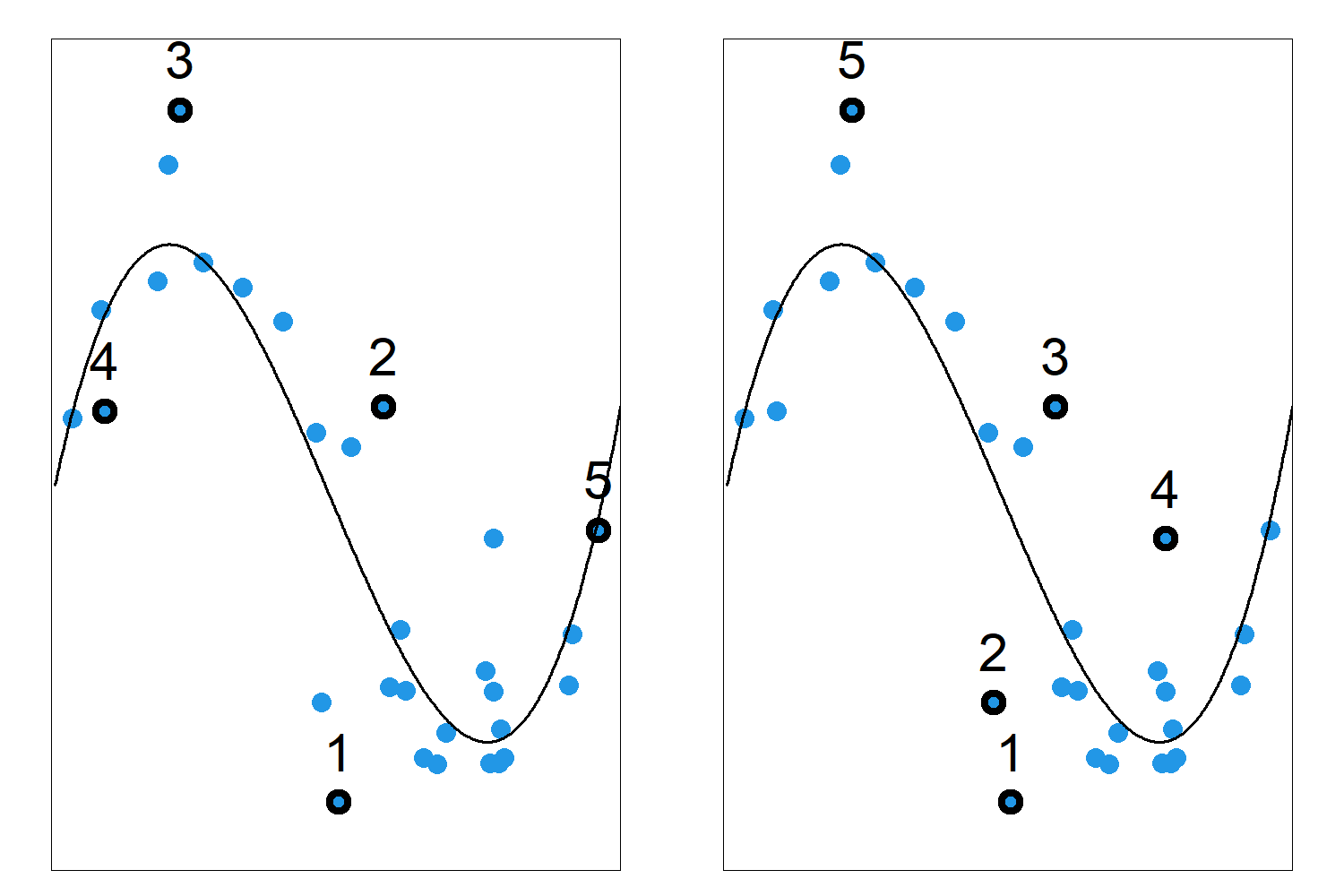} 
    \\
    \vspace{0.3cm}
    \includegraphics[width=4.5cm]{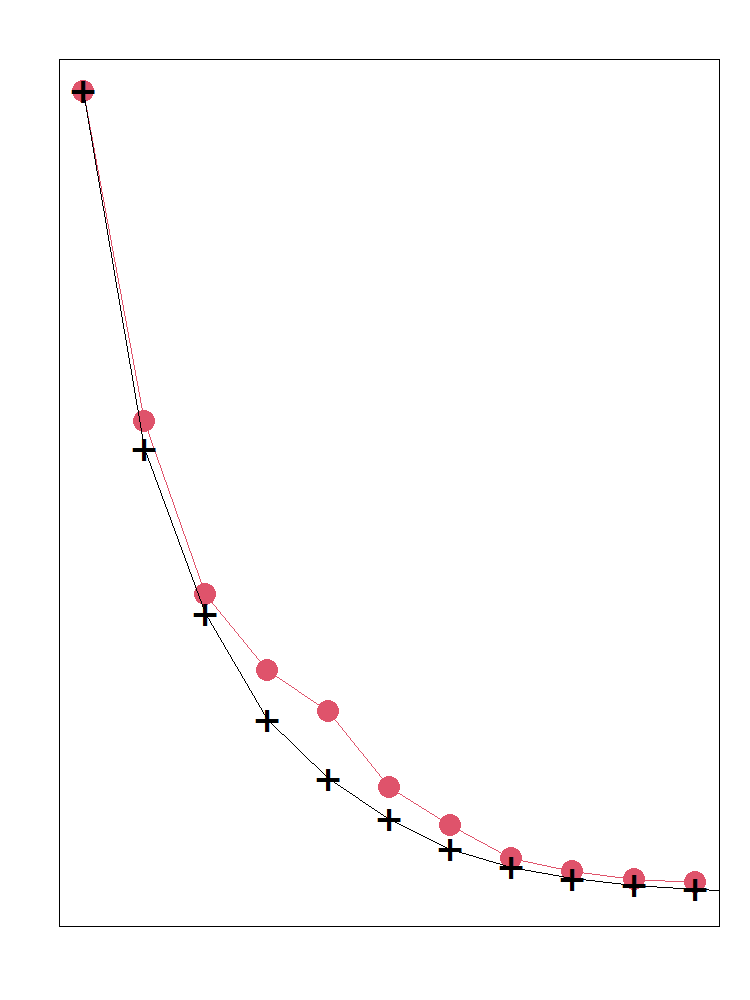} 
    \includegraphics[width=9cm]{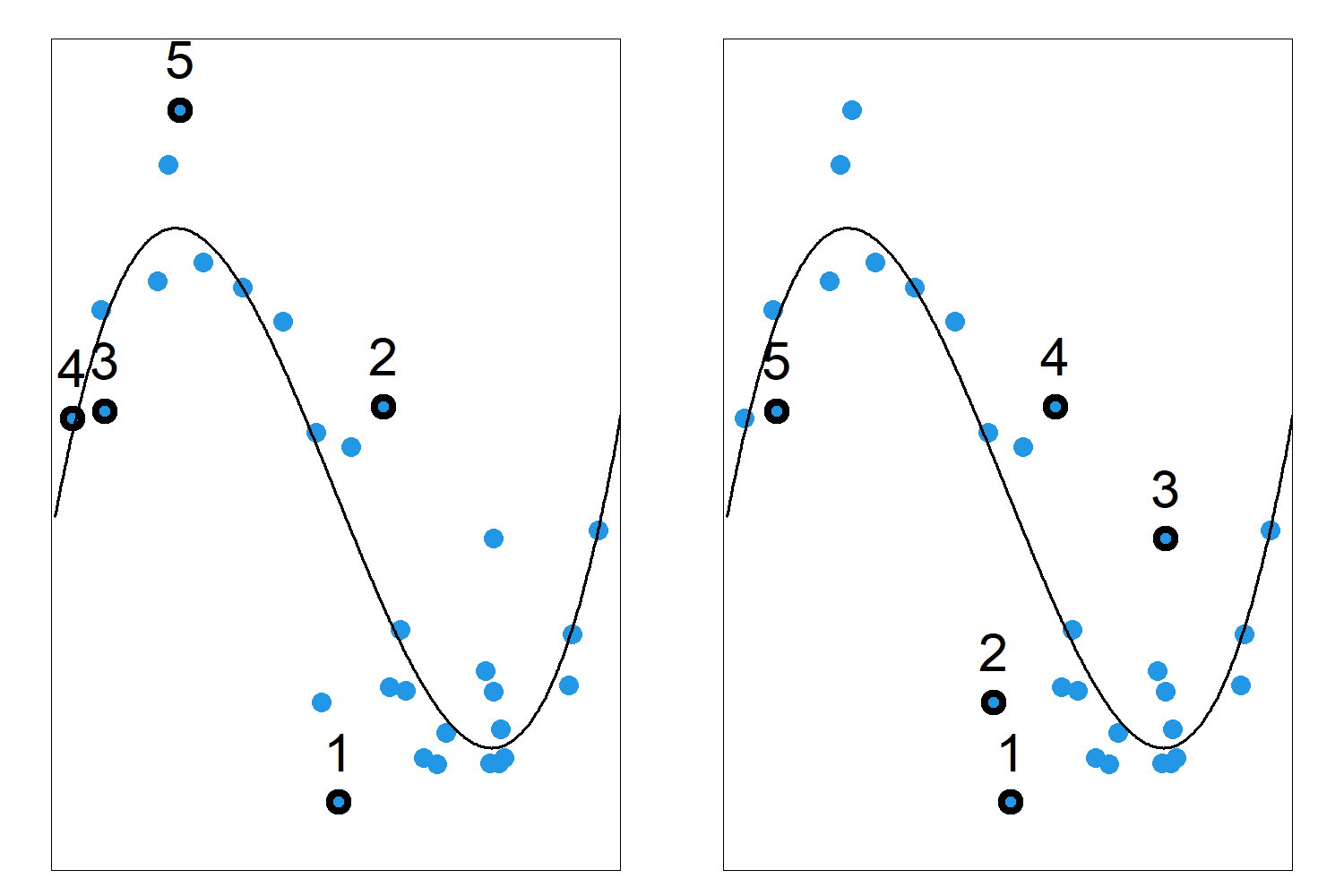}
    \caption{Residual variance and representing sets of the observations. The first and second rows of the figure correspond to the case comprising the normal and Student-t distribution (df=5) of the observational noise, respectively; in both cases, the dispersion of the noise is estimated from the data. In the leftmost panel of each row, the residual variances of the incomplete Cholesky decomposition (red dot) and the optimal selection of the subspace (black $+$) are compared; the horizontal and vertical axes indicate the dimension of the subspace and the residual variance, respectively. In the middle and right panels of each row, the set of selected points are presented: those selected by the incomplete Cholesky decomposition, and the result of an independent selection, respectively. The selected points are encircled and the number above the point indicates the order of the selection in each algorithm; it should be noted that these numbers are not the index $i$ of the observation. The number of the selected points shown in the panel is five in each case.}
    \label{fig:reresentive}
\end{figure}

\begin{figure}[thb]
    \centering
    \includegraphics[width=8cm]{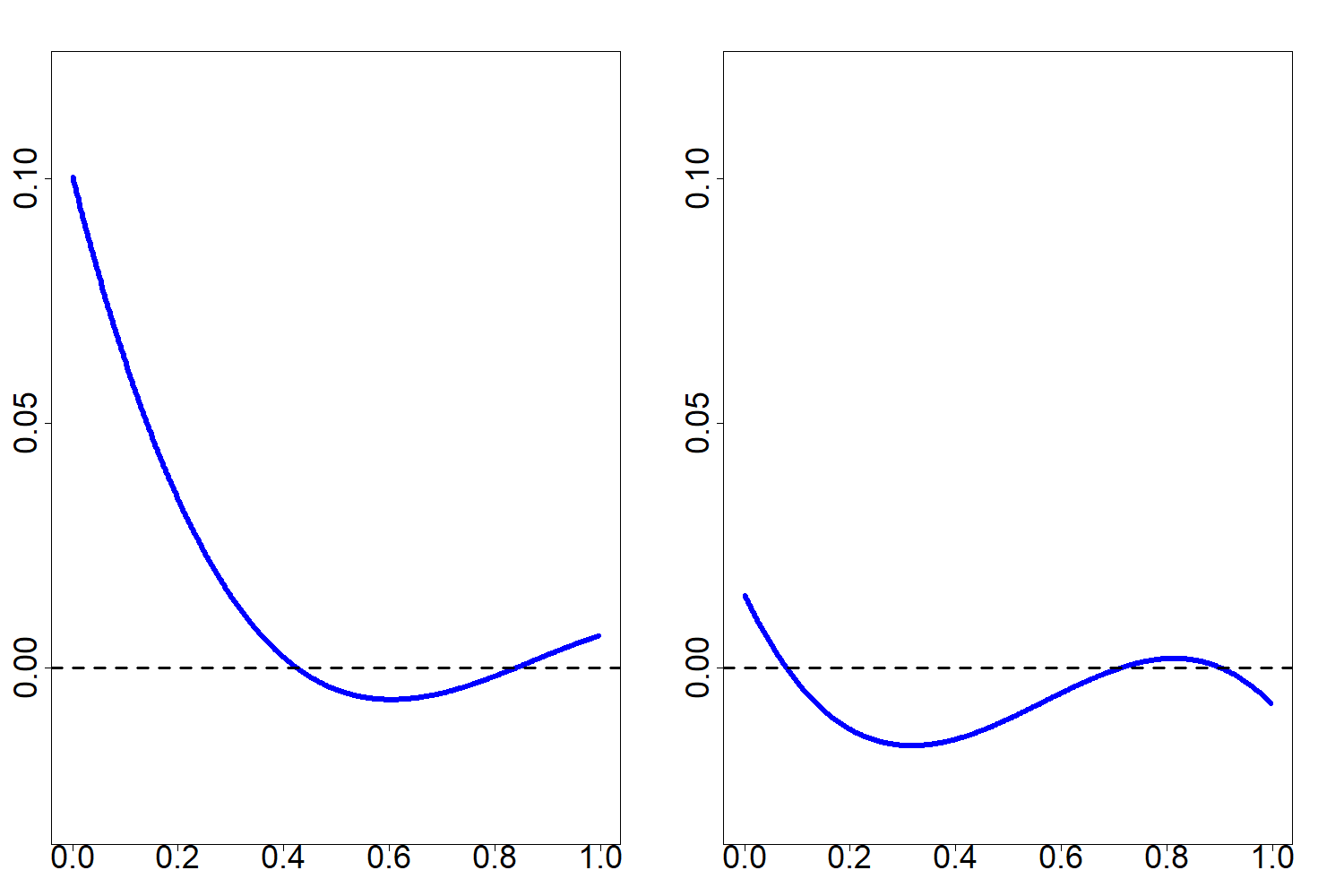}
    \caption{The shift of curves caused by merging observations. The horizontal axis indicates the value of the explanatory variable $z$, while the vertical axis indicates the amount of the shift of the fitted curve. The dotted horizontal line indicates zero. The left panel corresponds to merging the observation $i=3$ to $i=1$, while the right panel corresponds merging the observation $i=12$ to $i=11$. Here ``merging an observation A to another observation B'' is equivalent to removing the observation A from the data and doubling the weight of the observation B.}
    \label{fig:merge}
\end{figure}

Here, we exemplify the incomplete Cholesky decomposition and representative set of observations using the second example in Section~\ref{sec:essential_example}, Bayesian regression with a cubic polynomial. Two types of likelihood functions, the normal and Student-t (df=5), are assumed for the observational noise, where the dispersion of the noise is estimated from the data. The artificial data used in the experiment are the same as those in Section~\ref{sec:essential_example}. Results are presented in Fig.~\ref{fig:reresentive} and Fig.~\ref{fig:merge} and discussed in the following. 

The first column in Fig.~\ref{fig:reresentive} shows a comparison of the residual variances; $\Tr R$ by the incomplete Cholesky decomposition (red dot) and $\sum_{\alpha=m+1}^n \lambda_\alpha$ by the optimal selection of the subspace (black $+$) are compared. Both are normalized such that the total variance becomes unity; the vertical coordinates of the leftmost points in these panels are hence always unity. Here, the ``optimal selection of the subspace'' indicates that we fully diagonalize the matrix $W$ at the start, and eigenvectors are then selected sequentially in the order of the eigenvalues and added to the basis set; this efficiency corresponds to the case where we select principal {\it axes} in an optimal order and need not be realized with an optimal selection of {\it observations}.  

For the normal likelihood (first row), the result of the incomplete Cholesky decomposition closely follows the optimal selection of the subspace. For the Student-t likelihood (second row), the match is worse in the first several steps, but it converges when the number of selected observations increases to seven or more. 

The second and third columns in Fig.~\ref{fig:reresentive} present sets of selected observations; they compare observations selected by the incomplete Cholesky decomposition with those selected by an independent selection. The numbers $1, 2, \cdots$ above the observation points indicate the order of choice in each algorithm.  Here, ``independent selection'' means that we select the observation $i$ in the order of the diagonal element $W_{ii}$. From \eqref{eq:KL}, $W_{ii}$ is a measure of the change in the KL~divergence between the original and perturbed posterior, when the weight of the observation $i$ is changed while maintaining the other weights at unity. 

Let us see the case with normal likelihood (first row). The third (rightmost) column shows that nearby observations indexed by 2 and 1 are selected by the independent selection; they correspond to observations $i=11$ and $i=12$. Both points are selected despite close values of the explanatory variable $z$. In contrast, the panels in the second (center) column show that one observation is selected from $i=11,12$ and the other is dropped by the proposed method. This suggests that some ``repulsion'' among the selected points is automatically included in the proposed method, which makes a point close to an already selected point rarely selected. 

In the case with Student-t likelihood (second row), the proposed method selects nearby points indexed by 4 and 3 for the same data; they correspond to observations $i=1$ and $i=3$. These points have close values of the explanatory variable $z$, but are located on the extreme left and residuals from the fitted curve are considerably different. To examine this case, we perform a pair of experiments, wherein we measure the effect that (i)~the observation $i=3$ is merged into the observation $i=1$, (ii) the observation $i=12$ is merged into the observation $i=11$. Results are illustrated in Fig.~\ref{fig:merge}: the shift of curves caused by merging the observations is much larger in case (i) than in case (ii), especially at the leftmost part of the curve. This justifies the selection of both observations $i=1,3$ by the proposed algorithm.

\renewcommand{\thesubsection}{D~\arabic{subsection}}   
\setcounter{subsection}{0}

\clearpage
\section{Experiments on Approximate Bootstrap}
\label{sec:app:boot_example}

In this section, we discuss approximate bootstrap methods based on the idea of the Bayesian IJ. We consider the first-order algorithm \eqref{eq:boot1_IF1}, the second-order algorithm \eqref{eq:boot2_IF2}, and the importance sampling (IS) approach proposed by \cite{Lee_2017}. A common feature of these algorithms is that they circumvent the need for repeated MCMC runs on each resampled dataset. For evaluation, we compare these algorithms with the direct method, in which MCMC is applied to each bootstrap sample, treating it as the gold standard. 

In subsection~\ref{sec:app:boot_example_1}, we compare the performance of these approximate bootstrap algorithms for posterior means in a simple Weibull fitting example. 

In subsection~\ref{sec:app:boot_example_2}, we examine the idea proposed in Section~\ref{sec:essential_BIJK}, namely whether the principal space of the matrix $W$ can serve as a useful tool for reducing the computational burden when the sample size $n$ is large.

\subsection{An Example of the Approximate Bootstrap}
\label{sec:app:boot_example_1}

\subsubsection*{Importance sampling}

In addition to the first-order algorithm \eqref{eq:boot1_IF1} and the second-order algorithm \eqref{eq:boot2_IF2}, we also test an importance sampling (IS) approach to approximate bootstrapping proposed by \cite{Lee_2017}. It is based on the formula    
\begin{align}
\widehat{A^{(B)}}=\sum_{u=1}^M A(\Theta^{(u)})\mathsf{W}^{(u)}, \,\,\,
\mathsf{W}^{(u)}=  \frac{\prod_{i=1}^n \exp ((R_i-1) \, \logp{X_{i}}
{\Theta^{(u)}})}{\sum_{u=1}^M \prod_{i=1}^n \exp ((R_i-1) \, \logp{X_{i}}{\Theta^{(u)}})}.
  \label{eq:boot_IS}
\end{align}
Here, \( M \) represents the number of posterior samples obtained from an initial MCMC run using the original dataset \( X^n \). The terms \( A(\Theta^{(u)}) \) and \( \logp{X_{i}}{\Theta^{(u)}} \) denote the values of the statistic \( A(\theta) \) and the likelihood \( \logp{X_{i}}{\theta} \) computed for the \( u \)-th posterior sample \( \Theta^{(u)} \), respectively.

As \( M \) increases, the right-hand side of \eqref{eq:boot_IS} provides a consistent estimate of \( A^{(B)} \). However, with a finite \( M \), this method can suffer from large variance due to the concentration of weights on a small subset of posterior samples. In contrast, the approximations given by \eqref{eq:boot1_IF1} and \eqref{eq:boot2_IF2} offer more stable estimates, albeit at the cost of introducing moderate bias.

\subsubsection*{The experiment}

Here, we implement algorithms based on \eqref{eq:boot1_IF1}, \eqref{eq:boot2_IF2}, and \eqref{eq:boot_IS} for the Weibull analysis of life-span data, as discussed in Section~\ref{sec:essential_example}. These methods are evaluated against the results obtained from repeated MCMC-based bootstrapping, which serve as the gold standard.

The following are target statistics: (i)~The posterior mean $T^\gamma(X^n)=\Eppos[\gamma \mid X^n]$ of the shape parameter $\gamma$. (ii)~The predictive probability that $X>40$, which is represented as a posterior mean
\begin{align}
    T^{\,>40}(X^n)=\int_{40}^{\infty} p^{wb}(\tilde{x} \mid \gamma, \lambda)p(\gamma,\lambda \mid X^n) d\tilde{x} = \Eppos \left [\int_{40}^{\infty} p^{wb}(\tilde{x} \mid \gamma, \lambda) \, d\tilde{x}  \right].
    \label{eq:over40}
\end{align}
Here $p^{wb}(\cdot \mid \gamma, \lambda)$ is the likelihood defined by the density \eqref{eq:weibull_PDF} and $p(\gamma,\lambda \mid X^n)$ is the joint posterior density of  $\gamma$ and $\lambda$. 

\subsubsection*{Results}

\begin{figure}[hbt]
    \centering
    \includegraphics[width=14cm]{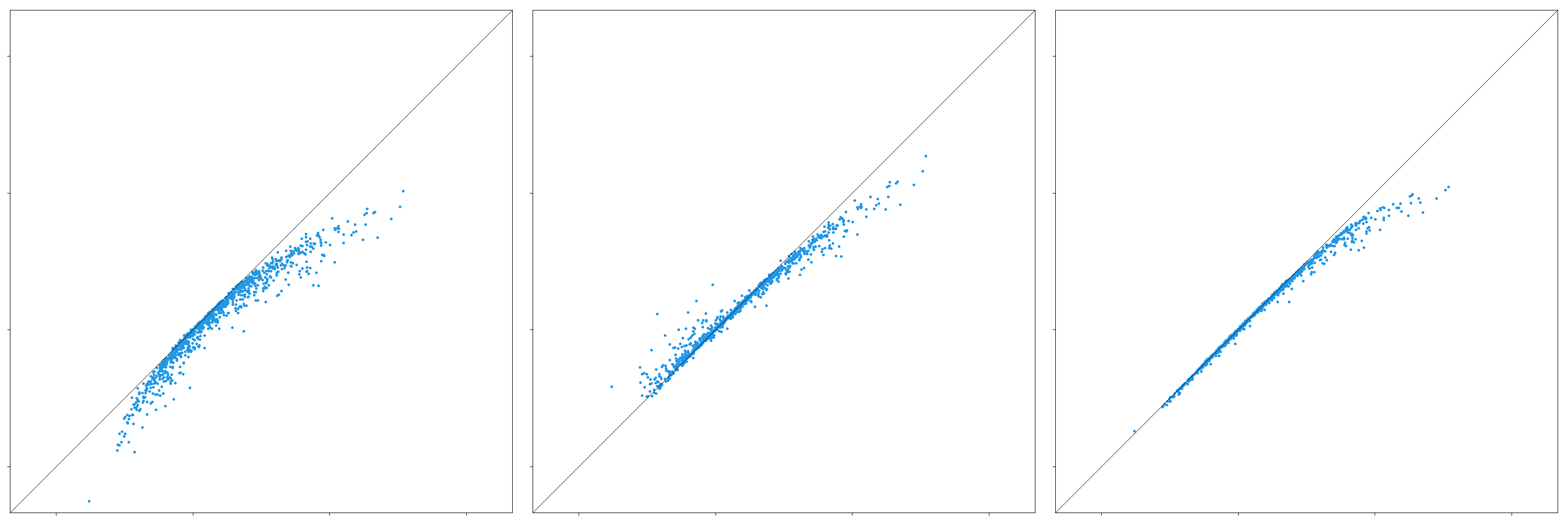}\\
    \vspace*{0.5cm}
    \includegraphics[width=14cm]{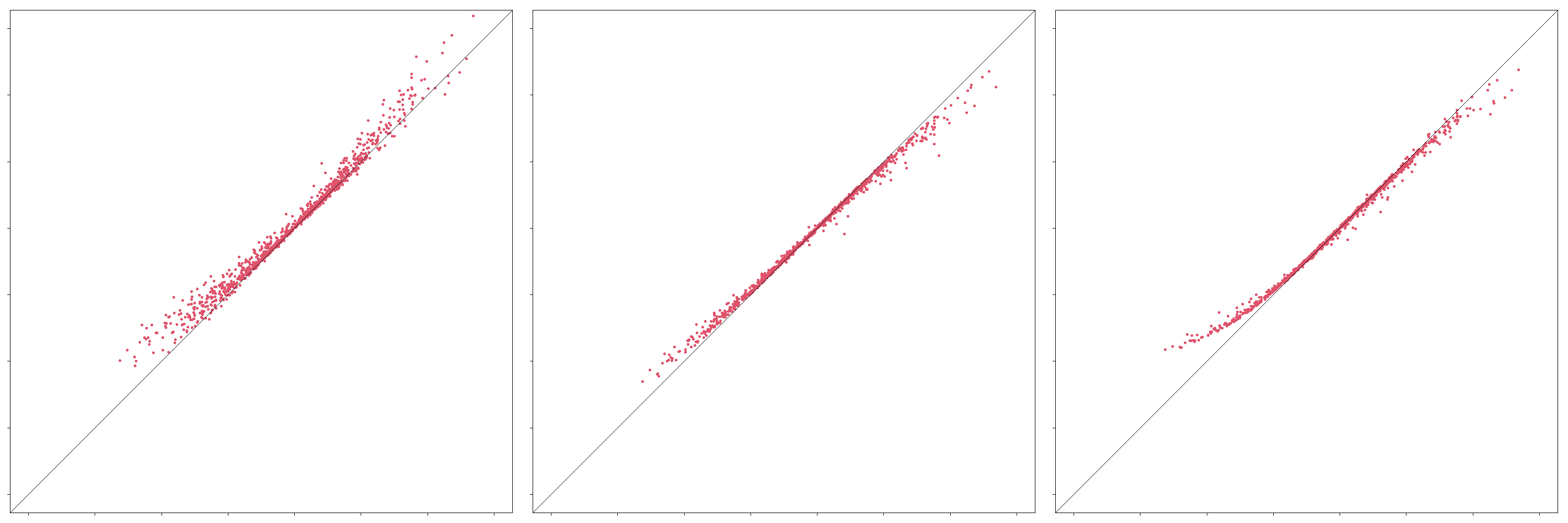}
    \caption{Approximate bootstrap by \eqref{eq:boot1_IF1}, \eqref{eq:boot2_IF2}, and \eqref{eq:boot_IS}. The upper and lower three panels present the results for $T^\gamma(X^n)$ with blue dots and those for $T^{\,>40}(X^n)$ by red dots, respectively. The results corresponding to \eqref{eq:boot1_IF1}, \eqref{eq:boot2_IF2}, and \eqref{eq:boot_IS} are presented from left to right. In each panel, the horizontal and vertical axes indicate the result of the full MCMC computation, and that of the approximate algorithms, respectively, for the same realization of $(R_1, \ldots, R_n)$. Each dot in a panel corresponds to a particular realization of $(R_1, \ldots, R_n)$. The number $N_b$ of the realization of $(R_1, \ldots, R_n)$ is 1000. 
    The diagonal line represents equality; both axes use the same scale in all panels.
    }
    \label{fig:age_boot}
\end{figure}

In Fig.~\ref{fig:age_boot}, the upper and lower sets of panels present the results for $T^\gamma(X^n)$ with blue dots, and the results for $T^{\,>40}(X^n)$ with red dots, respectively. In each of the upper and lower sets, three panels are presented, which correspond to the first-order approximation \eqref{eq:boot1_IF1}, second-order approximation \eqref{eq:boot2_IF2}, and the IS estimator \eqref{eq:boot_IS}, respectively. In each panel, the horizontal and vertical axes indicate the result of the full MCMC computation and the corresponding result of the algorithms based on \eqref{eq:boot1_IF1}, \eqref{eq:boot2_IF2}, or \eqref{eq:boot_IS}, respectively. Each dot in a panel corresponds to a particular realization of $(R_1, \ldots, R_n)$. Approximations discussed here are applied for each realization of $(R_1,\ldots,R_n )$. It is thus possible to compare the outputs of the algorithms for each set of $(R_1, \ldots, R_n)$. It should be noted, however, that the tails of the distribution may be over-emphasized in the following plots; usually, the majority of the 1000 points in a plot are in the central part and overlap. 

The first two columns in Fig.~\ref{fig:age_boot} show that the first-order approximation \eqref{eq:boot1_IF1} is outperformed by the second-order approximation \eqref{eq:boot2_IF2}, especially at the extremes of the bootstrap distributions. 

The IS estimator \eqref{eq:boot_IS} appears to be better than the second-order algorithm \eqref{eq:boot2_IF2} in the left tail of  $T^\gamma(X^n)$, but worse or comparable in other tail regions. The error of the IS algorithm at the distribution's extremes is likely due to the instability caused by the concentration of weights on a limited number of posterior samples; a more detailed analysis of this issue will be provided in Section~\ref{sec:app:boot_example_2} using a different example.

Table~\ref{tab:CPU_age} lists the computational times of these algorithms as measured using proc.time~() in base R. These computational times depend largely on the length of the utilized MCMC computations and the size of posterior samples, as well as other implementation details. Also, the elapsed time measured by proc.time~() fluctuates depending on the status of the computers. 

\begin{table}[htb]
    \centering
    \begin{tabular}{|l|r|r|}
    \hline
    Algorithm & $N_b=1000$ & $N_b=10000$ \\
    \hline
       repeated MCMC & 929 & NA \\
       first-order by Eq.\eqref{eq:boot1_IF1} & 4 & 5 \\ 
       second-order by Eq.\eqref{eq:boot2_IF2} & 6  & 7 \\ 
       IS by Eq.\eqref{eq:boot_IS} & 29 & 259 \\
    \hline
    \end{tabular}
    \caption{Computational time in seconds for approximate bootstraps in Weibull-fitting. $N_b$ indicates the number of realizations $(R_1, \ldots, R_n)$. The total length of the initial MCMC run for the original data is 30000 and the size of the generated samples is 13500. The repeated MCMC, which is only performed when $N_b=1000$, uses a run of length 9000 for each set of $(R_1, \ldots, R_n)$. The time for compiling the Stan codes is not included.
     }
    \label{tab:CPU_age}
\end{table}

\subsection{Evaluation of Dimensional Reduction in the Approximate Bootstrap}
\label{sec:app:boot_example_2}

\subsubsection*{Problem formulation}

In this subsection, we examine the dimensional reduction idea proposed in Section~\ref{sec:essential_BIJK}. A hierarchy of research questions is as follows:  
\begin{itemize}
\item[1.] How does the bootstrap method approximate sampling from the population? ?  
\item[2.] How accurately can the second-order approximation via the Bayesian IJ replicate the results of the bootstrap? 
\item[3.] How well does the projection onto the principal space of the matrix \( W \) approximate the results of the second-order approximation?  
\end{itemize}  

Here, we focus on Questions 2 and 3, along with comparisons of the computational time required by different algorithms. Question 3 is the primary focus, as it examines the adequacy of approximating higher-order posterior cumulants using the projection onto the principal space of the matrix \( W \).  We also discuss Question 2, extending our analysis beyond variance estimation to include the accuracy of percentile estimates.

\subsubsection*{Model and settings}

Let us consider the state space model for the Prussian horse data discussed in Section~\ref{sec:essential_example} as an example of the proposed algorithm. The states $(Z_t)$ are regarded as fixed but unknown variables and the likelihoods conditioned on $(Z_t)$ are considered. At each time $t, \, t=1,\ldots, 20$, there are 14 observations, each of which corresponds to one of the 14 coups in Prussia. 

Bootstrap replication is defined as sampling with replacement from the set of 14 observations at each time. This presents a rather unfavorable condition for both the bootstrap and the proposed approximations, as they rely on large-sample theory. Note that this definition may be inappropriate for representing frequentist uncertainty in the original problem setting, as the 14 observations are not considered independent draws from the same population. However, we do not explore this issue further here and simply treat this dataset as an example of overdispersed data.

Target statistic is the posterior mean $\Eppos[\exp(Z_t)]$ of the intensity of the Poisson distribution at each time $t$. 

\subsubsection*{The experiment}

In this experiment, we compare several approximation methods with the result of the repeated applications of MCMC. Specifically, we consider the following approaches:

\begin{itemize}
    \item The first-order algorithm \eqref{eq:boot1_IF1}.
    \item The original second-order algorithm \eqref{eq:boot2_IF2}.
    \item The second-order algorithm using \eqref{eq:second_alternative}.
    \item Second-order algorithms with projections onto the principal space with $a_M = 8,15$ and $20$.
    \item The importance sampling (IS) algorithm \eqref{eq:boot_IS}.
\end{itemize}

For all comparisons, we use the non-centered version of the matrix $W$. The first-order term in \eqref{eq:boot2_IF2} is always directly computed, and projection to the leading-$a_M$ principal space is applied only in the second-order term.

The total number of posterior samples is set to \( M = 4500 \)  
 (obtained from three MCMC runs of 4000 iterations each, with the initial 1000 iterations discarded, and 1:2 subsampling applied). When constructing the principal space of the matrix \( W \) using a subset of posterior samples, we set the subset size to \( M^* = 100 \).

\subsubsection*{Computational time} 

First, we examine the computational time reported in Table~\ref{tab:CPU_horse}. These computational times strongly depend on the details of the experiments, specifically, on the length of the MCMC runs used, the number of posterior samples, and the implementation of matrix multiplications.

As expected, the method involving repeated MCMC runs requires a substantial amount of computational time. The original second-order algorithm \eqref{eq:boot2_IF2} is significantly slower than other methods that avoid repeated MCMC. When \( N_b = 1000 \), the majority of the computational time is spent constructing the matrix \( \estKpos \). We construct  \( \estKpos \) for $20$ target statistics, resulting in a matrix of dimensions \( 20 \times 280 \times 280 \), which is computed using 4500 posterior samples. Additionally, evaluating \eqref{eq:boot2_IF2} for each realization of \( (R_1, \ldots, R_n) \) is computationally intensive, and this burden increases as \( N_b \) grows. Although the computational time may vary significantly depending on the implementation, the overall trend and qualitative behavior are expected to remain unchanged.

The second-order algorithm using \eqref{eq:second_alternative} and the IS algorithm \eqref{eq:boot_IS} exhibit similar behavior: they are computationally efficient when \( N_b = 1000 \), but their runtime increases significantly as \( N_b \) grows to 5000 and 10000.

In contrast, second-order algorithms utilizing the principal space approximation perform almost as efficiently as the first-order algorithm when \( N_b = 1000 \), with only a modest increase in computational time when \( N_b \) grows to 5000 or 10000. The initial MCMC step takes approximately 30 seconds, which constitutes the majority of the total computational time. Even when the subset size \( M^* \) for constructing \( W \) is increased from 100 to 4500, the additional computational cost remains limited in this example.

\begin{table}[htb]
\centering

    \begin{tabular}{|l|r|r|r|}
    \hline 
     & $N_b=1000$ & $N_b=5000$ & $N_b=10000$ \\
    \hline
       Repeated MCMC 
       & 12439 & NA & NA\\
       First-order by Eq.\eqref{eq:boot1_IF1} 
       & 30 & 35 &  41 \\ 
       Second-order by Eq.\eqref{eq:boot2_IF2} 
       & 156 & 252 & 356  \\
       Second-order using Eq.\eqref{eq:second_alternative}
       & 48  & 127 & 228 \\
       Projection ($a_M=20$, $M^*=100$) 
       & 33 & 39 &  47 \\ 
       IS by Eq.\eqref{eq:boot_IS} 
       & 48 &  125 & 224 \\ 
    \hline
    \end{tabular}
    \caption{
    Computational time in seconds for approximate bootstraps in the state space model. Here ``Projection'' means the second-order algorithm in which a projection to the principal space of dimension $a_M$ is used.  The total length of the initial MCMC run for the original data is 12000 and the size of the generated samples is 4500. The repeated MCMC, which is only performed when $N_b=1000$, uses a run of length 4000 for each set of $(R_1, \ldots, R_n)$. The time for compiling the Stan codes is not included. The elapsed time measured using proc.time~() fluctuates considerably depending on the system status. 
    }    
    \label{tab:CPU_horse}
\end{table}

\subsubsection*{Error caused by the projection onto the principal space}

We now turn to Question~3, evaluating how well the projection onto the principal space of the matrix \( W \) approximates the results of the second-order approximation, which serves as the gold standard in Question~3. 

\begin{figure}[htb]
\centering
\includegraphics[width=15cm]{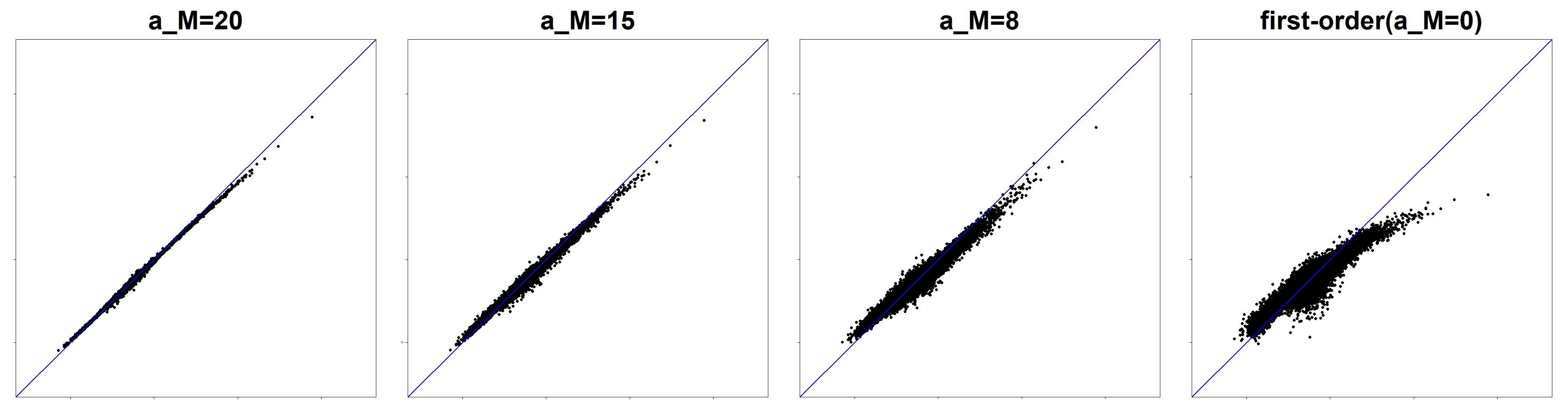}
\caption{
Effect of projection onto the principal space of $W$. In each panel, the horizontal axis shows the second-order approximation, while the vertical axis presents results from the approximations with $a_M = 20, 15, 8$, and the leftmost panel corresponds to the first-order approximation. Each point represents one of the 20 parameters across 3000 bootstrap replications, with substantial overlap. The principal space is computed using the incomplete Cholesky decomposition discussed in Section~\ref{sec:cholesky}. The diagonal line represents equality; both axes use the same scale in all panels.
}
\label{fig:proj4500}
\end{figure}

Figure~\ref{fig:proj4500} illustrates the effect of projection. The projection dimensions considered are $a_M = 20$, $15$, $8$ and the leftmost panel corresponds to the first-order approximation. Each point in the plots represents one of the 20 target statistics computed across 3000 bootstrap replications. As seen in the figure, the second-order approximation and the principal-space-based approximations show close agreement in the case of $a_M=20$ and $a_M=15$, except for slight systematic deviations in the upper extreme region. 

\begin{figure}[htb]
\centering
\includegraphics[width=15cm]{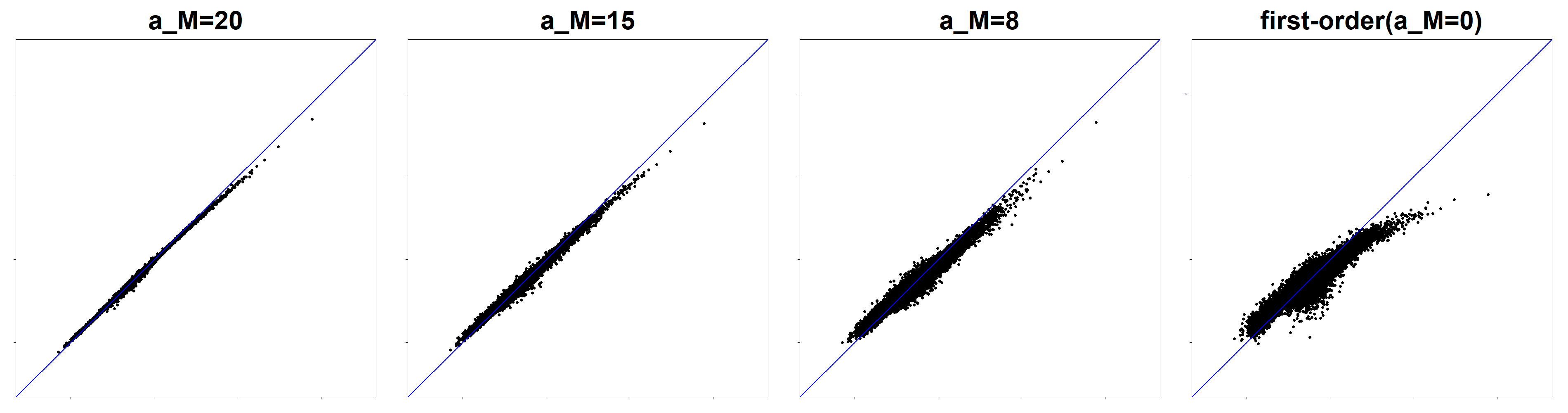}
\caption{Same as Figure~\ref{fig:proj4500}, but the principal space of $W$ is constructed using a subset of size $M^* = 100$ drawn from the full set of 4500 MCMC samples. As discussed in the main text, the results remain nearly unchanged.}
\label{fig:proj100}
\end{figure}

In Fig.~\ref{fig:proj4500}, the full set of 4500 MCMC samples is used to construct the principal space. In contrast, Fig.~\ref{fig:proj100} shows the result when a smaller subset of size $M^* = 100$ is used. Only minor differences are observed between the two figures, indicating that a crude approximation to the matrix $W$ is sufficient for the purpose of approximating the second-order term via projection.

\subsubsection*{Quantitative comparison of the algorithms}

Next, we proceed to Question 2, which concerns the accuracy of different approximations compared to the bootstrap with repeated MCMC computations. 

Here, we conduct a quantitative comparison of these four approximation algorithms. Since the contribution of the second-order terms to the bootstrap variance arises from terms of fourth order in $(R_i-1)$, which are of order $O(1/n^2)$ in total, we also include statistics that are sensitive to the tail of the distribution. Specifically, we consider the following six summary statistics of the bootstrap distribution: the mean, variance, and the 10\%, 25\%, 75\%, and 90\% quantiles.

Our procedure is defined as follows:
\begin{itemize}
    \item For each of the 20 target statistics (i.e., the values of a statistic at different time points) and the four candidate algorithms, we perform computations for 3000 bootstrap replications of the data and compute the six summary statistics defined above. We also compute these statistics using repeated MCMC, which serves as the gold standard in Question~2.

    \item For each approximation algorithm, each of the 20 target statistics, and each summary statistic, we calculate the difference from the gold standard.

    \item Now, we have multiple sets, each containing 20 difference values, with each value corresponding to a specific target statistic. Each set is defined by an approximation algorithm and a summary statistic. To visualize these sets, we use boxplots, where each box and whisker plot represents the distribution of 20 difference values within a set.
\end{itemize}

\begin{figure}[htb]
\centering
\includegraphics[width=18cm]{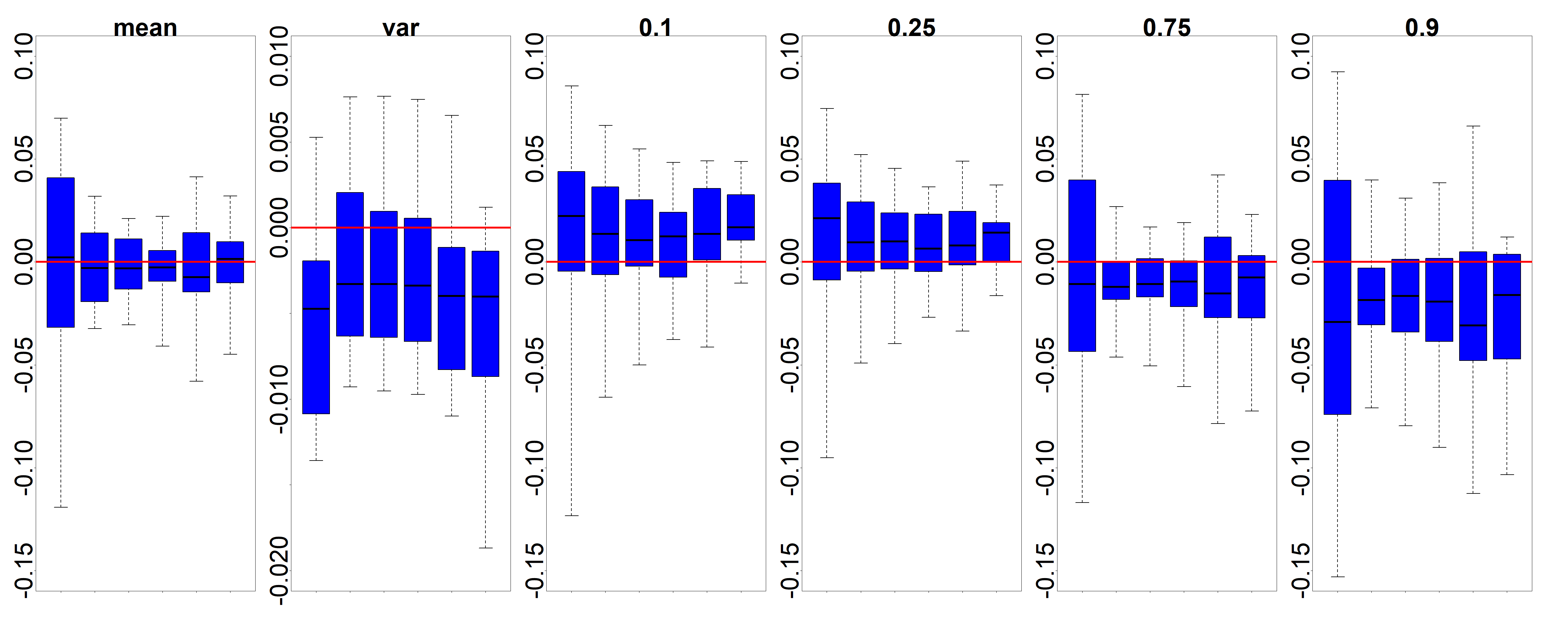}
\caption{Quantitative comparison of the algorithms. From left to right, the panels correspond to the mean, variance, and the 
10\%, 25\%, 75\%, and 90\% quantiles, respectively. In each panel, from left to right, the boxplots correspond to the first-order algorithm, the second-order algorithm using \eqref{eq:second_alternative}, the second-order algorithm using the projection onto the principal space with $a_M = 20,15,8$, and the IS algorithm \eqref{eq:boot_IS}, shown in this order. The \texttt{boxplot()} command in base R is used with the option \mbox{range=0}. All panels share the same y-axis range, except for the one showing the variance. The horizontal red line indicates zero.}

 \label{fig:box}
\end{figure}

The results are shown in Fig.~\ref{fig:box}. For the 75\% and 90\% quantiles, and the mean, the first-order approximation is outperformed by the other algorithms.

\subsubsection*{Stability of the results with respect to posterior samples}

Here, we examine the reproducibility of results under different sets of posterior samples, a factor that appears to be closely related to the observed fluctuations in the IS algorithm's performance.

\begin{figure}[htb]
\centering
\includegraphics[width=18cm]{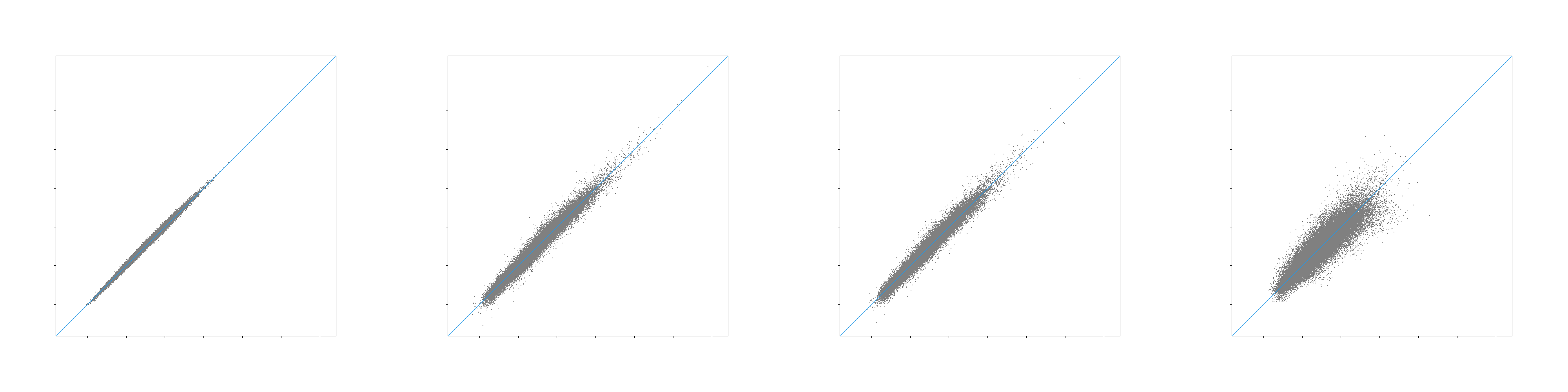}
\caption{
Comparison of results obtained using different posterior samples. Each axis represents estimates based on posterior samples from MCMC runs with different random number seeds. From left to right, the panels correspond to the first-order algorithm, the second-order algorithm, the second-order algorithm using the projection onto the principal space with $a_M = 15$ and $M^* = 100$, and the IS algorithm. All results are based on 4500 MCMC samples. The diagonal line represents equality; both axes use the same scale in all panels.
}
\label{fig:stability}
\end{figure}

In Fig.~\ref{fig:stability}, we compare pairs of results obtained from different posterior samples, each generated by independent MCMC runs with different random number seeds. For each of the 3000 bootstrap replicates $(R_1, \ldots, R_n)$, all 20 parameters are included in the comparison, resulting in a dense scatter of points. As shown in the figure, the dependence on posterior samples is smallest for the first-order algorithm (leftmost), moderate for the second-order algorithm (middle two), and most pronounced for the IS algorithm (rightmost). In this setting, the projection appears to cause a modest increase in variability at the left edge of the plot, although the effect is not substantial.

\begin{figure}[htb]
    \centering
    \includegraphics[width=11cm]{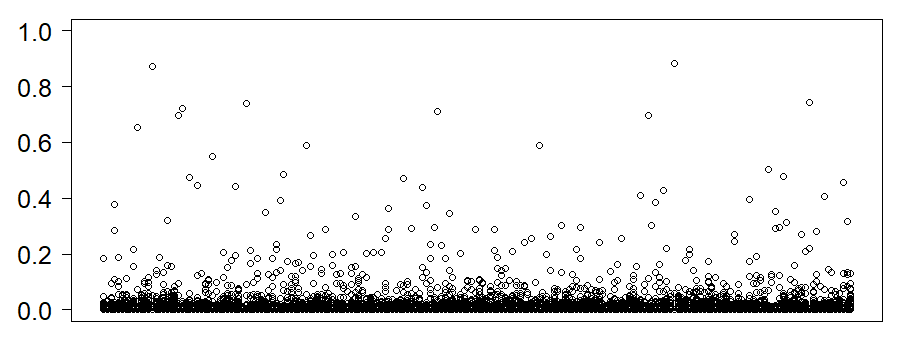}
    \caption{Relative size of weights $\mathsf{W}_i$ in the state space model for 200 sets of $(R_1, \ldots, R_n)$. The horizontal and vertical axes correspond to the indices $1,\cdots,200$ of the set and the values of the weights, respectively. For each set of $(R_1, \ldots, R_n)$, the sum of the weights is normalized to unity. The function boxplot () in base R graphics is used; the points in the figure are outliers detected by boxplot() and the boxes are almost hidden at the bottom of the figure. } 
    \label{fig:ISw}
\end{figure}

The erratic behavior and posterior sample dependence in the IS algorithm can be explained by the extreme concentration of weights into a small number of posterior samples that occur in some realizations of $(R_1, \ldots, R_n)$. To illustrate this, the values of the weights $\mathsf{W}_i$ are plotted in Fig.~\ref{fig:ISw} for 200 sets of $(R_1, \ldots, R_n)$. The maximum value of the weights exceeds 0.4 for a considerable number of realizations of $(R_1, \ldots, R_n)$, and in some cases, it even surpasses 0.7 or 0.8. Since the sum of the weights is normalized to unity within each set, $\mathsf{W}_i=0.4$ ($0.7$) means that 40\% (70\%) of the weight is concentrated on a single sample among 4500 posterior samples. 

A similar issue in importance sampling cross-validation (IS-CV) is discussed in \cite{P_1997}. Advanced techniques for cross-validation, such as those developed in \cite{Vehtari_PSIS_2024}, may also be useful for approximate bootstrap methods, but we leave this to future studies.

\subsubsection*{Summary and discussion} 

While this single example alone is not sufficient to draw a definitive conclusion, the obtained results suggest that the second-order algorithms utilizing the projection onto the principal space \( W \) provide a practical balance between computational time and accuracy. Specifically, in this example, the projection onto the principal space preserves accuracy even when the number of posterior samples used for constructing \( W \) is significantly reduced. 

The second-order algorithm using \eqref{eq:second_alternative} is another option when the required number \( N_b \) of bootstrap replicates is limited; however, its computational cost increases as \( N_b \) grows.

The IS algorithm shares several characteristics with the second-order algorithm based on \eqref{eq:second_alternative}, both in terms of computational cost and average performance. However, its strong dependence on the posterior samples can be a disadvantage, particularly in applications where reproducibility across initial MCMC runs is a critical concern.

An important issue not addressed in the present experiment concerns question~(1) in the hierarchy of research questions given at the beginning of this subsection.
It is not trivial that improving the bootstrap approximation necessarily improves the approximation to sampling from the population. 
This issue is crucial for practical applications, but it is left for future study.

Another possible criticism of this example and experiment is that, because the likelihood conditioned on state space variables belongs to the exponential family, the low-rank structure of the matrix \( W \) is readily suggested. In such a case, by applying the approach based on sufficient statistics, as described in \cite{Efron_2015}, it seems possible to implement the procedure directly in a space whose dimensionality matches the number of parameters from the outset.

To address this concern, we test Question~3 in a setting with a non-exponential-family likelihood. Specifically, we examine the error introduced by the projection onto the principal space when the Poisson likelihood is replaced by a negative binomial distribution with an unknown dispersion parameter. Since the original Prussian data yield only a small estimate of the dispersion parameter, we instead use an artificial dataset that has the same expectation as the Prussian data at each time point but is generated from a negative binomial distribution with a dispersion parameter $1.0$. 

\begin{figure}[htb]
\centering
\includegraphics[width=15cm]{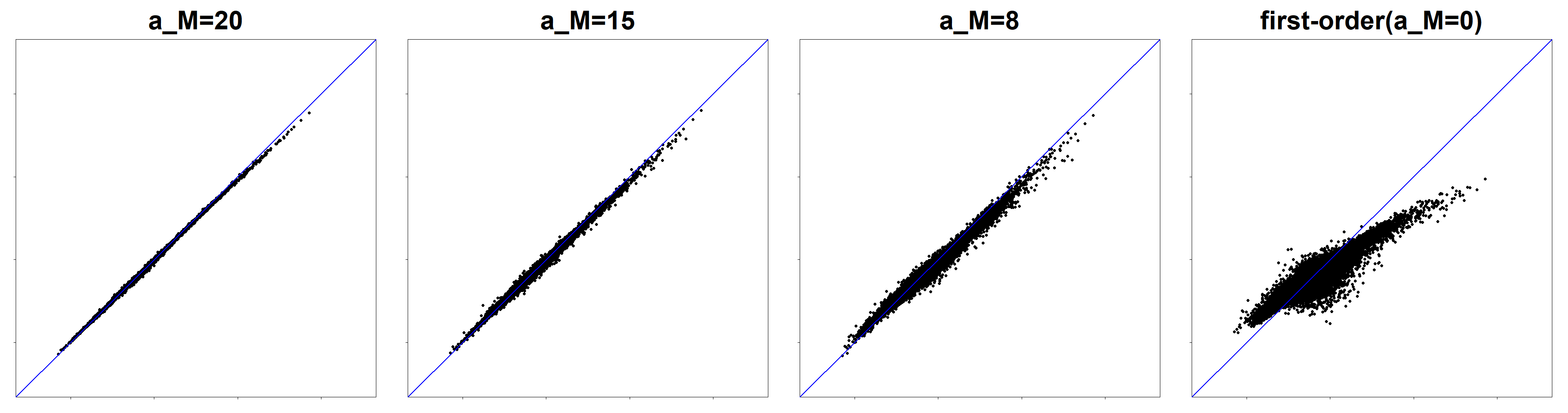}
\caption{
Same as Fig.~\ref{fig:proj4500}, but with the negative binomial likelihood applied to over-dispersed artificial data (details are provided in the main text). The diagonal line represents equality; the scale of the axes differs from Fig.~\ref{fig:proj4500}, but is consistent across the panels.
}

\label{fig:proj_negbinom_4500}
\end{figure}

\begin{figure}[htb]
\centering
\includegraphics[width=15cm]{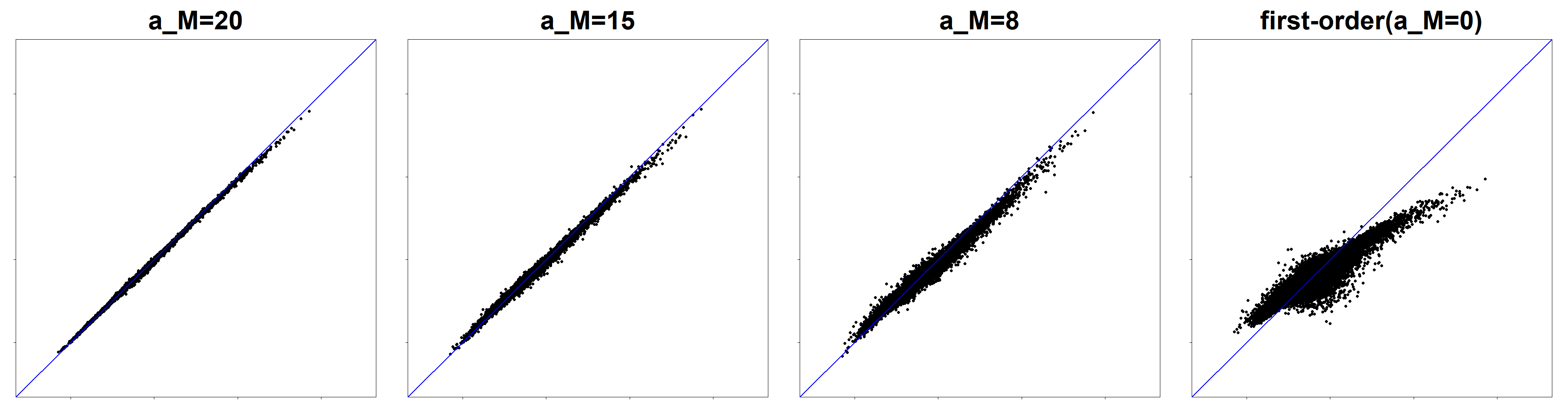}
\caption{Same as Fig.~\ref{fig:proj_negbinom_4500}, but with the principal space of $W$ constructed from a subset of size $M^* = 100$ drawn from the full set of 4500 MCMC samples.}
\label{fig:proj_negbinom_100}
\end{figure}

Our preliminary results, shown in Fig.~\ref{fig:proj_negbinom_4500}, suggest that the accuracy of the projection-based approximation with \( a_M = 15 \) and \( 20 \) remains comparable to that observed in the previous experiment using the Poisson likelihood and the original Prussian data. Although the precision of the eigenvalue estimates deteriorates when using a posterior subsample of size \( M^* = 100 \), the overall accuracy of the projection approximation is largely preserved, as shown in Fig.~\ref{fig:proj_negbinom_100}.  The diagonal line represents equality; both axes use the same scale in all panels.

\clearpage






\setcounter{figure}{0}
\renewcommand{\thesubsection}{S~\arabic{subsection}}   
\renewcommand{\thefigure}{S~\arabic{figure}}
\setcounter{subsection}{0}
\section*{Supplementary Material}


\makeatletter
\newcommand{\setrefnumber}[2]{%
  \expandafter\gdef\csname r@#1\endcsname{{#2}{}}%
}
\makeatother



\label{sec:app:strong}


This supplementary material focuses on the effect of the prior in the Bayesian IJ. 


\noindent The contents of each section are summarized as follows:

\begin{itemize}
\item
In Sec.~\ref{sec:centering}, the role and origin of ``centering'' (and hence double centering) in the Bayesian IJ are discussed. 

\item
In Sec.~\ref{sec:app:example}, a simple numerical example showcases the effect of informative priors on the frequentist covariance formulae \eqref{eq:cov_g_star} and \eqref{eq:cov_g}.

\item
In Sec.~\ref{sec:app:O(n)}--\ref{sec:app:centering}, 
we study the notion of the $O(n)$ prior, which is central in this supplement. 
It is contrasted with the usual $O(1)$ prior, in which the prior does not change in the limit as the sample size $n \to \infty$. 
We also provide heuristic derivations of properties that hold under the $O(n)$ setting.

\item
In Sec.~\ref{sec:app:model_selection}, we consider the subject of model selection in the case of $O(n)$ priors; it is shown that the bias correction term of $\WAIC$ is robust against even moderately informative priors.
\end{itemize}

\subsection{Rationale for Centering}
\label{sec:centering}

A natural method of understanding the meaning of centering in the definition \eqref{eq:1st_star} is in the context of influence functions. When we define the first-order influence functions $T^{(1)}(x,G)$, we usually assume $\EpX[T^{(1)}(x,G)]=0$, where $\EpX$ denotes an average over the population $G$. This condition allows simplifications of related formulae, such as from $\VarX[T^{(1)}(x,G)]$ to $\int \{T^{(1)}(x,G)\}^2 dG(x)$ for frequentist variance. The origin of the relation $\EpX[T^{(1)}(x,G)]=0$ may be traced back to the fact that influence functions are defined for the changes in a probability distribution $G$ that satisfy $\int dG(x)=1$; this corresponds to the condition $\sum_{i=1}^n w_i=1$ in the definition of the weighted posterior \eqref{eq:weighted_pos}.

In practice, it is important to note that centering is only effective in the case of considerably strong priors.  We discuss how and why this is the case in the rest of this supplement, which also elucidates the effect of informative priors in the Bayesian IJ. 

\subsection{Numerical Experiment for the Binomial Model with Informative Priors}
\label{sec:app:example}

To motivate the subsequent asymptotic analysis, we first consider a numerical example based on a binomial model with potentially over-dispersed data. Specifically, we assume a Bayesian model defined by a binomial likelihood and a beta prior, as described below.
\begin{align}
    X_i \sim Binom(N,q), \,\,\, q \sim Beta(\alpha,\beta).
    \label{eq:binomial_model}
\end{align}
In the following experiments, we assume that IID observations $X^n=(X_1,X_2,\ldots, X_n)$ are sampled from a beta binomial distribution, the probability mass function of which is
\begin{align}
p(x;q_0,\rho)={}_NC_x \frac{ B(x+\mathrm{a},N-x+\mathrm{b})}{B(\mathrm{a},\mathrm{b})}, \,\,\,
    \mathrm{a}=q_0 \frac{1-\rho}{\rho}, \,\,\, \mathrm{b}= (1-q_0) \frac{1-\rho}{\rho}. 
    \label{eq:beta_binomial}
\end{align}
$\EpX[X]/N=q_0$ when $X \sim p(x;q_0,\rho)$. The total number of trials is $nN$. 

Below, we focus on the frequentist variance of the posterior mean $\Eppos[q]$ and compare three types of estimators: (i) The posterior variance $\Varpos[q]$, (ii) $\hat{\Sigma}_{qq}$ defined by \eqref{eq:cov_g}, and (iii) $\hat{\Sigma}^*_{qq}$ defined by \eqref{eq:cov_g_star}. While (i) is a genuine Bayesian estimator, (ii) and (iii) are frequentist estimators obtained using the Bayesian IJ. These values are calculated using the MCMC method for each artificial dataset. The true value of the frequentist variance is estimated using the same set of artificial data.  

In all the experiments here, the number $N$ of trials in each observation is set to $5$, while the number $n$ of observations in each set of data is $20$; hence, the total number of trials in a set of data is $nN=100$. $q_0$ is 0.25. The number of the set of artificial data used to construct a boxplot is $800$.      

\begin{figure}[htb]
    \centering
    \includegraphics[width=5cm]{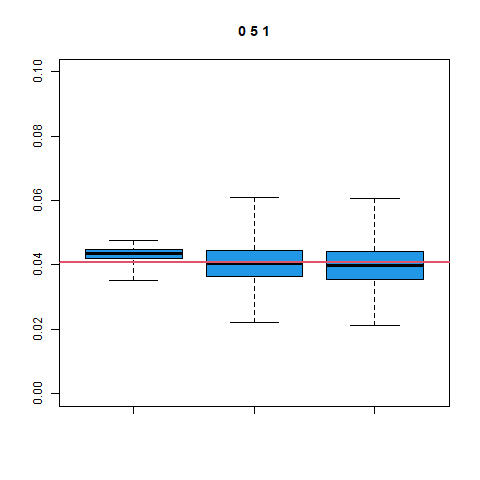}
    \includegraphics[width=5cm]{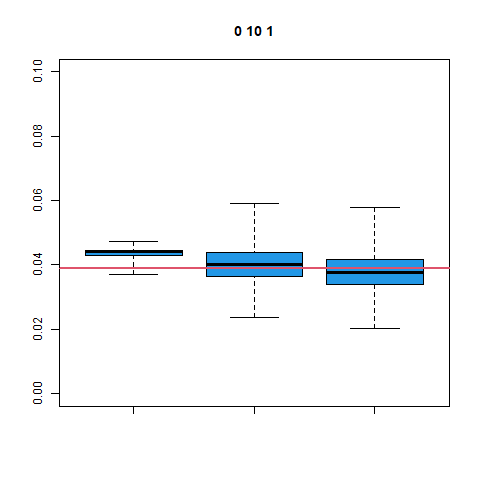}
    \includegraphics[width=5cm]{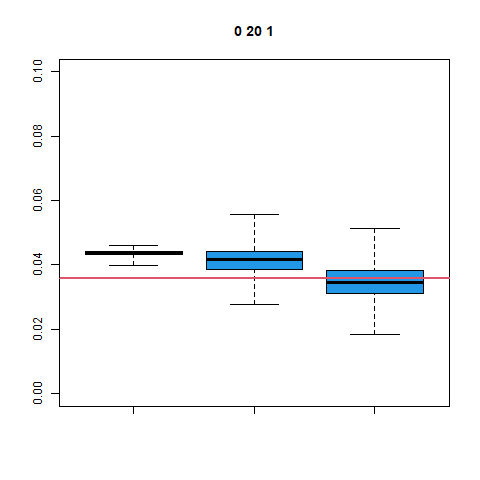}
    \caption{
    Effect of the informative prior in estimating frequentist variance in the $\rho=0$ case. The three panels present the results for $(\alpha,\beta)=(5,1),(10,1),(20,1)$. 
    In each panel, the Bayesian estimate $\Varpos[q]$, the simplified estimator $\hat{\Sigma}^{qq}$ given by \eqref{eq:cov_g}, and the original estimator $\hat{\Sigma}^*_{qq}$ given by \eqref{eq:cov_g_star} are presented from left to right. The horizontal red line indicates the true value of the frequentist variance. The boxplot is created using the base graphics of the R language with the ''range=0'' option;  the whisker in each boxplot spans the maximum and minimum of the plotted data.
    }
    \label{fig:variance_binom_strong}
\end{figure}

Fig.~\ref{fig:variance_binom_strong} presents the result of an experiment performed in the $\rho=0$ case, where the model can correctly represent the true distribution. We perform the experiments with the beta prior of $(\alpha, \beta)=(5,1),(10,1),(20,1)$ and make comparisons among $\Varpos[q]$, $\hat{\Sigma}^{qq}$, and $\hat{\Sigma}^*_{qq}$. 

When $\alpha=5$ (the left panel), all three estimators perform well, except for a small bias in $\Varpos[q]$. When $\alpha=10$ and $20$ (the middle and right panels), the genuine Bayesian estimator $\Varpos[q]$ overestimates the frequentist variance. It should be noted that the genuine Bayesian estimator may not reproduce the frequentist variance with informative priors, even when the likelihood is correctly specified. The simplified estimator $\hat{\Sigma}^{qq}$ given by \eqref{eq:cov_g} also overestimates the variance; it is minimal in the case of $\alpha=10$, but becomes more evident in the case of $\alpha=20$. In the case of $\alpha=20$,  the number of expected successes $nNq_0=25$ is almost comparable to $\alpha-1=19$; it is a fairly strong prior for this sample size.         

\begin{figure}[htb]
    \centering
    \includegraphics[width=5cm]{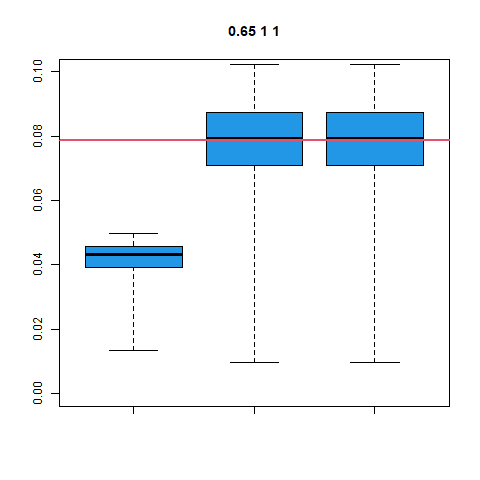}
    \includegraphics[width=5cm]{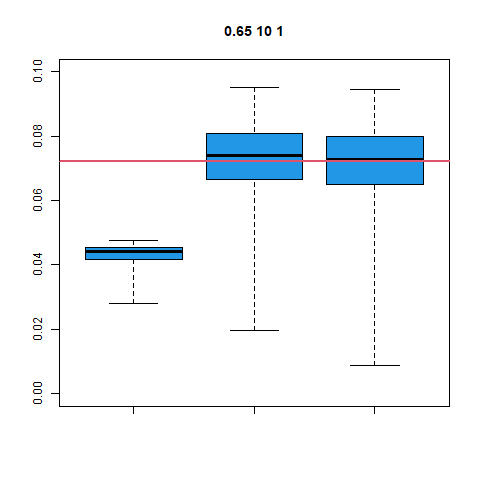}
    \includegraphics[width=5cm]{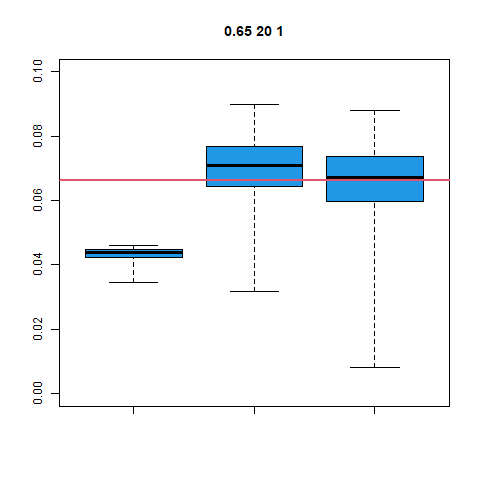}
    \caption{
    Effect of the informative prior in estimating frequentist variance in the $\rho=0.65$ case. The three panels present the results for $(\alpha,\beta)=(1,1),(10,1),(20,1)$. 
    In each panel, the Bayesian estimate $\Varpos[q]$, the simplified estimator $\hat{\Sigma}^{qq}$ given by \eqref{eq:cov_g}, and the original estimator $\hat{\Sigma}^*_{qq}$ given by \eqref{eq:cov_g_star} are presented from left to right. The horizontal red line indicates the true value of the frequentist variance. The other details are the same as those in Fig.~\ref{fig:variance_binom_strong}.}
    \label{fig:variance_binom_strong_misspecified}
\end{figure}

Fig.~\ref{fig:variance_binom_strong_misspecified} presents the result of an experiment performed in the over-dispersed $\rho=0.65$ case, where the model cannot correctly represent the true distribution. We perform the experiments with the beta prior of $(\alpha, \beta)=(1,1),(10,1),(20,1)$ and again compare $\Varpos[q]$, $\hat{\Sigma}^{qq}$, and $\hat{\Sigma}^*_{qq}$. The results are similar to the case $\rho=0$, except that the genuine Bayesian estimator $\Varpos[q]$ massively underestimates the variance in all cases of $\alpha$. 

\subsection{$O(n)$ priors}
\label{sec:app:O(n)}

Here, we compare two asymptotic settings that differ in how the prior scales with the sample size: the standard case with an \( O(1) \) prior and the case with an \( O(n) \) prior.

\subsubsection*{\protect{$O(1)$} Priors}
First, we consider the case where the prior is independent of the sample size $n$. Hereafter we refer to this setting as {\it $O(1)$ priors}.  In this case, the effect of the centering is asymptotically negligible for a large $n$, that is, the relation \eqref{eq:cov_zero}
\begin{align}
    \frac{1}{n} \sum_{i=1}^n \Covpos[A(\theta),\,\logp{X_{i}}{\theta}]
    = \Covpos \left [A(\theta), \,\frac{1}{n} \sum_{i=1}^n \logp{X_{i}}{\theta} \right ] 
    = o_p(1/n)
\end{align}
holds in \eqref{eq:1st_star}, while the first term in \eqref{eq:1st_star} is $O_p(1/n)$. An initiative argument to justify \eqref{eq:cov_zero} is provided in Appendix~\ref{sec:app:cov_zero}; in a nutshell, it is caused by the posterior concentration around the MLE $\hat{\theta}$ that satisfies $\sum_{i=1}^n \logp{X_{i}}{\hat{\theta}} = 0$. 

\subsubsection*{\protect{$O(n)$} Priors}

Under the setting of $O(1)$ prior, the effect of the prior usually vanishes in $n \rightarrow \infty$ for regular models. To quantify the effect of the prior more precisely, a ``$O(n)$  setting'' is introduced in the literature (\cite{Konishi_Kitagawa_book,Ninomiya_2021}), where the magnitude of the prior is intensified as $O(n)$ in the $n \rightarrow \infty$ limit; hereafter we refer to it as {\it $O(n)$ priors}. To be explicit, we define
\begin{align}
    \log p(\theta)= n \Lambda \log \tilde{p}(\theta)+\log p_0(\theta)+ c
\end{align}
with densities $ \tilde{p}(\theta), p_0(\theta)$, and constants $\Lambda$ and $c$. The first term $n \Lambda \log \tilde{p}(\theta)$ is the main term that represents an informative prior.  We also introduce a constant $\Lambda$ to control the strength of the informative prior. $p_0(\theta)$ is a ``base prior'' included as a formality and is often neglected hereafter. The constant $c$ is used for normalizing $p(\theta)$. We maintain the number of parameters constant when $n \rightarrow \infty$ as in the classical setting of $O(1)$ priors.   

Under the assumption of the $O(n)$ prior, we observe the following:
\begin{itemize}
    \item [(a)] The original formula \eqref{eq:cov_g_star} with centering provides an asymptotically correct estimate of the frequentist covariance even in the $O(n)$ setting. 
    \item [(b)] The simplified formula \eqref{eq:cov_g} without centering does not provide an asymptotically correct estimate of the frequentist covariance in the $O(n)$ setting; in other words, $n$ times \eqref{eq:cov_g} is not a consistent estimator of $n$ times the frequentist covariance.
    \item [(c)] However, the difference between \eqref{eq:cov_g_star} and \eqref{eq:cov_g} is $O_p(\Lambda^2)$ when we ignore $o_p(1/n)$ terms; it is different from $O_p(\Lambda)$, which we had expected. When $\Lambda$ is small, it results in a considerably smaller correction compared with cases wherein the difference is $O_p(\Lambda)$.
\end{itemize}
The property (a) is a natural consequence of the concentration of the posterior measure in the $O(n)$ setting; a generalization of the heuristic argument in Section~2.1 of \cite{Bayesian_IJK} to $O(n)$ priors is provided in Section~\ref{sec:app:covariance}. Derivations of the properties (b) and (c) are presented in Section~\ref{sec:app:centering}. 

The properties (a) and (b) are widely applied in applications of the Bayesian IJ that are based on \eqref{eq:1st_star}.  However, the third property (c) is specific to the application to the frequentist covariance. A similar property also holds for the bias correction term of \WAIC (Section~\ref{sec:app:model_selection}), but it may not hold in the case of other outcomes of the Bayesian IJ. 

These properties provide qualitative support for the intuition that the Bayesian IJ includes the effect of the prior, although it depends on the prior only through the definition of the posterior mean. The centering in \eqref{eq:1st_star} improves the results for an informative prior, but it captures a considerable amount of the prior information even without centering, as the difference between \eqref{eq:cov_g_star} and \eqref{eq:cov_g} is $O_p(\Lambda^2)$.   

\subsection{An alternative form of frequentist covariance formula}

In the following, we present an alternative form of the frequentist covariance formula \eqref{eq:cov_g_star}. This alternative expression will be used throughout this supplement.

Under the setting of \( O(n) \) priors, the right-hand side of \eqref{eq:1st_star} is asymptotically equivalent to that of the following expression:
\begin{align}
\label{eq:1st_star_star}
\Covpos^{**}[A(\theta), \logp{X_{i}}{\theta}]  
& = \Covpos \left [A(\theta), \logp{X_{i}}{\theta}  
+ \frac{1}{n} \log p(\theta) \right ],
\end{align}
which leads to an alternative estimator of the frequentist covariance:
\noeqref{eq:cov_g_star_star}
\begin{align}
\label{eq:cov_g_star_star}
   \hat{\Sigma}^{**}_{AB} & =
   \sum_{i=1}^n \, {
   \Covpos \left [A(\theta), \logp{X_{i}}{\theta}+\frac{1}{n} \log p(\theta) \right  ] \, 
   \Covpos \left [B(\theta), \logp{X_{i}}{\theta}+\frac{1}{n} \log p(\theta)  \right ] }.
\end{align}
The definition \eqref{eq:1st_star_star} effectively adds \( 1/n \) of the log prior to the log-likelihood of each observation, which appears to be a natural modification when informative priors are used.

To derive the asymptotic equivalence, we begin from a relation that holds for $O(n)$ priors.
For $O(n)$ priors, the expansion \eqref{eq:lpos_expand} no longer holds. However, if we define 
\begin{align}
l^*_{pos}(\theta)
= \frac{1}{n}\sum_{i=1}^n \left\{ \logp{X_{i}}{\theta}  +  \frac{1}{n} \log p(\theta) \right\}
=\frac{1}{n}\sum_{i=1}^n \{ \logp{X_{i}}{\theta}  + \Lambda \log \tilde{p}(\theta) \} +o(1/n)
\end{align}
and make $\theta^*$ a MAP estimate that satisfies  $(l^*_{pos})'(\theta^*)=0$, we can proceed in the same manner, which gives us the relation 
\begin{align}
\label{eq:cov_star_zero}
 \Covpos \left [A(\theta), \sum_{i=1}^n \left\{ \logp{X_{i}}{\theta}  +  \frac{1}{n} \log p(\theta) \right\} \right ]
 =
 \Covpos \left [A(\theta), nl^*_{pos}(\theta) \right ]  =o_p(1).
\end{align}
The relation \eqref{eq:cov_star_zero} also plays an important role throughout this supplement.

Using \eqref{eq:cov_star_zero}, we obtain 
\begin{align}
- \frac{1}{n} \sum_{i=1}^n \Covpos \left [A(\theta), \logp{X_{i}}{\theta}  
\right ] =\Covpos \left [A(\theta), \frac{1}{n}\log p(\theta) \right ]+o_p(1/n), 
\end{align}
which results in an asymptotic equivalence between \eqref{eq:1st_star} and \eqref{eq:1st_star_star}.

\begin{remark}
    The modification \eqref{eq:1st_star_star} is already discussed in \cite{Iba_Yano_arXiv2} in the context of information criteria. In general, the effect of $O(n)$ priors on model selection is studied in \cite{Konishi_Genshiro_1996}, \cite{Vehtari_etal_2017}, and \cite{Ninomiya_2021}. We discuss this issue in greater detail in Section~\ref{sec:app:model_selection}. 
\end{remark}

\subsection{Consistency of {\protect \eqref{eq:cov_g_star} with $O(n)$ Priors}}
\label{sec:app:covariance}

An informal argument for the alternative form \eqref{eq:cov_g_star_star} of the frequentist covariance formula is provided. This argument shows that the expression \eqref{eq:cov_g_star_star}, and therefore \eqref{eq:cov_g_star} in the main text, is asymptotically valid under the setting of \( O(n) \) priors. Equivalently, \( n \) times \eqref{eq:cov_g_star} serves as a consistent estimator of \( n \) times the frequentist covariance in this setting.

\subsubsection*{Preliminary}

In the definition
$
    \log p(\theta)= n \Lambda \log \tilde{p}(\theta)+\log p_0(\theta)+ c
$
of an $O(n)$ prior, we set $p_0(\theta)=1$ and ignore the normalizing constant $c$, which does not affect the result. Then, the prior and resultant posterior are written as
\begin{align}
    \log p(\theta)=n \Lambda \log \tilde{p}(\theta),
\end{align}
and
\begin{align}
    p(\theta \mid X^{n} ) = 
    \frac{ \exp\{\sum_{i=1}^{n} \logp{X_{i}}{\theta} +n \Lambda \log \tilde{p}(\theta) \}}
    {\int \exp\{\sum_{i=1}^{n} \logp{X_{i}}{\theta'} +n \Lambda \log \tilde{p}(\theta') \} d\theta'},
    \label{eq:pos_On}
\end{align}
respectively. We also define the population version of the posterior as
\begin{align}
    p(\theta; G, n) = 
    \frac{\exp\{n \int \logp{x}{\theta} \, dG(x)+n \Lambda \log \tilde{p}(\theta) \}}
    {\int \exp\{n \int \logp{x}{\theta'} \, dG(x)+n \Lambda \log \tilde{p}(\theta') \} \, d\theta'},
\end{align}
with $G$ representing the population. Hereafter, $\EpposP$ denotes the expectation with respect to $p(\theta; G, n)$.

We define $\theta_0^*=(\theta_{0,\alpha}^*)$ as a solution of 
\begin{align} 
\label{eq:theta_star_zero}
    \EpX \left [ \left. 
    \frac{\partial}{\partial\theta_\alpha} \left \{ \logp{X}{\theta}  + \Lambda \log \tilde{p}(\theta) \right  \}\right |_{\theta=\theta_0^*} \right ]=0.
\end{align} 
Using $\theta_0^*$, the information matrices $\mathcal{I}^*$ and $\mathcal{J}^*$ are defined as 
\begin{align}
\label{eq:def_I_star}
    \mathcal{I}^*_{\alpha\beta}=
    \EpX \left [
    \left. 
    \frac{\partial}{\partial\theta_\alpha} \left \{\logp{X}{\theta}+ \Lambda \log \tilde{p}(\theta) \right  \} \right |_{\theta=\theta_0^*}
    \left. 
    \frac{\partial}{\partial\theta_\beta} \left \{ \logp{X}{\theta}+ \Lambda \log \tilde{p}(\theta) \right  \} \right |_{\theta=\theta_0^*}
    \right ],
\end{align}
\begin{align}
\label{eq:def_J_star}
    \mathcal{J}^*_{\alpha\beta}
    = - \EpX \left [ \left.
    \frac{\partial^2}{\partial\theta_\alpha \partial\theta_\beta} \left \{  \logp{X}{\theta}+ \Lambda \log \tilde{p}(\theta) \right  \} \right |_{\theta=\theta_0^*} 
    \right   ].
\end{align}
The derivatives in \eqref{eq:def_I_star} and \eqref{eq:def_J_star} are evaluated at $\theta_0^*$ defined by \eqref{eq:theta_star_zero}, while the derivatives in the population version of \eqref{eq:def_I} and \eqref{eq:def_J} are evaluated at the projection $\theta_0$ to the model. 

We also define empirical versions of $\mathcal{J}^*_{\alpha\beta}$ and $\mathcal{I}^*_{\alpha\beta}$. Let $\theta^*=(\theta^*_\alpha)$ be a MAP estimate with the observations $X^n=(X_i)$, which satisfies
\begin{align}
\label{eq:theta_star}
    \sum_{i=1}^n  
    \left. 
    \frac{\partial}{\partial\theta_\alpha} 
    \left \{ \logp{X_i}{\theta}+ \Lambda \log \tilde{p}(\theta) \right \} \right |_{\theta=\theta^*}=0.
\end{align}
We then define
\begin{align}
\label{eq:def_I_star_emp}
    \hat{\mathcal{I}}^*_{\alpha\beta}=
    \frac{1}{n}
    \sum_{i=1}^n
    \left \{
    \left. 
    \frac{\partial}{\partial\theta_\alpha} \left \{\logp{X_i}{\theta}+ \Lambda \log \tilde{p}(\theta) \right  \} \right |_{\theta=\theta^*}
    \left. 
    \frac{\partial}{\partial\theta_\beta} \left \{ \logp{X_i}{\theta}+ \Lambda \log \tilde{p}(\theta) \right  \} \right |_{\theta=\theta^*}
    \right \},
\end{align}
\begin{align}
\label{eq:def_J_star_emp}
    \hat{\mathcal{J}}^*_{\alpha\beta}
    = - \frac{1}{n} \sum_{i=1}^n  \left.
    \frac{\partial^2}{\partial\theta_\alpha \partial\theta_\beta} \left \{  \logp{X_i}{\theta}+ \Lambda \log \tilde{p}(\theta) \right  \} \right |_{\theta=\theta^*}.
\end{align}
under the setting of $O(n)$ priors. 

Finally, the relation~\eqref{eq:J_pos_basic} is generalized to
\begin{align}
\label{eq:J_pos_basic_star}
\EpposP[(\theta_\alpha-\theta_{0,\alpha}^*)(\theta_\beta-\theta_{0,\beta}^*)]
 & = \frac{(\mathcal{J}^{*})^{-1}_{\alpha\beta}}{n} + o_p(1/n), \\
\label{eq:J_pos_basic_star_2}
\Eppos[(\theta_\alpha-\theta^*_\alpha)(\theta_\beta-\theta^*_\beta)]
 & = \frac{(\hat{\mathcal{J}}^{*})^{-1}_{\alpha\beta}}{n} + o_p(1/n),
\end{align}
under the setting of $O(n)$ priors.
It should be noted that $\Eppos[\theta] = \theta^* + O_p(1/n)$ in our setting, but the error induced by replacing $\Eppos[\theta]$ with $\theta^*$ on the left-hand side of~\eqref{eq:J_pos_basic_star_2} is of order $O_p(1/n^2)$. 
The same observation also applies to~\eqref{eq:J_pos_basic_star}.


\subsubsection*{Sandwiched representation of frequentist covariance}

Based on the equation \eqref{eq:theta_star} for the MAP estimate, we can derive an asymptotic representation of frequentist covariance under the setting of $O(n)$ priors. It is a straightforward generalization of the classical result on the MLE, which corresponds to $\Lambda=0$. First, using the frequentist central limit theorem under the IID assumption of $X_i$s, we obtain
\begin{align}
\label{eq:CLT_theta_star}
    \frac{1}{\sqrt{n}}\sum_{i=1}^n  
    \left. 
    \frac{\partial}{\partial\theta_\alpha} 
    \left \{ \logp{X_i}{\theta}+ \Lambda \log \tilde{p}(\theta) \right \} \right |_{\theta=\theta_0^*}
    \sim  N(0,\mathcal{I}^*),
\end{align}
using \eqref{eq:theta_star_zero} and \eqref{eq:def_J_star}.

On expanding the left hand side of the equation \eqref{eq:theta_star} around $\theta_0^*$ and transposing a term, we obtain 
\begin{align}
\label{eq:s_theta_star}
    & \frac{1}{\sqrt{n}}\sum_{i=1}^n  
    \left. 
    \frac{\partial}{\partial\theta_\alpha} 
    \left \{ \logp{X_i}{\theta}+ \Lambda \log \tilde{p}(\theta) \right \} \right |_{\theta=\theta_0^*}
    \\ & =
    -\frac{1}{n}\sum_{\beta=1}^k \left [ \left.
    \frac{\partial^2}{\partial\theta_\alpha \partial\theta_\beta}  \left \{ \sum_{i=1}^n \left \{ \logp{X_i}{\theta}+ \Lambda \log \tilde{p}(\theta) \right  \} \right \} \right |_{\theta=\theta_0^*}  \!\!\! \sqrt{n}(\theta^*-\theta_0^*)_\beta \right ] + o_p(1).
\end{align}
On taking the variance of both sides of \eqref{eq:s_theta_star}
using \eqref{eq:CLT_theta_star}, the frequentist covariance of $\theta^*$ is computed as
\begin{align}
    \CovX[(\theta^*-\theta_0^*)_\alpha,(\theta^*-\theta_0^*)_\beta] =
    \frac{1}{n}\sum_{\mathstrut\gamma,\gamma'=1}^k
    ({\mathcal{J}}^*)^{-1}_{\alpha\gamma} \,
    {\mathcal{I}}^*_{\gamma\gamma'} \,
    ({\mathcal{J}}^*)^{-1}_{\gamma'\beta} \,
    +o(1/n).
\end{align}
Thus, for arbitrary statistics $A(\theta)$ and $B(\theta)$, the delta method gives us 
\begin{align}
\label{eq:conventional_covariance}
    \CovX[A(\theta^*),B(\theta^*)] =
    \frac{1}{n}
    \sum_{\mathstrut\alpha,\beta=1}^k
    \sum_{\mathstrut\gamma,\gamma'=1}^k
    (\mathcal{J}^*)^{-1}_{\alpha\gamma} \,
    \mathcal{I}^*_{\gamma\gamma'} \,
    (\mathcal{J}^*)^{-1}_{\gamma'\beta} \,
    \left. \frac{\partial A}{\partial \theta_\alpha}
    \right |_{\theta=\theta_0^*}
    \left. \frac{\partial B}{\partial \theta_\beta}
    \right |_{\theta=\theta_0^*}
    +o(1/n).
\end{align}
On replacing the population versions $\mathcal{I}^*$ and $\mathcal{J}^*$  with their empirical counterparts $\hat{\mathcal{I}}^*$ and $\hat{\mathcal{J}}^*$, and using the empirical values of the partial derivatives of $A$ and $B$, we obtain 
\begin{align}
\label{eq:conventional_covariance_emp_0}
    \CovX[A(\theta^*),B(\theta^*)] =
    \frac{1}{n}
    \sum_{\mathstrut\alpha,\beta=1}^k
    \sum_{\mathstrut\gamma,\gamma'=1}^k
    (\hat{\mathcal{J}}^*)^{-1}_{\alpha\gamma} \,
    \hat{\mathcal{I}}^*_{\gamma\gamma'} \,
    (\hat{\mathcal{J}}^*)^{-1}_{\gamma'\beta} \,
    \left. \frac{\partial A}{\partial \theta_\alpha}
    \right |_{\theta=\theta^*}
    \left. \frac{\partial B}{\partial \theta_\beta}
    \right |_{\theta=\theta^*}
    +o_p(1/n). \qquad
\end{align}
In our settings, $\Eppos[\theta]=\theta^*+O_p(1/n)$, but the error caused by the substitution of $\theta^*$ with $\Eppos[\theta]$ on the left hand side of \eqref{eq:conventional_covariance_emp_0} is shown to be on the order of $O_p(1/n^2)$. We then obtain
\begin{align}
\label{eq:conventional_covariance_emp}
\qquad
    \CovX[\Eppos[A],\Eppos[B]] =
    \frac{1}{n}
    \sum_{\mathstrut\alpha,\beta=1}^k
    \sum_{\mathstrut\gamma,\gamma'=1}^k
    (\hat{\mathcal{J}}^*)^{-1}_{\alpha\gamma} \,
    \hat{\mathcal{I}}^*_{\gamma\gamma'} \,
    (\hat{\mathcal{J}}^*)^{-1}_{\gamma'\beta} \,
    \left. \frac{\partial A}{\partial \theta_\alpha}
    \right |_{\theta=\theta^*}
    \left. \frac{\partial B}{\partial \theta_\beta}
    \right |_{\theta=\theta^*}
    +o_p(1/n).
\end{align}

\subsubsection*{Frequentist covariance formula in the settings of $O(n)$ priors}

On expanding \eqref{eq:1st_star_star} around $\theta^*$ and using \eqref{eq:J_pos_basic_star_2}, we obtain
\begin{align}
   & \Covpos [  A(\theta),  \logp{X_{i}}{\theta}+\Lambda \log \tilde{p}(\theta) ] \\ 
   & = 
   \Covpos \left [ 
   \sum_{\alpha=1}^k
   \left. \frac{\partial A}{\partial \theta_\alpha}
   \right |_{\theta=\theta^*} \!\!\!\!
   (\theta-\theta^*)_\alpha
   , \,\,\,
   \sum_{\beta=1}^k 
   \left. \left \{ \frac{\partial}{\partial\theta_\beta} \logp{X_i}{\theta} 
   +\Lambda \log \tilde{p}(\theta) \right \}
   \right |_{\theta=\theta^*} 
   (\theta-\theta^*)_\beta
   \right ]
   +o_p(1/n) \\
   & = 
   \frac{1}{n} \sum_{\alpha, \beta=1}^k
   \left. \frac{\partial A}{\partial \theta_\alpha}
   \right |_{\theta=\theta^*} 
   \left. \left \{ \frac{\partial}{\partial\theta_\beta} \logp{X_i}{\theta} 
   +\Lambda \log \tilde{p}(\theta) \right \}
   \right |_{\theta=\theta^*} 
   (\hat{\mathcal{J}}^*)^{-1}_{\alpha\beta}
   +o_p(1/n).
\shortintertext{Let us substitute this into \eqref{eq:cov_g_star_star} and use the definition \eqref{eq:def_J_star_emp}, then we have} \label{eq:cov_g_star_star_expand}
\hat{\Sigma}^{**}_{AB}
    & = \sum_{i=1}^n  \Covpos[A(\theta),  \logp{X_{i}}{\theta}+\Lambda \log \tilde{p}(\theta)
    ] \, \Covpos[B(\theta),  \logp{X_{i}}{\theta}+\Lambda \log \tilde{p}(\theta)]
    \\ 
    & =\frac{1}{n} \sum_{\mathstrut\alpha,\beta=1}^k
    \sum_{\mathstrut\gamma,\gamma'=1}^k
    (\hat{\mathcal{J}}^*)^{-1}_{\alpha\gamma} \,
    \hat{\mathcal{I}}^*_{\gamma\gamma'} \,
    (\hat{\mathcal{J}}^*)^{-1}_{\gamma'\beta} \,
    \left. \frac{\partial A}{\partial \theta_\alpha}
    \right |_{\theta=\theta^*}
    \left. \frac{\partial B}{\partial \theta_\beta}
    \right |_{\theta=\theta^*}
    +o_p(1/n),
\end{align}
which coincides with \eqref{eq:conventional_covariance_emp}.

\subsection{The Effect of the Centering and Inconsistency of {\protect \eqref{eq:cov_g} with $O(n)$ Priors}}
\label{sec:app:centering}

In the above, we argued that the frequentist covariance formulae \eqref{eq:cov_g_star} and \eqref{eq:cov_g_star_star}  are correct in the order of $o_p(1/n)$ under the setting of $O(n)$ priors. Now, we investigate the order of the change from the simplified version \eqref{eq:cov_g} to the original version \eqref{eq:cov_g_star} in the strength $\Lambda$ of the prior and sample size $n$. 

Let us start from \eqref{eq:cov_star_zero} and rewrite it as
\begin{align}
\label{eq:sum_lambda}
 \Covpos \left [A(\theta), \sum_{i=1}^n \logp{X_{i}}{\theta}  
 \right ] =-n\Lambda \Covpos[A(\theta),\log \tilde{p}(\theta)] +o_p(1).
\end{align}
Then, a cross term in \eqref{eq:cov_g_star_star} is evaluated as
\begin{align}
\label{eq:cross}
\sum_{i=1}^n
\Covpos[A(\theta), & \logp{X_{i}}{\theta} ] \,\, 
\Covpos[B(\theta), \Lambda \log \tilde{p}(\theta) ] 
\\
& =
\Lambda \Covpos \left [A(\theta), \sum_{i=1}^n \logp{X_{i}}{\theta} \right ] 
\Covpos[B(\theta), \log \tilde{p}(\theta) ] \,\, 
\\
& =
\bigg \{ \!\! -\Lambda^2 n \, \Covpos[A(\theta), \log \tilde{p}(\theta) ]+  o_p(1) \bigg\} \,\,
\Covpos[B(\theta), \log \tilde{p}(\theta) ],
\shortintertext{where the $o_p(1)$ term in $\{\,\}$ is $O(\Lambda)$. Another cross term, where the role of $A(\theta)$ and $B(\theta)$ is reversed, is also evaluated in a similar way. On the other hand, the fourth term that does not contains the likelihood is evaluated as}
\label{eq:diag}
\sum_{i=1}^n  
\Covpos[A(\theta), & \Lambda \log \tilde{p}(\theta) ] \, \,
\Covpos[B(\theta), \Lambda \log \tilde{p}(\theta) ] \\
& = \Lambda^2 n 
\Covpos[A(\theta), \log \tilde{p}(\theta) ]
\Covpos[B(\theta), \log \tilde{p}(\theta) ].
\end{align}
On calculating $2\times \eqref{eq:cross}+\eqref{eq:diag}$ and adding it to \eqref{eq:cov_g}, we have
\begin{align}
\label{eq:Lambda2}
   \hat{\Sigma}^{**}_{AB} =
   & \sum_{i=1}^n  
   \Covpos[A(\theta), \logp{X_{i}}{\theta} ] \, \,
   \Covpos[B(\theta), \logp{X_{i}}{\theta} ]
\\  
   & \qquad \qquad -
   \Lambda^2 n 
   \Covpos[A(\theta), \log \tilde{p}(\theta) ]
   \Covpos[B(\theta), \log \tilde{p}(\theta) ]
   + o_p(1/n).
\end{align}
When we focus on the order in $\Lambda$, the above formula indicates that the difference between $\hat{\Sigma}^{**}_{AB}$ and $\hat{\Sigma}_{AB}$ is $O_p(\Lambda^2)$, when we ignore $o_p(1/n)$ contributions. This means that the difference between $\hat{\Sigma}^{*}_{AB}$ given by \eqref{eq:cov_g_star} and $\hat{\Sigma}_{AB}$ is also $O_p(\Lambda^2)$. It is different from $O_p(\Lambda)$, which might be expected.

However, when we focus on the order in $n$, the difference between $\hat{\Sigma}^{**}_{AB}$ and $\hat{\Sigma}_{AB}$ is $O_p(1/n)$, because both $\Covpos[A(\theta), \log \tilde{p}(\theta) ]$ and $\Covpos[A(\theta), \log \tilde{p}(\theta)]$ are on the order of $O_p(1/n)$ for generic statistics $A$ and $B$. 

These results suggest that the difference between \eqref{eq:cov_g} and \eqref{eq:cov_g_star} is only relevant for highly informative priors. The majority of the effect of the informative priors is already contained in a simplified form \eqref{eq:cov_g} through the definition of the posterior covariance. From another perspective, however, the same result also indicates that the difference between \eqref{eq:cov_g_star} and \eqref{eq:cov_g} is on the order of $O_p(1/n)$, while \eqref{eq:cov_g_star} and \eqref{eq:cov_g} are also on the order of $O_p(1/n)$. This means that $n$ times \eqref{eq:cov_g} is not a consistent estimator of $n$ times the frequentist covariance under the setting of $O(n)$ priors, when we consider the limit of $n \rightarrow \infty$.

\subsection{Model Selection}
\label{sec:app:model_selection}

\subsubsection*{Bias correction term of \WAIC}

Finally, we touch on related issues in model selection. As discussed in \cite{Vehtari_etal_2017} and \cite{Ninomiya_2021}, \WAIC is not an unbiased estimator of the predictive loss under the setting of $O(n)$ prior. \cite{Iba_Yano_arXiv2} proposed a modification
\begin{align}
\label{eq:PCIC}
\sum_{i=1}^n & \Covpos \left [\logp{X_{i}}{\theta}, \logp{X_{i}}{\theta}  
+ \frac{1}{n} \log p(\theta) \right ]
\\
& = \sum_{i=1}^n \Covpos[\logp{X_{i}}{\theta}, \logp{X_{i}}{\theta}  
+ \Lambda \log \tilde{p}(\theta) ]
\end{align}
of the bias correction term of \WAIC, so as to provide a unbiased estimator up to the order of $o_p(1)$ under the setting of $O(n)$ priors; it has the form of the bias correction term of \PCIC (\cite{Iba_Yano_arXiv, Iba_Yano_arXiv2}), and can also be considered an analog of the form \eqref{eq:cov_g_star_star} of the frequentist covariance formula. In \cite{Iba_Yano_arXiv2}, the Gibbs predictive loss for a specific example is discussed, but it is generalized to the predictive loss and also other models including singular models.  

However, the difference between \eqref{eq:PCIC} and the bias correction term of \WAIC
\begin{align}
\label{eq:WAIC}
\sum_{i=1}^n \Varpos [\logp{X_{i}}{\theta}]
= \sum_{i=1}^n \Covpos [\logp{X_{i}}{\theta}, \logp{X_{i}}{\theta}]
\end{align}
is usually small, except that the effect of the informative prior is very strong. To explain this, we argue that the difference between \eqref{eq:PCIC} and \eqref{eq:WAIC} is on the order of $O(\Lambda^2)$.
If we set $A(\theta)=\Lambda \log \tilde{p}(\theta)$ in \eqref{eq:sum_lambda}, we can express the difference between \eqref{eq:PCIC} and \eqref{eq:WAIC} as
\begin{align}
\Covpos \left [ \sum_{i=1}^n \logp{X_{i}}{\theta}, \Lambda \log \tilde{p}(\theta) \right ] 
& =
-n\Lambda \Covpos[\Lambda \log \tilde{p}(\theta),\log \tilde{p}(\theta)] +o_p(1)
\\
& =
- \Lambda^2 n \Varpos[\log \tilde{p}(\theta)]+o_p(1),
\end{align}
which gives us the desired result.

The difference between \TIC and \WAIC in regular models is usually on the order of $O(\Lambda)$ under the setting of $O(n)$ priors. Combined with the above result, this indicates that a considerable amount of the effect of the prior is already contained in the bias correction term of \WAIC through the definition of the posterior variance.   

\begin{remark} 
The relations \eqref{eq:cov_star_zero} and \eqref{eq:sum_lambda} for the general $A(\theta)$ do not hold in singular models. However, we conjecture that a specialized form of \eqref{eq:cov_star_zero}
\begin{align}
\label{eq:cov_star_zero_like1}
 \Covpos \left [\logp{X_{j}}{\theta} , \sum_{i=1}^n \{ \logp{X_{i}}{\theta}  
+ \Lambda \log \tilde{p}(\theta) \} \right ] =o_p(1),
\end{align}
\begin{align}
\label{eq:cov_star_zero_like2}
 \Covpos \left [\log \tilde{p}(\theta) , \sum_{i=1}^n \{ \logp{X_{i}}{\theta}  
+ \Lambda \log \tilde{p}(\theta) \} \right ] =o_p(1)
\end{align}
will hold for singular models, and it is sufficient to show that the difference between \eqref{eq:PCIC} and \eqref{eq:WAIC} is $O_p(\Lambda^2)$. 
\end{remark}

\subsubsection*{A view from information matrices}

In the above derivation, we directly treat the problem in a posterior covariance form and do not use an expansion in terms of the parameters; hence, we do not use information matrices. Here we briefly discuss it from the perspective of modifications in information matrices.

Let us consider the population version for simplicity. We define a set of modified information matrices $\mathcal{I}^\dagger$ and $\mathcal{J}^\dagger$ as
\begin{align}
\label{eq:def_I_dagger}
    \mathcal{I}^\dagger_{\alpha\beta}=
    \EpX \left [
    \left. \frac{\partial}{\partial\theta_\alpha}  
    \logp{X}{\theta} \right |_{\theta=\theta_0^*}
    \left. \frac{\partial}{\partial\theta_\beta}  
    \logp{X}{\theta} \right |_{\theta=\theta_0^*}
    \right ],
\end{align}
\begin{align}
\label{eq:def_J_dagger}
    \mathcal{J}^\dagger_{\alpha\beta}=- \EpX \left [ \left.
    \frac{\partial^2}{\partial\theta_\alpha \partial\theta_\beta}  \logp{X}{\theta} \right |_{\theta=\theta_0^*}
    \right ].
\end{align}
They have the same form as the usual information matrices $\mathcal{I}$ and $\mathcal{J}$, 
which are defined as the population versions of $\hat{\mathcal{I}}$ and $\hat{\mathcal{J}}$, 
but are evaluated at $\theta = \theta_0^*$ defined by \eqref{eq:theta_star_zero} instead of $\theta = \theta_0$.

After some computation, we obtain
\begin{align}
\mathcal{I}^*_{\alpha\beta}=\mathcal{I}^\dagger_{\alpha\beta}
    - \Lambda^2
    \left. \frac{\partial}{\partial\theta_\alpha}\log \tilde{p}(\theta)\right |_{\theta=\theta_0^*}
    \left. \frac{\partial}{\partial\theta_\beta}\log \tilde{p}(\theta)\right |_{\theta=\theta_0^*},
\end{align}
and we can directly obtain
\begin{align}
    \mathcal{J}^*_{\alpha\beta}=
    \mathcal{J}^\dagger_{\alpha\beta}
    - \Lambda
    \left. \frac{\partial^2}{\partial\theta_\alpha \partial\theta_\beta} \log \tilde{p}(\theta) \right |_{\theta=\theta_0^*}.
\end{align}
These results show that the difference between $\mathcal{I}^*$ and $\mathcal{I}^\dagger$ is on the order of $O(\Lambda^2)$, while the difference between $\mathcal{J}^*$ and $\mathcal{J}^\dagger$ is on the order of $O(\Lambda)$.

Using a Taylor expansion with the parameters, we can show that the difference between \PCIC with a bias correction term \eqref{eq:PCIC} and \TIC is given by the following modifications: (i) change from $\hat{\theta}$ to $\theta^*$, (ii) change from $\hat{\mathcal{J}}^\dagger$ to $\hat{\mathcal{J}}^*$, and (iii) change from $\hat{\mathcal{I}}^\dagger$ to $\hat{\mathcal{I}}^*$, where $\hat{\mathcal{I}}^\dagger$ and $\hat{\mathcal{J}}^\dagger$ are empirical versions of $\mathcal{I}^\dagger$ and $\mathcal{J}^\dagger$ defined around the MAP estimate $\theta=\theta^*$. The effect of (i) and (ii) is already contained in the bias correction term \eqref{eq:WAIC} of \WAIC ; it is automatically included through the definition of the posterior means. Thus, the modification caused by (iii) is the only source of the difference between \WAIC and \PCIC; it is of the order of $O(\Lambda^2)$, as indicated by the above argument.

\end{document}